\documentclass[twocolumn,tighten,twocolappendix]{aastex631}
\usepackage{CJKutf8}
\hypersetup{linkcolor=red,citecolor=blue,filecolor=cyan,urlcolor=blue}

\submitjournal{\apj}

\usepackage{graphics}
\usepackage{amsmath}
\usepackage{physics}
\usepackage{enumitem}
\usepackage{placeins}
\usepackage{makecell}

\definecolor{lightblue}{RGB}{172, 215, 230}
\definecolor{lightpink}{RGB}{255, 182, 193}
\definecolor{lightgreen}{RGB}{144, 237, 144}
\definecolor{crimson}{RGB}{220, 20, 60}

\shorttitle{Origins of Outflowing Cold Clouds}
\shortauthors{PENG et al.}
\graphicspath{{./}{figures/}}

\begin{document}
\begin{CJK*}{UTF8}{gbsn}

\title{Physical Origins of Outflowing Cold Clouds in Local Star-forming Dwarf Galaxies}

\author[0000-0003-3467-6810]{Zixuan Peng (彭子轩)}
\affiliation{Department of Physics, Broida Hall, University of California at Santa Barbara, Santa Barbara, CA 93106, USA}
\correspondingauthor{Zixuan Peng (彭子轩)}
\email{zixuanpeng@ucsb.edu}

\author[0000-0001-9189-7818]{Crystal L. Martin}
\affiliation{Department of Physics, Broida Hall, University of California at Santa Barbara, Santa Barbara, CA 93106, USA}

\author[0000-0001-8755-3836]{Zirui Chen}
\affiliation{Department of Physics, Broida Hall, University of California at Santa Barbara, Santa Barbara, CA 93106, USA}

\author[0000-0003-3806-8548]{Drummond B. Fielding}
\affiliation{Center for Computational Astrophysics, Flatiron Institute, 162 Fifth Avenue, New York, NY 10010, USA}

\author[0000-0002-9217-7051]{Xinfeng Xu}
\affiliation{Department of Physics and Astronomy, Northwestern University, 2145 Sheridan Road, Evanston, IL 60208, USA}

\author[0000-0001-6670-6370]{Timothy Heckman}
\affiliation{Center for Astrophysical Sciences, Department of Physics \& Astronomy, Johns Hopkins University, Baltimore, MD 21218, USA}
\affiliation{School of Earth and Space Exploration, Arizona State University, Tempe, AZ 85287, USA}

\author[0000-0002-9190-9986]{Lise Ramambason}
\affiliation{Institut fur Theoretische Astrophysik, Zentrum für Astronomie, Universität Heidelberg, Albert-Ueberle-Str. 2, D-69120 Heidelberg, Germany}

\author[0000-0001-8237-1515]{Yuan Li (李远)}
\affiliation{Department of Physics, Broida Hall, University of California at Santa Barbara, Santa Barbara, CA 93106, USA}

\author[0000-0003-4166-2855]{Cody Carr}
\affiliation{Center for Cosmology and Computational Astrophysics, Institute for Advanced Study in Physics, Zhejiang University, Hangzhou 310058, China}
\affiliation{Institute of Astronomy, School of Physics, Zhejiang University, Hangzhou 310058, China}

\author[0000-0003-3424-3230]{Weida Hu}
\affiliation{Department of Physics and Astronomy, Texas A\&M University, College Station, TX 77843-4242, USA}
\affiliation{George P. and Cynthia Woods Mitchell Institute for Fundamental Physics and Astronomy, Texas A\&M University, College Station, TX}

\author[0000-0002-2178-5471]{Zuyi Chen}
\affiliation{Steward Observatory, University of Arizona, 933 N Cherry Ave, Tucson, AZ 85721, USA}

\author[0000-0002-9136-8876]{Claudia Scarlata}
\affiliation{Minnesota Institute for Astrophysics, University of Minnesota, 116 Church Street SE, Minneapolis, MN 55455, USA}

\author[0000-0002-6586-4446]{Alaina Henry}
\affiliation{Center for Astrophysical Sciences, Department of Physics \& Astronomy, Johns Hopkins University, Baltimore, MD 21218, USA}
\affiliation{Space Telescope Science Institute, 3700 San Martin Drive, Baltimore, MD 21218, USA}






\begin{abstract}
We study the physical origins of outflowing cold clouds in a sample of 14 low-redshift dwarf ($M_{\ast} \lesssim 10^{10} \ M_{\odot}$) galaxies from the COS Legacy Archive Spectroscopic SurveY (CLASSY) using Keck/ESI data.
Outflows are traced by broad (FWHM $\sim 260 \ \rm{km \ s^{-1}}$) and very-broad (VB; FWHM $\sim 1200 \ \rm{km \ s^{-1}}$) velocity components in strong emission lines like [O\,{\footnotesize III}] $\lambda 5007$ and $\rm{H}\alpha$.
The maximum velocities ($v_{\rm{max}}$) of broad components correlate positively with SFR, unlike the anti-correlation observed for VB components, and are consistent with superbubble models.
In contrast, supernova-driven galactic wind models better reproduce the $v_{\rm{max}}$ of VB components. 
Direct radiative cooling from a hot wind significantly underestimates the luminosities of both broad and VB components. 
A multi-phase wind model with turbulent radiative mixing reduces this discrepancy to at least one dex for most VB components. 
%
Stellar photoionization likely provides additional energy since broad components lie in the starburst locus of excitation diagnostic diagrams.
We propose a novel interpretation of outflow origins in star-forming dwarf galaxies$-$broad components trace expanding superbubble shells, while VB components originate from galactic winds. 
One-zone photoionization models fail to explain the low-ionization lines ([S\,{\footnotesize II}] and [O\,{\footnotesize I}]) of broad components near the maximal starburst regime, which two-zone photoionization models with density-bounded channels instead reproduce.
These two-zone models indicate anisotropic leakage of Lyman continuum photons through low-density channels formed by expanding superbubbles.
Our study highlights extreme outflows ($v_{\rm{max}} \gtrsim 1000 \ \rm{km \ s^{-1}}$) in 9 out of 14 star-forming dwarf galaxies, comparable to AGN-driven winds.
\end{abstract}


\section{Introduction} \label{sec:intro}
\end{CJK*}

Cosmological simulations require galactic winds to produce the observed stellar mass, star formation rate (SFR), and metallicity of galaxies, as well as to match the global scaling relations, such as the mass-metallicity relation \citep{dave_2011_b,dave_2011_a,Torrey_2014,SD_2015}. 
Galactic winds have been ubiquitously detected in starburst galaxies across cosmic time through both absorption and emission lines \citep{Martin_2006, Steidel_2010, Martin_2012}.
These winds manifest in the extremely hot ionized gas ($\sim$ $10^7$ K), traced by X-ray emission \citep[e.g.,][]{Lopez_2020}, and in the cold and denser phase ($\sim$ $10^4$ K) as indicated by $\rm{H}\alpha$ emission \citep[e.g.,][]{Martin_1999}. In edge-on galaxies, emission lines with asymmetric line profiles are easily identified along their minor axis due to the often bipolar morphology of the outflowing gas \citep{heckman_nature_1990}. As a complementary approach, blueshifted UV and optical absorption lines such as Na\,{\footnotesize I}, Mg\,{\footnotesize II}, and Si\,{\footnotesize II} \citep{Martin_2006, Prochaska_2011}, which are observed in face-on systems with absorbing material in front of the continuum source, have been frequently employed to detect galactic winds. 
The different dependencies on gas densities ($n$) for emission ($\propto n^2$) and absorption ($\propto n$) lines complicate the interpretation of outflow properties \citep[e.g., scaling relations with global galaxy properties;][]{Heckman_2015} derived from these profiles, as they may probe distinct regions of the same outflow \citep{Xu_2024}.   

\defcitealias{CC85}{CC85}
From the extremely hot and diffuse phase to the cold and dense phase, emission and absorption lines reveal the multiphase nature of galactic winds. 
This multiphase nature significantly challenges observers when validating existing galactic wind models. Early single-phase models \citep[e.g.,][hereafter CC85]{CC85} predominantly focus on the hot wind, but many observations primarily trace the cold gas. 
Observers often record an outflow mass rate that is orders of magnitude smaller than theoretical models predict since a large fraction of the undetected outflow might reside in other gaseous phases \citep{Concas_2022, TF_2023}.

Comparing theoretical models with observations requires a detailed understanding of the radial structures of density, velocity, pressure, and temperature, as these structures determine the shapes of observed spectral line profiles and their corresponding luminosities.
These radial structures are also intrinsically tied to the long-standing question regarding the physical nature of the observed cold gas, which can be accelerated to $\sim 1000 \ \rm{km \ s^{-1}}$ or even larger \citep{Martin_2015}. 

\defcitealias{Fielding_2022}{FB22}
\defcitealias{Thompson_2016}{T16}
There are two main theoretical perspectives on the physical origins of the outflowing cold clouds:
(1) one emphasizes the entrainment of cold clouds in hot winds, where the acceleration might arise from radiation pressure \citep{Murray_2005, Murray_2011}, ram pressure \citep{Heckman_2011}, or the turbulent radiative mixing between the hot and cold phases \citep[][hereafter FB22]{Binette_2009, GO18, GO20, Schneider_2020, Fielding_2022}; (2) the other attributes the cold clouds to direct radiative cooling from the hot winds \citep{Thompson_2016, Lochhaas_2018, Gray_2019, Danehkar_2021, Danehkar_2022}, which can naturally explain the high velocity of these clouds. 
In both perspectives, radiative cooling plays a pivotal role in determining the ultimate fate of the observed cold clouds. 
Within the entrainment context, the survival and potential growth of cold clouds in the hot winds depend on whether the cooling timescale of the mixed gas between the two phases is shorter than the cloud-crushing timescale \citep{GO18, GO20}. 
From the direct cooling perspective, the radiative cooling time should be shorter than the advection timescale for the hot gas to enter the catastrophic cooling regime and then produce the observed emission and absorption lines in the cold phase \citep[][hereafter T16]{Thompson_2016}.

These outflowing cold clouds might also trace expanding superbubble shells before these shells become Rayleigh-Taylor unstable as they blow out to generate galactic winds \citep{Mac_Low_1988, BB_2013}. 
Recent work on J1044+0353 \citep*{Martin_2024}, a local analog of Reionization-era galaxies (one of our targets, see Section \ref{sec:observation_and_data}), finds the measured outflow velocities of broad components ($\sim 200 \ \rm{km \ s^{-1}}$ FWHM) in strong optical emission and UV absorption lines are consistent with the predicted expansion velocities of superbubble shells \citep{Weaver_1977}.
This finding aligns with \cite{TT_1997}, who argue that the faint and broad wings (broad/narrow intensity ratio $\lesssim 30 \%$) in optical emission lines can result from expanding supershells in the low-metallicity interstellar medium (ISM) where massive OB clusters deposit significant amounts of energy ($E \geq 10^{39} \ \rm{erg \ s^{-1}}$), which delays the fragmentation due to Rayleigh-Taylor instabilities. 

With the launch of JWST, these studies further question the physical origins of outflowing cold clouds in high-redshift galaxies (e.g., whether they are expanding superbubble shells or galactic winds), where recent works find ubiquitous broad velocity components in rest-frame optical emission lines \citep{Carniani_2024, Zhang_2023, Xu_JWST_2023}.  


This work aims to alleviate the physical-origin puzzle of outflowing cold clouds in star-forming galaxies. 
\defcitealias{Xu_2022}{X22}
For this purpose, CLASSY \citep{Berg_2022}, including 45 compact, low-redshift ($0.002 < z < 0.182$) starburst galaxies, is ideal for an in-depth investigation of the origins of outflowing gas for two reasons. 
First, previous studies by \cite{Xu_2022} (hereafter X22) and \cite{Xu_2023} have already obtained many crucial outflow properties, such as outflow velocities and mass outflow rates, by analyzing UV absorption lines across a wide range of ionization states (e.g., C\,{\footnotesize II}, Si\,{\footnotesize II}, Si\,{\footnotesize III}, Si\,{\footnotesize IV}), with ionization potential energies ranging from $\sim$ 10 to 45 eV. 
The physical properties derived from UV absorption lines can be juxtaposed with those inferred from strong optical lines, allowing for a direct comparison to discern whether they share similar physical origins. 
Second, CLASSY spans a substantial range in key galaxy observable properties such as stellar mass ($6.22 < \mathrm{log} (M_{\ast} / M_{\odot}) < 10.06$), gas-phase metallicity (6.98 $<$ 12 + log(O/H) $<$ 8.77), and SFR ($-2 \lesssim \log \rm{SFR / M_{\odot} \ \rm{yr^{-1}}} \lesssim 2$), which is advantageous for determining potential scaling relationships with these global properties and inferring the underlying physical mechanisms. 

The structure of this paper is summarized as follows: 
In Section \ref{sec:observation_and_data}, we present our observations using the Keck Echellette Spectrograph and Imager \citep[ESI;][]{keck_esi}, outline the data reduction procedures we adopted, describe the line-fitting models, define each detected velocity component, and summarize the spectroscopically derived physical properties.
In Section \ref{sec:bpt_diagram}, we utilize strong line ratios to infer the underlying excitation mechanisms. The excitation diagnostic diagrams can help determine whether each velocity component of our targets is dominated by stellar photoionization, shock, or AGN. 
In Section \ref{sec:broad_origins}, using galactic wind models from \citetalias{Thompson_2016} and \citetalias{Fielding_2022}, 
we rigorously evaluate the theoretical perspective that attributes these outflowing clouds to direct condensation from hot winds or turbulent radiative mixing layers (TRMLs) between hot winds and cold clouds. 
The implications of our findings are discussed within the Sections and also in Section \ref{sec:discussion}. We summarize our key results in Section \ref{sec:conclusion}. Throughout this paper, we use $\mathrm{Z_{\odot}} = 0.0134$ for solar metallicity, corresponding to 12 + log(O/H) = 8.69 \citep{Asplund_2021}, and adopt a Flat $\Lambda$CDM cosmology with $\Omega_m = 0.3$, $\Omega_{\Lambda} = 0.7$, and $H_0 = 70 \ \rm{km \ s^{-1} \ Mpc^{-1}}$.

\setlength{\tabcolsep}{6pt} 
\begin{deluxetable*}{cccccccccc}
\tablecaption{Physical Properties of Galaxies from the Literature and SED fittings\label{table:physical_properties}}
\tablehead{
\colhead{Object} & \colhead{$z$} 
& \colhead{$L_{D}$} & \colhead{$r_{50}$} & \colhead{log $M_{\ast}$} & \colhead{12 + log(O/H)} & $E(B-V)_{\rm{MW}}$ & 
\colhead{$v_{\rm{cir}}$} & 
\colhead{log $\dot{\Sigma}_{\ast}$} &
\colhead{$t_{\rm{cos}}$}
\\
\colhead{} & \colhead{} & \colhead{($\rm{Mpc}$)} &
\colhead{($\rm{kpc}$)} &
\colhead{($M_{\odot}$)} &  \colhead{} & \colhead{} & \colhead{($\rm{km \ s^{-1}}$)} & 
\colhead{($\rm{M_{\odot} \ yr^{-1} \ kpc^{-2}}$)} &
\colhead{Myr}
\\
\colhead{(1)} & \colhead{(2)} & \colhead{(3)} & \colhead{(4)} & \colhead{(5)} & \colhead{(6)} & \colhead{(7)} & \colhead{(8)} & \colhead{(9)} &
\colhead{(10)}
}
\startdata
\colhead{J0127--0619} & 0.0054 & 23 & 0.02 & 8.74$^{+ 0.18}_{- 0.15}$ & $7.68 \pm 0.02$ & $0.0372 \pm 0.0008$  & 57.0$^{+ 6.0}_{- 6.3}$ & 2.00$_{- 0.17}^{+ 0.16}$ & 369$^{+ 867}_{-260}$ \\
\colhead{J0144+0453} & 0.0052 & 23 & 0.37 & 7.65$^{+ 0.24}_{- 0.29}$ & $7.76 \pm 0.02$ & $0.0302 \pm 0.0015$ & 27.5$^{+ 5.8}_{- 4.1}$ & -0.75$_{- 0.41}^{+ 0.51}$ & 32$^{+ 47}_{-18}$ \\
\colhead{J0337--0502} & 0.01352 & 58 & 0.45 & 7.06$^{+ 0.24}_{- 0.21}$ & $7.46 \pm 0.04$ & $0.0402 \pm 0.0015$ & 18.5$^{+ 2.7}_{- 2.8}$ & -0.42$_{- 0.84}^{+ 0.43}$ & 31$^{+ 47}_{- 21}$ \\
\colhead{J0808+3948} & 0.09123 & 417 & 0.13 & 9.12$^{+ 0.30}_{- 0.17}$ & $8.77 \pm 0.12$ & $0.0399 \pm 0.0006$ & 73.1$^{+ 8.8}_{- 13.2}$ & 2.25$_{- 0.20}^{+ 0.27}$ & 104$^{+ 224}_{-62}$  \\
\colhead{J0926+4427} & 0.18067 & 875 & 0.66 & 8.76$^{+ 0.30}_{- 0.26}$ & $8.08 \pm 0.02$ & $0.0156 \pm 0.0008$ & 57.8$^{+ 10.9}_{- 10.5}$ & 0.59$_{- 0.14}^{+ 0.16}$ & 69$^{+ 108}_{- 39}$ \\
\colhead{J0938+5428} & 0.10210 & 470 & 0.51 & 9.15$^{+ 0.18}_{- 0.29}$ & $8.25 \pm 0.02$ & $0.0152 \pm 0.0012$ & 75.0$^{+ 15.8}_{- 8.3}$ & 0.84$_{- 0.21}^{+ 0.20}$ & 86$^{+ 95}_{- 48}$ \\
\colhead{J1025+3622} & 0.12650 & 592 & 0.79 & 8.87$^{+ 0.25}_{- 0.27}$ & $8.13 \pm 0.01$ & $0.0097 \pm 0.0006$ & 62.1$^{+ 12.2}_{- 9.5}$ & 0.44$_{- 0.17}^{+ 0.20}$ & 123$^{+ 176}_{-71}$ \\
\colhead{J1044+0353} & 0.01287 & 55 & 0.10 & 6.80$^{+ 0.41}_{- 0.26}$ & $7.45 \pm 0.03$ & $0.0367 \pm 0.0006$ & 15.6$^{+ 3.0}_{- 3.7}$ & 0.62$_{- 0.14}^{+ 0.17}$ & 3$^{+ 15}_{- 1}$ \\
\colhead{J1129+2034} & 0.0047 & 21 & 0.04 & 8.09$^{+ 0.37}_{- 0.27}$ & $8.28 \pm 0.04$ & $0.0173 \pm 0.0013$ & 36.8$^{+ 7.4}_{- 8.1}$ & 1.70$_{- 0.39}^{+ 0.56}$ & 32$^{+ 61}_{-20}$ \\
\colhead{J1200+1343} & 0.06675 & 300 & 0.23 & 8.12$^{+ 0.47}_{- 0.42}$ & $8.26 \pm 0.02$ & $0.0240 \pm 0.0007$  & 37.6$^{+ 12.0}_{- 10.1}$ & 1.25$_{- 0.21}^{+ 0.18}$ & 8$^{+ 29}_{-4}$  \\
\colhead{J1359+5726} & 0.03383 & 148 & 0.74 & 8.41$^{+ 0.31}_{- 0.26}$ & $7.98 \pm 0.01$ & $0.0099 \pm 0.0007$ & 45.5$^{+ 8.6}_{- 8.4}$ & -0.12$_{- 0.20}^{+ 0.14}$ & 97$^{+ 199}_{-64}$ \\
\colhead{J1429+0643} & 0.17350 & 836 & 0.44 & 8.80$^{+ 0.35}_{- 0.21}$ & $8.10 \pm 0.03$ & $0.0216 \pm 0.0008$ & 59.0$^{+ 9.0}_{- 12.3}$ & 1.33$_{- 0.14}^{+ 0.19}$ & 15$^{+ 25}_{-7}$ \\
\colhead{J1444+4237} & 0.0023 & 11 & 7.42 & 6.48$^{+ 0.17}_{- 0.17}$ & $7.64 \pm 0.02$ & $0.0110 \pm 0.0004$ & 12.6$^{+ 1.5}_{- 1.3}$ & -4.48$_{- 0.11}^{+ 0.08}$ & 21$^{+ 40}_{-10}$  \\
\colhead{J1525+0757} & 0.07579 & 343 & 0.36 & 10.06$^{+ 0.28}_{- 0.42}$ & $8.33 \pm 0.04$ & $0.0355 \pm 0.0011$ & 137.3$^{+ 44.6}_{- 23.5}$ & 1.09$_{- 0.70}^{+ 0.26}$ & 243$^{+ 734}_{-171}$  \\
\enddata
\tablecomments{Description for each column: (2) Redshift, \(z\), (3) luminosity distance, $L_{D}$, (5) total stellar mass, $\log M_{\ast}$, and (6) gas-phase metallicity, 12 + log(O/H), of each galaxy are extracted from Tables 5 and 6 in \cite{Berg_2022}. (7) The Galactic foreground extinction, $E(B-V)_{\rm{MW}}$, of each galaxy is obtained from the \cite{schlafly_measuring_2011} dust maps. 
(4) Half-light radius, \(r_{50}\), (8) circular velocity, \(v_{\rm{cir}}\), and (9) SFR surface density, \(\log \dot{\Sigma}_{\ast}\), for each galaxy are taken from Table A2 of \citetalias{Xu_2022}. 
(10) Stellar population age of each galaxy within the HST/COS aperture, $t_{\rm{cos}}$, is derived from \texttt{BEAGLE} SED fitting (see Appendix \ref{sec:sn_rate} for details).
}
\end{deluxetable*}

\begin{figure*}[h!]
\centering
\includegraphics[width=1.0\textwidth]{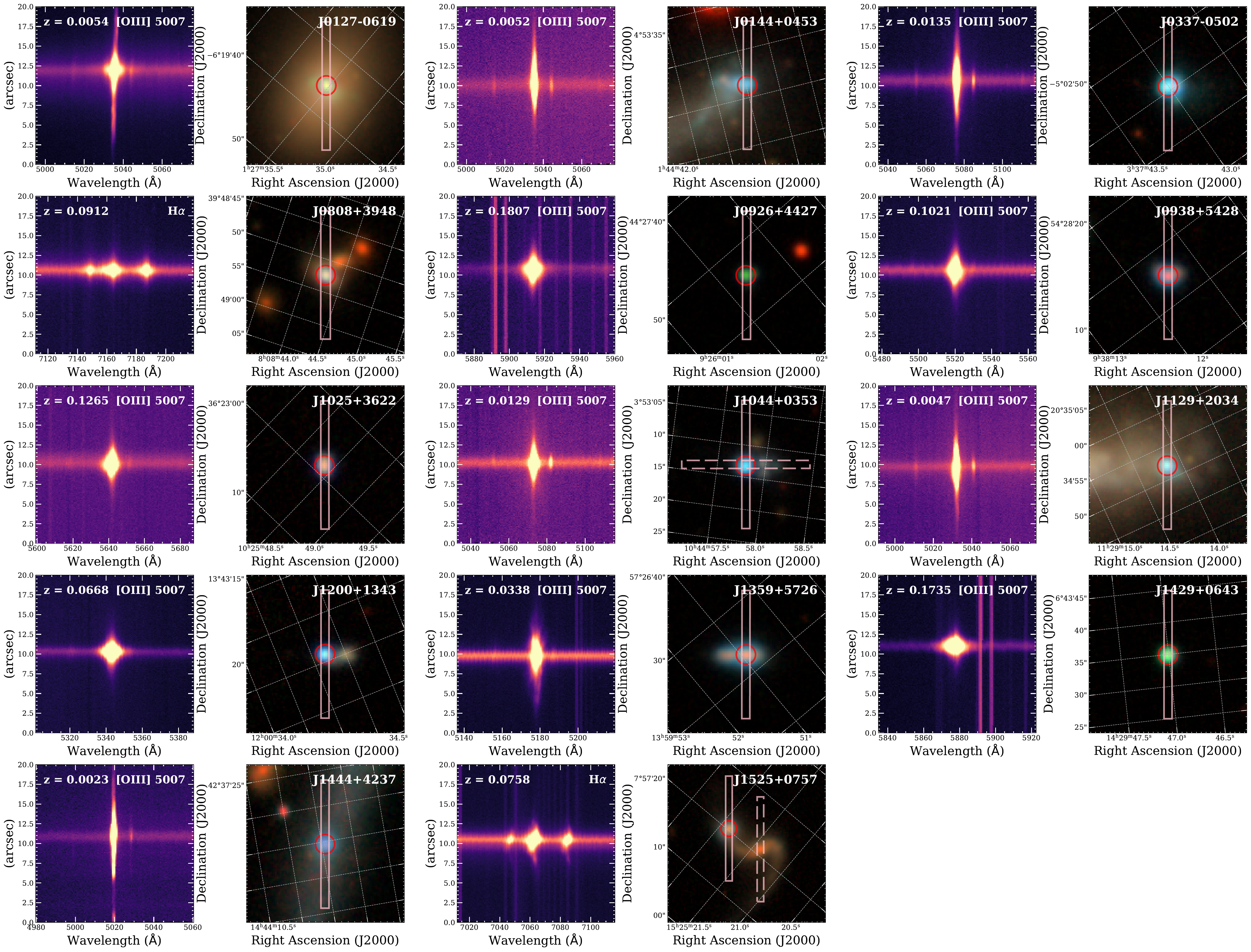}
\caption{Rectified 2D spectra and color-composite plot of each galaxy. For each column, the left panel shows the rectified 2D spectra centered on the strongest emission line ($\rm{H}\alpha$ or [O\,{\footnotesize III}] $\lambda 5007$) of each galaxy. The redshift and the strongest line are labeled in each left panel. The right panel shows the color-composite plot derived from DECalS images. Colors blue, green, and red correspond to the g, r, and z bands, respectively. The flux values are stretched using the \cite{Lupton_2004} method (please refer to their footnote 7). The red circle shows the SDSS circular aperture with a radius of 1\farcs5, while the pink rectangular aperture (both solid and dashed) represents the ESI $1\farcs25 \times 20\farcs0$ slit used in observations. The pink dashed aperture is not used for analysis in this study.
} \label{fig:esi_aperture}
\end{figure*}

\section{Observation and Data Processing} \label{sec:observation_and_data}

\subsection{Keck ESI Observation}\label{subsec:esi_observation}

We used the echelle mode of Keck/ESI, with a wavelength coverage of 390 to 1090 nm, to observe 14 of the 45 CLASSY galaxies (10 targets on January 14, 2023, and 4 more on April 19, 2023; see Table \ref{table:physical_properties}). Based on their outflow velocities measured from UV absorption lines \citepalias{Xu_2022}, this sample includes 8 fast outflow galaxies ($v_{\rm{out}} \ \rm{(AL)} \gtrsim 150 \ \mathrm{km \ s^{-1}}$) and 6 slow outflow galaxies ($v_{\rm{out}} \ \rm{(AL)} \lesssim 60 \ \mathrm{km \ s^{-1}}$; see Table \ref{table:velocity_info}).


For our ESI observations, we chose a $1\farcs25$ $\times$ $20\farcs0$ slit to match the seeing, obtaining a FWHM resolution of $93 \ \mathrm{km \ s^{-1}}$, and a 1 (spatial) $\times$ 2 (spectral) binning to avoid oversampling in the dispersion direction. 
Each slit was centered upon the most luminous stellar complex of each observed galaxy, consistent with the SDSS's (Sloan Digital Sky Survey) PLUG position. 
For targets observed at airmasses of $\lesssim 1.25$, the slit was oriented at a position angle (PA) aligned with the photometric minor axis if the difference between the PA and the parallactic angle \citep{Filippenko_1982}, $\eta$, at half the total integration time is $|\eta - \rm{PA}| \lesssim 45^{\circ}$.
Otherwise, the slit was aligned with $\eta$ to minimize relative light losses due to atmospheric differential refraction.
Each slit is shown as the solid pink rectangular aperture in the color-composite image constructed from the g, r, and z band DECalS images \citep{decals_16, desi_2019} in Figure \ref{fig:esi_aperture}.
For J1044+0353 and J1525+0757, both of which exhibit merger-like morphologies with multiple stellar complexes, we also positioned or oriented the slit to cover one of the off-center stellar complexes, as indicated by the dashed pink rectangular aperture in Figure \ref{fig:esi_aperture}. A detailed analysis of these off-center slits will be addressed in future work. 
Each galaxy was observed with an integration time of approximately 45 minutes, plus an additional 10 minutes on the off-field sky.
Standard star observations of Feige 67 and BD+284211 are taken at the end of the January and April observations, respectively, for absolute flux calibrations.

\subsection{ESI Data Reduction}\label{subsec:esi_data_reduction}
We use the \textsc{Python} package \texttt{PypeIt} \citep{pypeit:zenodo} for basic data processing, including bias subtraction, flat fielding, wavelength calibration, order tracing, and flux sensitivity function generation. 
The residuals between the flux-calibrated ESI spectra of the standard stars and the \texttt{PypeIt} reference spectra are within $\pm 5 \%$ across all ESI orders.
To efficiently remove cosmic rays from each galaxy's 2D spectra, we utilize the package \texttt{astroscrappy} \citep{curtis_mccully_2018_1482019}.

We employ modified \texttt{PypeIt} functions in conjunction with customized \textsc{Python} routines for the remaining reduction steps. 
Each 2D spectrum is rectified to correct for spatial distortions within the prisms and ensure that every dispersion axis follows the same wavelength solution. We adopt the standard extraction method proposed by the VLT/X-Shooter pipeline \citep{Xshooter_pipeline} to avoid ``gaps'' resulting from 2D rectification, which manifest as ``ripples'' in the 1D extracted spectra.
Finally, we extract 1D spectra from each ESI order, stitch all orders to form a continuous spectrum over the entire wavelength range, and coadd all 1D spectra using the \textsc{ech\_combspec} method in \texttt{PypeIt}. 

After obtaining the merged 1D spectra for the object and the sky, we employed the ESO \texttt{SkyCorr} \citep{Noll_2014} package for 1D sky subtraction. This process involves iteratively determining the scaling factors between the 1D sky and object spectra, taken at different times and sky positions, across various groups of sky lines. Here we favor the 1D sky subtraction method over the conventional 2D subtraction due to our targets being spatially extended (see the left panels in Figure \ref{fig:esi_aperture}), filling the entire ESI slit. Additionally, the infrequent sky chopping does not permit reliable direct 2D sky subtraction. 

We scaled the error spectra to account for the correlated noises introduced by the standard extraction method.
The variance rescaling factor is determined by comparing the median variance with the square of the root-mean-square noise of the continuum near strong emission lines. The typical scaling factor is $\sim 1.5$ for the extracted 1D spectra.

To correct for the instrumental broadening $\sigma_{\rm{int}}$, we determine $\sigma_{\rm{int}}$ from the master arc spectrum as $38 \pm 4 \ \rm{km \ s^{-1}}$, representing the median and one-sigma scatter across all orders, consistent with Figure 3 in \cite{keck_esi}. All reported velocity measurements are corrected for $\sigma_{\rm{int}}$ in this work.

\setlength{\tabcolsep}{6pt} 
\begin{deluxetable*}{ccccccc}
\tablecaption{Derived $v_{\rm{max}}$ and FWHM from Optical Emission and UV Absorption Lines (AL)\label{table:velocity_info}}
\tablehead{
\colhead{Object} & \colhead{$v_{\rm{out}}$ (AL)} &  \colhead{$v_{\rm{max}}$ (Broad)} & \colhead{$v_{\rm{max}}$ (VB)} & \colhead{$\rm{FWHM_{out}}$ (AL)} & \colhead{FWHM$_{\rm{out}}$ (Broad)} & \colhead{FWHM$_{\rm{out}}$ (VB)} \\
\colhead{(1)} & \colhead{(2)} & \colhead{(3)} & \colhead{(4)} & \colhead{(5)} & \colhead{(6)} & \colhead{(7)} 
}
\startdata
J0127--0619 & no outflow & $377 \pm 4$ & $2263 \pm 149$ & no outflow & $374 \pm 4$ & $2176 \pm 147$ \\
J0144+0453 & $48 \pm 16$ & \nodata & $815 \pm 231$ & $182 \pm 61$ & \nodata & $789 \pm 229$ \\
J0337--0502\tablenotemark{a} & no outflow & $86 \pm 11$ & $1516 \pm 159$ & no outflow & $86 \pm 11$ & $1516 \pm 159$ \\
J0808+3948 & $646 \pm 65$ & $586 \pm 7$ & \nodata & $382 \pm 127$ & $543 \pm 6$ & \nodata \\
J0926+4427 & $353 \pm 52$ & $370 \pm 5$ & $1141 \pm 88$ & $402 \pm 130$ & $367 \pm 5$ & $1132 \pm 88$ \\
J0938+5428 & $215 \pm 72$ & $311 \pm 9$ & $1051 \pm 248$ & $272 \pm 91$ & $296 \pm 9$ & $1053 \pm 243$ \\
J1025+3622 & $155 \pm 24$ & $204 \pm 6$ & $693 \pm 67$ & $409 \pm 76$ & $203 \pm 6$ & $669 \pm 66$ \\
J1044+0353\tablenotemark{a} & $52 \pm 12$ & $135 \pm 11$ & $2146 \pm 347$ & $123 \pm 24$ & $136 \pm 11$ & $2146 \pm 347$ \\
J1129+2034 & $51 \pm 17$ & $235 \pm 16$ & $1822 \pm 317$ & $144 \pm 15$ & $221 \pm 15$ & $1775 \pm 312$ \\
J1200+1343 & $192 \pm 13$ & $320 \pm 6$ & $1224 \pm 171$ & $306 \pm 62$ & $313 \pm 6$ & $1180 \pm 169$ \\
J1359+5726 & $161 \pm 23$ & $217 \pm 6$ & $1468 \pm 225$ & $359 \pm 70$ & $210 \pm 6$ & $1462 \pm 223$ \\
J1429+0643 & $230 \pm 51$ & $311 \pm 11$ & $614 \pm 30$ & $461 \pm 152$ & $276 \pm 11$ & $604 \pm 30$ \\
J1444+4237\tablenotemark{a} & $54 \pm 18$ & \nodata & $1805 \pm 317$ & $112 \pm 37$ & \nodata & $1765 \pm 316$ \\
J1525+0757 & $408 \pm 28$ & $274 \pm 5$ & $693 \pm 22$ & $495 \pm 59$ & $243 \pm 5$ & $669 \pm 22$ 
\enddata
\tablenotetext{a}{We apply the dGL model to fit [O\,{\footnotesize III}] $\lambda 5007$ and H$\alpha$ for J0337-0502, J1044+0353, and J1444+4237, as these galaxies exhibit non-Gaussian broad wings for their VB components. Their FWHM$_{\rm{out}}$ (VB) are reported as IPV$_{24}$ (VB) instead (see Section \ref{subsubsec:vel_comp_def} for details).}
\tablecomments{Columns (3) and (4) show $v_{\rm{max}}$ (see Equation \ref{eq:2.3.2}) for the broad and VB components. Columns (6) and (7) display the measured FWHMs for the broad and VB components. Columns (1) and (5) show the UV absorption-based outflow velocity, \(v_{\rm{out}} \ \rm{(AL)}\), and \(\rm{FWHM_{out}} \ \rm{(AL)}\) extracted from Table A1 in \citetalias{Xu_2022}. 
\citetalias{Xu_2022} derive $v_{\rm{out}} \ \rm{(AL)}$ ($\rm{FWHM_{out}} \ \rm{(AL)}$) from the median central velocity (FWHM) of the blueshifted second Gaussian from UV absorption troughs that pass the F-test (i.e., double-Gaussian performs statistically better than the single Gaussian model). 
The first Gaussian has a fixed velocity centroid at $0 \ \rm{km \ s^{-1}}$ \citepalias[see Section 4.2 in][]{Xu_2022}.
\citetalias{Xu_2022} define $v_{\rm{max}} \ (\rm{AL}) = v_{\rm{out}} \ (\rm{AL}) + 0.5 \ \rm{FWHM_{out}} \ (\rm{AL})$ $\sim 2 \ v_{\rm{out}} \ (\rm{AL})$. 
This \(v_{\rm{max}} \ \rm{(AL)}\) definition is used to compare the maximum outflow velocities derived from UV absorption and optical emission lines in Section \ref{sec:discussion}. 
J0127-0619 and J0337-0502 are labeled as ``no outflow'' in \citetalias{Xu_2022}.}
\end{deluxetable*}

\subsection{Emission-line Modeling}\label{subsec:line_fitting}

We utilize \texttt{VerEmisFitting} \citep{veremisfit}, a user-friendly \textsc{Python} package, to constrain the gas kinematics of each galaxy based on emission-line fittings. 
This package allows users to fit emission lines using up to three Gaussian functions per line in velocity space. Users can also interactively define the line-fitting window, set local-continuum regions, and mask spectral regions. We outline the essential fitting steps below and direct the reader to the \texttt{VerEmisFitting} documentation\footnote{https://github.com/jasonpeng17/VerEmisFitting} for a detailed explanation of each step.
\begin{enumerate}[leftmargin=*]
    \item \textbf{\underline{Fitting Algorithms}}: The fitting algorithm employs the Levenberg-Marquardt (\textsc{leastsq}) and Nelder-Mead (\textsc{nelder}) methods  from the package {\tt\string LMFIT} \citep{lmfit}. To reduce the likelihood that the best-fitting result gets trapped in a local minimum within the multi-dimensional $\chi^2$ space, \texttt{VerEmisFitting} adjusts the initial guesses for each iteration based on the specified range of variation for each parameter (defined by users). The typical number of iterations for finding the best-fitting result is $\geq 1000$ for each galaxy. The fitting results from the \textsc{leastsq} and \textsc{nelder} methods usually converge at this large number of iterations.
    \item \textbf{\underline{Fitted Emission Lines}}: Our line-fitting procedure involves fitting multiple emission lines in the velocity space. These lines ([O\,{\footnotesize II}] $\lambda\lambda 3726, 3729$, [O\,{\footnotesize III}] $\lambda 4363$, $\rm{H}\beta$, [O\,{\footnotesize III}] $\lambda 5007$, [O\,{\footnotesize I}] $\lambda 6300$, [N\,{\footnotesize II}] $\lambda\lambda  6548, 6583$,  $\rm{H}\alpha$, [S\,{\footnotesize II}] $\lambda\lambda 6717, 6731$, and [O\,{\footnotesize II}] $\lambda\lambda 7319+, 7330+$\footnote{The notation [O\,{\footnotesize II}] $\lambda\lambda 7319+, 7330+$ includes two auroral doublets, [O\,{\footnotesize II}] $2s^22p^3~^2D^o_{3/2} - 2s^22p^3~^2P^o_{1/2,3/2}$~$\lambda\lambda 7319,7320$ and [O\,{\footnotesize II}] $2s^22p^3~^2D^o_{5/2} - 2s^22p^3~^2P^o_{1/2,3/2}$~$\lambda\lambda 7330,7331$ \citep{Sharpee_2004}. Since both doublets are not spectrally resolved, we measure the combined flux of each doublet, where the flux of each doublet represents the sum of two individual lines. These doublets are utilized to determine the $\rm{O}^{+}$ ionic abundance and are therefore only fitted when the galaxies do not have detectable [O\,{\footnotesize II}] $\lambda\lambda 3726, 3729$ lines.}) are pivotal for determining nebular conditions like dust attenuation and gas-phase metallicity (Section \ref{subsubsec:derived_phys_prop}), the excitation mechanism of each velocity component (Section \ref{sec:bpt_diagram}), and the physical origins of fast-outflowing cold clouds (Section \ref{sec:broad_origins}).
    \item \textbf{\underline{Line Fitting Models}}: 
    We apply either a single-component (single-Gaussian) or a multi-component model to the local-continuum-subtracted line profile. 
    The multi-component model can be a two-component (double-Gaussian) or a three-component model (triple-Gaussian or double-Gaussian plus Lorentzian, dGL).
    The Lorentzian profile is utilized when residuals in the broad line wings, especially for [O\,{\footnotesize III}] $\lambda 5007$ and $\rm{H}\alpha$, from triple-Gaussian fits indicate a non-Gaussian nature. 
    In the scenario that the dGL model is preferred, the triple-Gaussian model will underestimate the line flux of broad wings by as much as a factor of 2.
    \item \textbf{\underline{Validation of Multi-component Models}}: The need for a multi-component model is validated using an F-test \citepalias{Xu_2022} at the 3-sigma significance level. 
    If a line achieves only 1-sigma or 2-sigma significance in the F-test, yet other lines within the same ionization zone \citep{berg_characterizing_2021} do meet the 3-sigma significance level F-test, we still apply the same number of velocity components for its fitting. 
    For instance, the double-Gaussian fit for [O\,{\footnotesize II}] $\lambda\lambda 3726, 3729$ in J0808+3948 only passes a 2-sigma significance level F-test, whereas [N\,{\footnotesize II}] $\lambda\lambda 6548, 6583$ passes at the 3-sigma level.
    Since O$^{+}$ and N$^{+}$ species belong to the low-ionization zone, we fit both the [O\,{\footnotesize II}] and [N\,{\footnotesize II}] doublets using the double-Gaussian model for J0808+3948.
    \item \textbf{\underline{Balmer Absorption Troughs}}: For Balmer lines with observable underlying stellar absorption lines, it is essential to account for their impacts on Balmer decrement measurements and strong line ratios sensitive to excitation-diagnostic analysis. 
    Following \cite{Peng_2023}, we use the Lorentzian profile to fit the Balmer absorption trough. 
    We adopt a constraint on the amplitudes of the absorption troughs, with the H$\alpha$ stellar absorption equivalent width being two-thirds that of H$\beta$. 
    This equivalent width constraint is derived from \texttt{STARBURST99} models \citep{Leitherer_1999, Leitherer_2014} in \citet{Henry_2021} and is consistent with \texttt{BEAGLE} models \citep{chevallard_modelling_2016}.
\end{enumerate}

\subsubsection{Constraints on Line Fittings} \label{subsubsec:line_fitting_constraints}
Since one of the primary objectives of emission line fittings is to consistently identify the same velocity components across different lines, we first constrain the gas kinematics by assuming that each component's velocity centroid and width are identical for all fitted emission lines (\textit{``Fitting Strategy I''}). 
Although emission lines from distinct ionization zones might have slightly different gas kinematics, this fitting approach, which is weighted towards the gas kinematics of the strongest line (i.e., [O\,{\footnotesize III}] $\lambda 5007$ or H$\alpha$), generally performs well when galaxies have relatively weak low-ionization lines (e.g.,  integrated signal-to-noise ratio, SNR, $\lesssim 20$ for [O\,{\footnotesize I}] $\lambda 6300$ and SNR $\lesssim 50$ for [O\,{\footnotesize II}] $\lambda\lambda 3726, 3729$ or [S\,{\footnotesize II}] $\lambda\lambda 6717, 6731$). These galaxies include J0144+0453, J0808+3948, J0926+4427, and J1444+4237.
    
Nevertheless, \textit{Fitting Strategy I} generally fails to reproduce the velocity centroids of other emission lines than [O\,{\footnotesize III}] $\lambda 5007$ and H$\alpha$ in two cases. (1) When galaxies exhibit these low-ionization lines with relatively high SNR, \textit{Fitting Strategy I} can lead to  ``P Cygni'' or ``inverted P Cygni'' profile-like residuals around their central peaks due to slight velocity offsets compared to [O\,{\footnotesize III}] or H$\alpha$. (2) Similarly, such residuals may also appear in H$\beta$ when a triple-component model is necessary for H$\alpha$, but an F-test justifies only a single or double-component model for H$\beta$.
    
Consequently, we set the velocity centroids of H$\beta$, [S\,{\footnotesize II}] $\lambda\lambda 6717, 6731$, [O\,{\footnotesize II}] $\lambda\lambda 3726, 3729$ and/or [O\,{\footnotesize I}] $\lambda 6300$ as free parameters, while tying their velocity widths to match those of the other intended lines (\textit{``Fitting Strategy II''}) for the following galaxies: J0127-0619 (H$\beta$ and [S\,{\footnotesize II}]), J0337-0502 (H$\beta$ and [S\,{\footnotesize II}]), J0938+5428 (H$\beta$, [S\,{\footnotesize II}], [O\,{\footnotesize II}], and [O\,{\footnotesize I}]), J1025+3622 (H$\beta$ and [S\,{\footnotesize II}]), J1044+0353 (H$\beta$, [S\,{\footnotesize II}], and [O\,{\footnotesize I}]), J1129+2034 (H$\beta$, [S\,{\footnotesize II}], and [O\,{\footnotesize I}]), J1200+1343 (H$\beta$ and [O\,{\footnotesize I}]), J1359+5726 (H$\beta$, [S\,{\footnotesize II}], and [O\,{\footnotesize I}]), J1429+0643 ([O\,{\footnotesize II}] and [O\,{\footnotesize I}]), and J1525+0757 (H$\beta$). 
The velocity centroids derived from different emission lines generally differ by less than $20 \ \rm{km^{-1}}$; however, \textit{Fitting Strategy II} is essential to accurately measure each component's line flux for these galaxies.

The fitting strategies described above are preferred over ``fitting by elements'' (e.g., Hydrogen, Oxygen, and Sulfur) or ``fitting by ionization zones'' \citep[e.g., low and high ionization zones;][]{berg_characterizing_2021}, as the best-fitting gas kinematics for each element or zone is highly sensitive to SNR, which can vary significantly across these elements or zones. Therefore, we argue that our chosen fitting strategies are crucial for achieving consistent and physical gas kinematics for each galaxy. 
We hereafter refer to the outflow properties of each velocity component (e.g., the maximum velocity $v_{\rm{max}}$) as constrained by the strongest lines, [O\,{\footnotesize III}] $\lambda 5007$ and H$\alpha$.

In addition to the constraints on gas kinematics, we fix the line ratio for the [N\,{\footnotesize II}] $\lambda\lambda 6548, 6583$ doublet to 3.05 \citep{DKP_2023} to ensure physical fitting results. The fitting result of each galaxy is shown in Appendix \ref{sec:line_fitting_results}. 
Utilizing the best-fitting parameter values, we compute the flux of each component in every line profile. The uncertainty of each fitting parameter is obtained from the {\tt\string LMFIT} covariance matrix and then is propagated to the flux error of each velocity component.

\begin{figure*}[htb]
\centering
\includegraphics[width=1.0\textwidth]{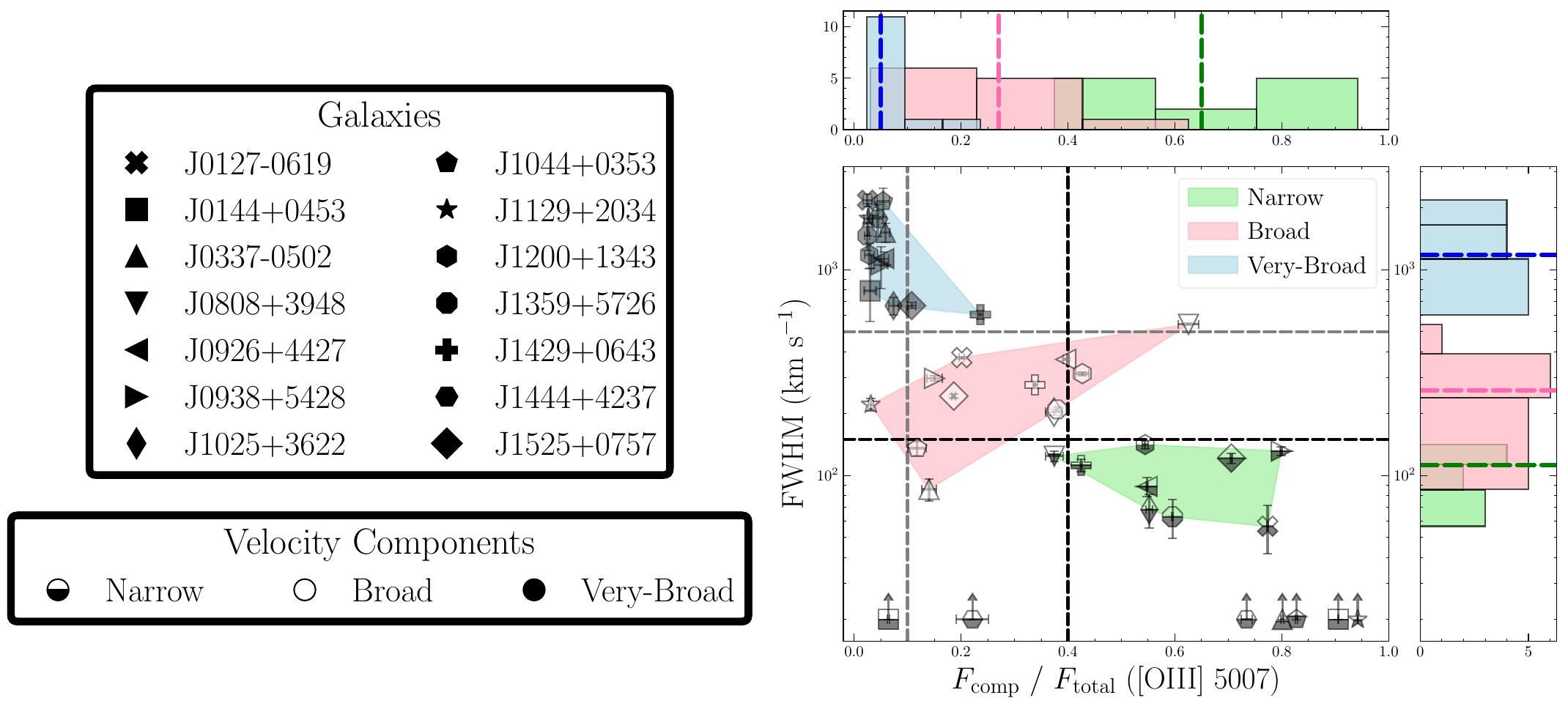}
\caption{Summary of velocity components in the FWHM--[O\,{\footnotesize III}] $\lambda 5007$ flux ratio space. The regimes of narrow, broad, and VB velocity components in this space are represented by the green, pink, and blue-shaded polygon regions, respectively. 
The black (grey) dashed lines roughly demarcate the narrow and broad (broad and VB) components.
The marker style for each galaxy and velocity component, as specified in the left-hand legend, will be used consistently throughout the rest of the paper.
The velocity widths of several galaxies' narrow components are unresolved, including J0144+0453 and J1444+4237 which show double narrow components (see Section \ref{subsubsec:vel_comp_def}), so we assign a lower limit of $\rm{FWHM} = 20 \ \rm{km \ s^{-1}}$ for including them in the plot.
Galaxies that exhibit Lorentzian-like broad wings in their VB components are reported with their $\rm{IPV_{24}}$ values instead (see Table \ref{table:velocity_info} and Section \ref{subsubsec:vel_comp_def} for details).
The histograms and the median values (shown as dashed lines) of each velocity component's FWHM and $F_{\rm{comp}} / F_{\rm{total}}$ values are plotted on the right and top panel, respectively. \label{fig:vel_comp_def}}
\end{figure*}
\subsubsection{Velocity Component Definitions} \label{subsubsec:vel_comp_def}
We hereafter attribute the detected first, second, and third emission velocity components as the ``narrow'' (bottom-filled markers), ``broad'' (open markers), and ``very-broad (VB)'' (filled markers) components (Figure \ref{fig:vel_comp_def}).
These components can be roughly categorized using the median and one-sigma scatter of the FWHM
and the component/total flux ratio of [O\,{\footnotesize III}] $\lambda 5007$, $F_{\rm{comp}} / F_{\rm{total}}$, as summarized below.
\begin{itemize}[leftmargin=*]
    \item ``\textit{narrow}'': $\rm{FWHM} \lesssim 150 \ \rm{km \ s^{-1}}$ and $F_{\rm{comp}} / F_{\rm{total}} \gtrsim 0.40$
    \item ``\textit{broad}'': $150 \ \rm{km \ s^{-1}} \lesssim \rm{FWHM} \lesssim 500 \ \rm{km \ s^{-1}}$ and $ 0.10 \lesssim F_{\rm{comp}} / F_{\rm{total}} \lesssim 0.40$
    \item ``\textit{VB}'': $\rm{FWHM} \gtrsim 500 \ \rm{km \ s^{-1}}$ and $F_{\rm{comp}} / F_{\rm{total}} \lesssim 0.10$
\end{itemize}
The median $F_{\rm{comp}} / F_{\rm{total}}$ (FWHM) values are around 0.65 (110 $\rm{km \ s^{-1}}$), 0.30 (260 $\rm{km \ s^{-1}}$), and 0.05 (1200 $\rm{km \ s^{-1}}$) for the narrow, broad, and VB components, respectively. 

This categorization above excludes galaxies exhibiting a double-narrow component (DNC) feature (J0144+0453 and J1444+4237).
We employ the double-Gaussian model to fit their asymmetric line cores, revealing evidence of multiple nuclei, as suggested by their merger-like morphologies discussed in Section \ref{subsec:merger_flows}.
For the two DNC galaxies, in addition to the two unresolved narrow components, we classify the other component as the VB component, as it has FWHM $\gtrsim 800 \ \rm{km \ s^{-1}}$ with a component/total flux ratio $\lesssim 10\%$, consistent with the VB component characteristics described above.
Besides DNC galaxies, outliers of these demarcation lines are mostly AGN candidates (see Section \ref{sec:bpt_diagram}) or galaxies with merger-like morphology.


We consider the narrow component as gas tracing virial motions \citep{TM_81} within the examined region and the broad or VB component as outflowing gas (or equivalently, the ``outflowing'' components). 
To estimate $v_{\rm{max}}$ of each outflowing component, we adopt a definition similar to the form proposed in \cite{Zhang_2023}:
\begin{equation}
\label{eq:2.3.2}
v_{\rm{max}} =  \frac{\left|\Delta v \right|}{2} + \rm{IPV_{24, out}},
\end{equation}
where $\left|\Delta v \right|$\footnote{The VB component's $\left|\Delta v \right|$ for the DNC galaxies is calculated relative to the stronger narrow component. Since the velocity centroid differences between the two narrow components are $\lesssim 50 \ \rm{km \ s^{-1}}$, which is negligible compared to their $\rm{IPV_{24, out}}$ (VB), this choice does not impact our subsequent arguments.} represents the absolute velocity centroid difference between the narrow and the broad or VB component.
$\rm{IPV_{24, out}}$ represents the inter-percentile velocity \citep{Martin_2015} that corresponds to 76\% of the line profile area for the broad or VB component.

We prefer using IPV over $\rm{FWHM}$ in the original definition of $v_{\rm{max}}$ in \cite{Zhang_2023} because our line profile modeling of outflowing components also uses Lorentzian profiles (see Table \ref{table:velocity_info}).
Measuring velocity widths using FWHM may underestimate the true velocity spread when profiles, like Lorentzian, have ``heavier'' tails that fall off more slowly than Gaussian profiles.
Specifically, we use IPV$_{24}$ since it equals the FWHM for a Gaussian profile.
IPV$_{24}$ is also used to compare the velocity widths of simulated line profiles from galactic wind models, which can exhibit non-Gaussian shapes, to observations as presented in Section \ref{sec:broad_origins}.


\subsubsection{Derived Physical Properties} \label{subsubsec:derived_phys_prop}

To correct for Galactic foreground extinction, we obtain the $E(B-V)_{\rm{MW}}$ value (see Table \ref{table:physical_properties}) of each galaxy from the dust maps of \cite{schlafly_measuring_2011} and apply the correction using the Galactic extinction curve interpolated from Table 3 of \cite{Fitzpatrick_1999}. 
The intrinsic extinction $A_V$ for each velocity component is determined from the Balmer decrement, utilizing each component's $\rm{H}\alpha$/$\rm{H}\beta$ ratio and applying the SMC extinction curve from Table 4 (SMC Bar) of \cite{gordon_2003}.
For galaxies without detections of broad or VB components in $\rm{H}\alpha$ or $\rm{H}\beta$, we use the $A_V$ derived from the narrow components.
The intrinsic Case B $\rm{H}\alpha$/$\rm{H}\beta$ line ratios at different electron temperature $T_e$ and density $n_e$ are extracted from \citet{storey_recombination_1995}. 
The extinction-corrected line fluxes are used to determine $n_e$, $T_e$, and oxygen abundance (in terms of 12 + log(O/H)) of each velocity component.

Using the \textsc{getCrossTemDen} method from \texttt{PyNeb} \citep{pyneb,pyneb_2015}, $n_e$ and $T_e$ can be simultaneously measured from the density-sensitive doublet [O\,{\footnotesize II}] $\lambda \lambda 3726, 29$ or [S\,{\footnotesize II}] $\lambda \lambda 6716, 31$ and temperature-sensitive line ratio between [O\,{\footnotesize III}] $\lambda 4363$ and [O\,{\footnotesize III}] $\lambda 5007$. 7 out of 14 galaxies exhibit both [O\,{\footnotesize II}] and [S\,{\footnotesize II}] doublet detections, yielding consistent $n_e$ values within uncertainties. 
As [S\,{\footnotesize II}] doublets are detected in all galaxies, [S\,{\footnotesize II}]-derived densities are used for consistency in subsequent analysis.
[O\,{\footnotesize III}] $\lambda 4363$ is typically detectable in low-metallicity galaxies with an oxygen abundance of $\rm{12+log(O/H)} \lesssim 8.0$. 10 out of 14 galaxies have valid [O\,{\footnotesize III}] $\lambda 4363$ measurements. For galaxies without $T_e$ measurements, we determine $n_e$ assuming a fiducial $T_e$ of $10, 000$ K and estimate its systematic uncertainty by also calculating $n_e$ at $T_e$ values of $5, 000$ and $15, 000$ K.



Given the $n_e$ and $T_e$ measurements, we can use the direct $T_e$ method to determine the oxygen abundance \citep[e.g.,][]{P_rez_Montero_2017}. In the absence of detected temperature-sensitive [O\,{\footnotesize II}] lines, we infer $T_e$([O\,{\footnotesize II}]) from $T_e$([O\,{\footnotesize III}]) via the relation from \cite{arellano-cordova_ten_2020}, and apply the \textsc{getIonAbundance} method from \texttt{PyNeb} to derive $\rm{O^+}$ and $\rm{O^{2+}}$ ionic abundances based on the line ratios [O\,{\footnotesize II}] $\lambda\lambda 3726, 3729$/$\rm{H}\beta$ and [O\,{\footnotesize III}] $\lambda 5007$/$\rm{H}\beta$, respectively. If the doublet [O\,{\footnotesize II}] $\lambda\lambda 3726, 3729$ is not detectable, the multiplet [O\,{\footnotesize II}] $\lambda\lambda 7319+, 7330+$ is utilized instead. Given that J0144+0453 and J1444+4237 exhibit two narrow velocity components that are not spectrally resolved in the multiplet [O\,{\footnotesize II}] $\lambda\lambda 7319+, 7330+$, their oxygen abundances are not determined in this work.

Adopting this approach, metallicity measurements are available for 8 out of the 14 galaxies \citep[consistent with the reported metallicity values from][shown in Table \ref{table:physical_properties}]{Berg_2022}. 
Two galaxies have oxygen abundance measurements for both narrow and broad components. Their broad-component metallicities are similar to those of the narrow components within two standard deviations. 
Table \ref{table:derived_physical_properties} summarizes these spectroscopically derived physical properties.

\section{Excitation Mechanisms of Different Velocity Components} \label{sec:bpt_diagram}
\begin{figure*}[htb]
    \includegraphics[width=1.0\textwidth]{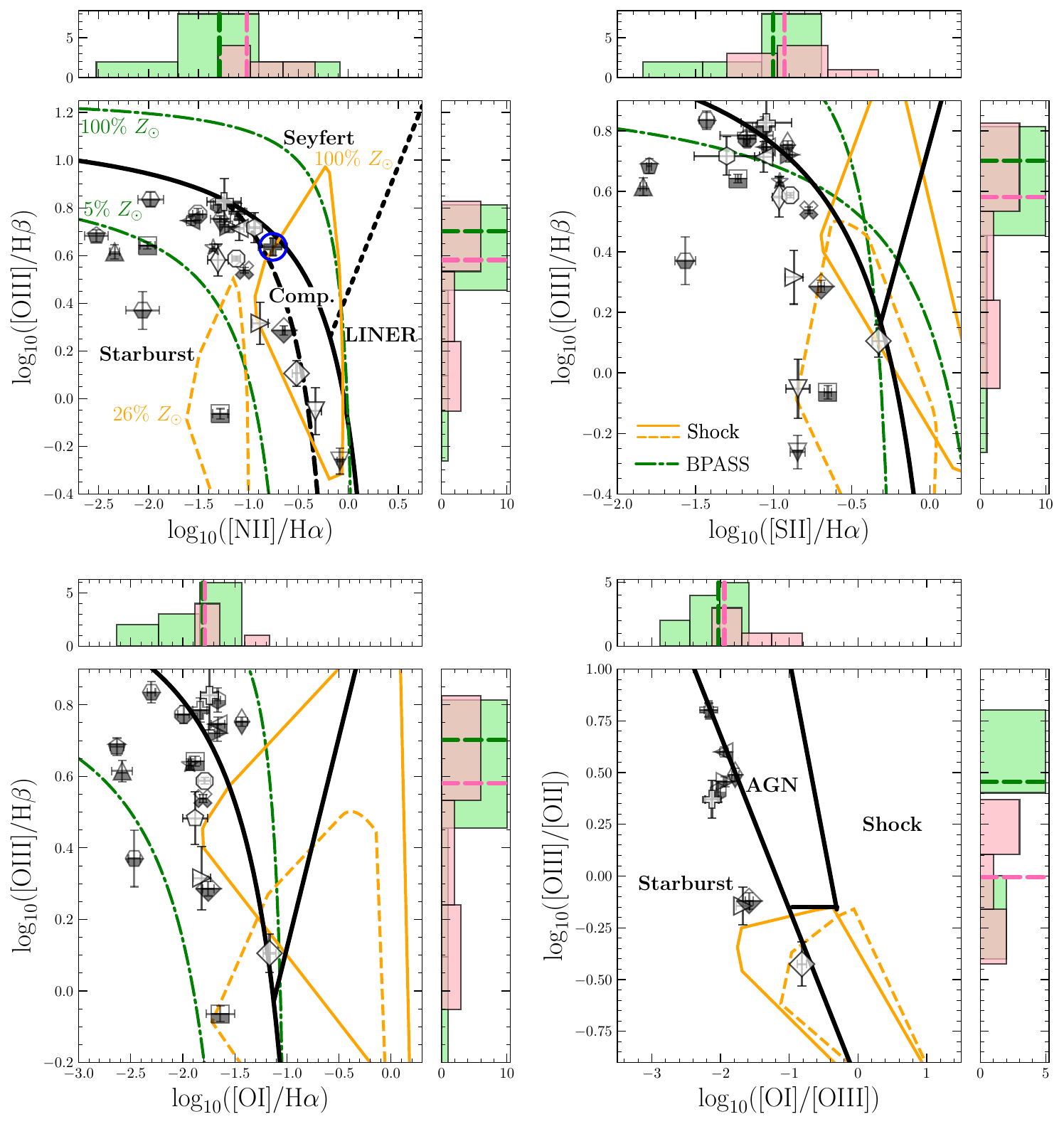}
\caption{[N\,{\footnotesize II}] (top-left), [S\,{\footnotesize II}] (top-right), and [O\,{\footnotesize I}] (bottom-left) BPT diagrams as well as the [O\,{\footnotesize I}]-[O\,{\footnotesize II}]-[O\,{\footnotesize III}] diagram (bottom-right) for each velocity component. 
The black solid and dashed curves indicate the starburst-AGN demarcations by \cite{Ke06} and \cite{Ka03} in the top-left, top-right, and bottom-left panels. In the bottom-right panel, black solid lines show the demarcation between starburst, shock, and AGN models \citep{mingozzi_classy_2024}.
The division line between Seyferts and LINERs is shown as the black solid line following \cite{Ke06} for the upper right and lower left panels and the black dotted line following \cite{Schawinski_2007} for the upper left panel. 
The green dash-dotted lines represent the maximal starburst prediction at metallicity $5\% \ Z_{\odot}$ and $100\% \ Z_{\odot}$ including massive star binaries through the BPASS models \citep{Xiao18}. 
The pure-shock models \citep{AM_2019}, featuring varying magnetic field strengths ($10^{-4} \leq B / \mu \rm{G} \leq 10$) and shock velocity values ($100 \leq v_{\rm{sh}} / (\rm{km \ s^{-1}})\leq 1000$) at a preshock density of $1 \ \mathrm{cm^{-3}}$, are shown for metallicities of $Z = 0.26 \ Z_{\odot}$ and $Z_{\odot}$.
J1429+0643 is the only galaxy with detectable VB components in $\rm{H}\beta$ and [N\,{\footnotesize II}], highlighted by a blue circle on the [N\,{\footnotesize II}]-BPT panel.   
The histograms and the median values (shown as dashed lines) of the narrow (green) and the broad (pink) component's line ratio values are plotted on the top and right sub-panels.
\label{fig:bpt_diagram}}
\end{figure*}
Recently, \cite{mingozzi_classy_2024} analyze optical and UV excitation diagnostic diagrams of the whole CLASSY sample to identify the excitation mechanism, including stellar photoionization, shock, or AGN, of each velocity component found in excitation-sensitive lines. They employ up to two Gaussian functions for fitting these lines in 24 out of 45 galaxies when single Gaussian fits proved inadequate. 
The FWHMs of their narrow ($\rm{FWHM} \sim 150 \ \rm{km \ s^{-1}}$) and broad ($\rm{FWHM} \sim 450 \ \rm{km \ s^{-1}}$) components are roughly consistent with our categorization illustrated in Section \ref{subsubsec:vel_comp_def}. Their optical diagnostics include the classic BPT diagrams \citep[i.e., {[}O\,{\footnotesize III}{]}/$\rm{H}\beta$ vs. {[}N\,{\footnotesize II}{]}/$\rm{H}\alpha$, {[}S\,{\footnotesize II}{]}/$\rm{H}\alpha$, and {[}O\,{\footnotesize I}{]}/$\rm{H}\alpha$;][]{Ke01, Ka03, Ke06}, in addition to the [O\,{\footnotesize I}]--[O\,{\footnotesize II}]--[O\,{\footnotesize III}] and [He\,{\footnotesize II}]/$\rm{H}\beta$ vs.\ [N\,{\footnotesize II}]/$\rm{H}\alpha$ diagrams. Hereafter, the [O\,{\footnotesize III}]/$\rm{H}\beta$ vs. [N\,{\footnotesize II}]/$\rm{H}\alpha$, [S\,{\footnotesize II}]/$\rm{H}\alpha$, and [O\,{\footnotesize I}]/$\rm{H}\alpha$ diagrams will be abbreviated as the [N\,{\footnotesize II}], [S\,{\footnotesize II}], and [O\,{\footnotesize I}] BPT diagrams, respectively. 
While some galaxies, such as J0808+3948 and J1429+0643, displayed inconsistent results across diagrams, \cite{mingozzi_classy_2024} primarily identify the components as stellar photoionized dominated. 
Nevertheless, our line fitting results suggest that strong lines like [O\,{\footnotesize III}] $\lambda 5007$ and $\rm{H}\alpha$ require a triple-Gaussian or a dGL model for most of our targets. 
Since \cite{mingozzi_classy_2024} do not consider these faint VB components, we should re-evaluate the diagnostic diagrams to ascertain potential impacts on the strong line ratios and the implied excitation mechanisms. 

While we utilize the three-component model for fitting VB components in some galaxies, classical demarcations between starburst galaxies and AGN \citep{Ke01, Ka03, Ke06} indicate that most galaxies fall within the starburst locus in the BPT diagrams (Figure \ref{fig:bpt_diagram}), aligning with \cite{mingozzi_classy_2024}. However, several exceptions can be classified as either composite starburst-AGN (underlined) or AGN objects. These exceptions are summarized in the following lists, which detail their velocity components and corresponding excitation diagnostic diagrams.
\begin{itemize}
    \itemsep -3pt
    \item \text{[N\,{\footnotesize II}] BPT}: J0808+3948 (\underline{narrow} and \underline{broad}), J1200+1343 (\underline{broad}), J1429+0643 (\underline{narrow} and \underline{VB})
    \item \text{[S\,{\footnotesize II}] BPT}: J0938+5428 (narrow), J1025+3622 (narrow), J1200+1343 (narrow), J1429+0643 (broad)
    \item \text{[O\,{\footnotesize I}] BPT}: J0926+4427 (narrow), J0938+5428 (narrow), J1025+3622 (narrow), J1200+1343 (narrow), J1429+0643 (narrow and broad), J1525+0757 (broad)
    \item \text{[O\,{\footnotesize I}]-[O\,{\footnotesize II}]-[O\,{\footnotesize III}}]: None
\end{itemize}
J1429+0643 is the only galaxy with potential AGN contributions across all three velocity components.

\begin{figure*}[htb]
    \includegraphics[width=1.0\textwidth]{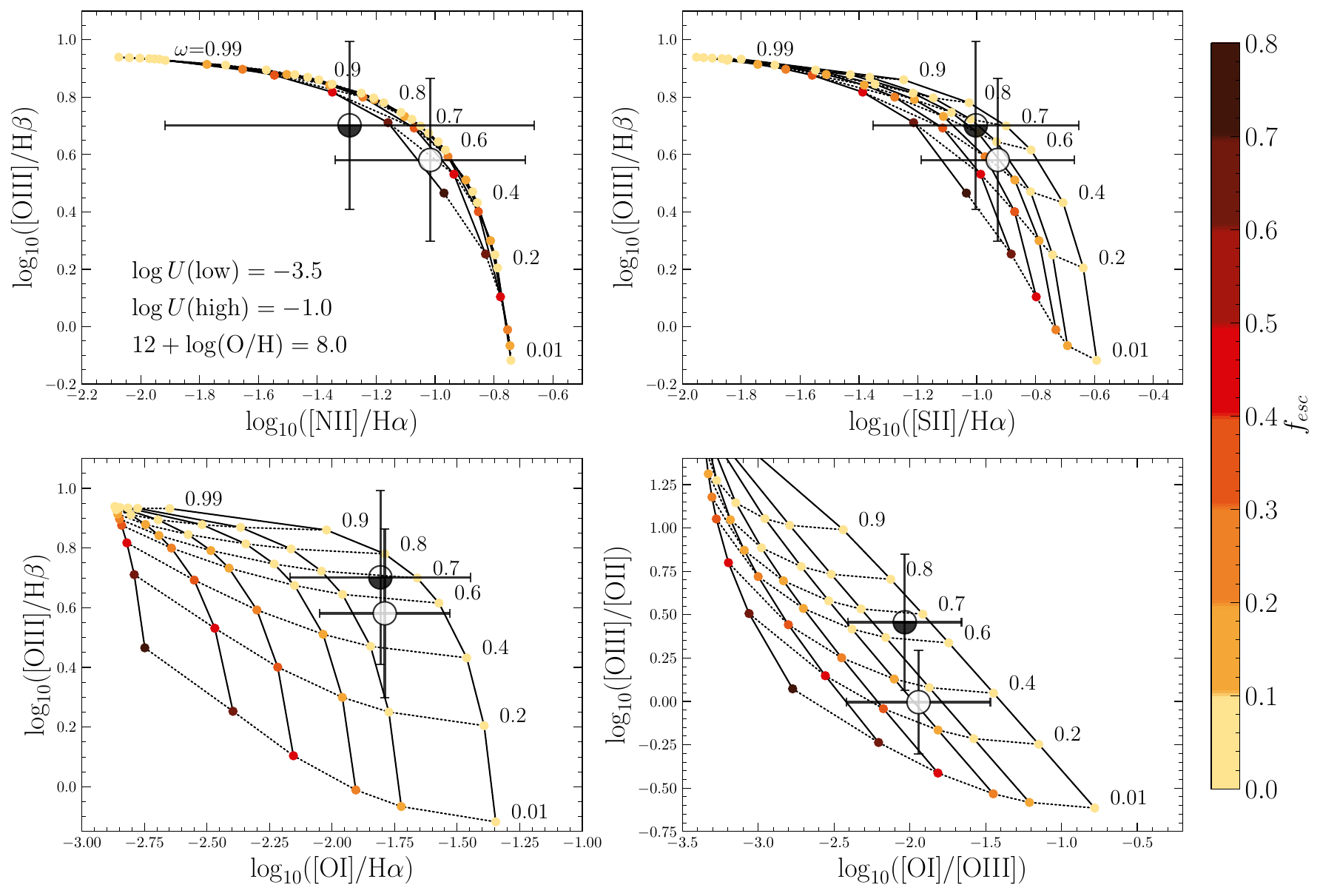}
\caption{The same four-panel excitation-diagnostics diagram as Figure \ref{fig:bpt_diagram} overlays with the two-zone ``density-bounded channels'' (DBC) photoionization models from \citet{Ramambason_2020}. These photoionization models are based on the BOND \citep[Bayesian Oxygen and Nitrogen abundance Determinations;][]{Asari_2016} model grids, which are available on the Mexican Million Models dataBase \citep[3MdB;][]{Morisset_2015}. 
Model points are color-coded by their escape fraction values. The high and low-ionization zones are set with $\log U = -1.0$ and -3.5, respectively.
The weighting factor $\omega$ of the high-ionization zone varies from 0.01 to 0.99, labeled alongside the grid point. The assumed metallicity of these model grids is 12 + log(O/H) = 8.0, which is roughly consistent with the median metallicity of our sample. 
The median and the standard deviation of the narrow and broad components are shown as the bottom-filled and open circular markers in each panel, respectively. 
\label{fig:bpt_diagram_ram20}}
\end{figure*}

The classical approach, assuming single stellar populations, does not account for the influences of metallicity and the presence of binary stellar populations in determining maximal starburst demarcation lines. \cite{Xiao18} address this by calculating binary maximal starburst lines as a function of metallicity using the \texttt{BPASS} (Binary Population and Spectral Synthesis) code \citep{BPASS_2017}. 
We present the binary maximal starburst lines at $Z = Z_{\odot}$ and $Z = 0.05 \ Z_{\odot}$, which approximately cover the metallicity range in our sample, as green dash-dotted lines in each BPT diagram in Figure \ref{fig:bpt_diagram}. The binary models typically yield higher [O\,{\footnotesize III}]/$\rm{H}\beta$ and low-ionization line ratios relative to $\rm{H}\alpha$, except at the high-end of [S\,{\footnotesize II}]/$\rm{H}\alpha$, where the low-metallicity binary line may surpass both the high-metallicity line and the classical model. Given the metallicity effects on maximal starburst lines, we argue that stellar photoionization predominantly drives most of our sample.

Figure \ref{fig:bpt_diagram} also reveals that the broad components exhibit marginally higher [N\,{\footnotesize II}]/$\rm{H}\alpha$, [S\,{\footnotesize II}]/$\rm{H}\alpha$, and [O\,{\footnotesize I}]/$\rm{H}\alpha$ ratios, along with lower [O\,{\footnotesize III}]/$\rm{H}\beta$ ratios compared to the narrow components. This observation aligns with findings from \cite{mingozzi_classy_2024}. 
The elevated low-ionization ratios in the broad components could indicate the presence of shocked gas.
For example, broad line profiles in ultraluminous infrared galaxies (ULIRGs) are often associated with shocked gas from galactic winds \citep[e.g.,][]{SM_2012_2}.
To test this hypothesis, we plot pure-shock models \citep{AM_2019}, incorporating a range of magnetic field strengths ($10^{-4} \leq B / \mu \rm{G} \leq 10$) and shock velocities ($100 \leq v_{\rm{sh}} / (\rm{km \ s^{-1}})\leq 1000$) for two distinct metallicity values: 
$Z = Z_{\odot}$ (orange solid polygons)
and $Z = 0.26 \ Z_{\odot}$ (orange dashed polygons).
The preshock density, $n_{\rm{pre}}$, is set at a fiducial value of $1 \ \mathrm{cm^{-3}}$ since line ratios are not as sensitive to $n_{\rm{pre}}$ as other shock parameters. 
We find that most galaxies' narrow and broad-line ratios generally do not align with the pure-shock model grids. 
This conclusion remains consistent when we consider ``shock plus precursor'' models in the BPT diagrams. Therefore, the contribution of shocked gas should be subdominant in most galaxies.
The broad component of J1525+0757 stands out as the only exception potentially dominated by shock-related mechanisms, as its line ratios roughly align with the shock models in all excitation-diagnostic diagrams. 

If most broad components are photoionized, then the observed trend of higher low-ionization ratio and lower [O\,{\footnotesize III}]/$\rm{H}\beta$ ratios might indicate their smaller ionization parameter values, $\log U$, compared to the narrow components \citep{Xiao18, Ramambason_2020}. 
The $\log U$ values that best match the median line ratios of the broad components range between $-3.0 \lesssim \log U \lesssim -2.5$ for the one-zone photoionization models \citep[single $U$;][]{Ramambason_2020}. This $\log U$ range for broad components is consistent with the median $\log U$ values measured from the low ($-2.71 \pm 0.07$) and intermediate ($-2.64 \pm 0.02$) ionization regions using [S\,{\footnotesize III}]/[S\,{\footnotesize II}] and [O\,{\footnotesize III}]/[O\,{\footnotesize II}] ratios, respectively \citep{Mingozzi_2022}.

However, one-zone photoionization models from \cite{Ramambason_2020} struggle to accurately reproduce the observed [S\,{\footnotesize II}]/$\rm{H}\alpha$ and [O\,{\footnotesize I}]/$\rm{H}\alpha$ ratios near the maximal starburst line of \cite{Ke06}. 
Hence, we apply the two-zone (two $U$) photoionization models from \cite{Ramambason_2020} to mimic the contribution from diffuse ionized gas or low column density channels with lower $\log U$ to boost [S\,{\footnotesize II}] and [O\,{\footnotesize I}] emissions. 
The high and low-ionization zones are defined with $\log U = -1.0$ and $\log U = -3.5$, respectively, to approximate the range of $\log U$ values observed across the entire CLASSY sample \citep{mingozzi_classy_2024}.
The line luminosities of the high and low-ionization zones are weighted by factors $\omega$ (similar to covering fraction) and $1 - \omega$, respectively. 
Compared to the two-zone ``density-bounded galaxy'' (DBG; leakage through high-$U$ regions) models, we find that the median line ratios of the broad components (pink circular markers in Figure \ref{fig:bpt_diagram_ram20}) can be better reproduced by the two-zone ``density-bounded channels'' (DBC; leakage through low $U$ density-bounded channels) models with the ionizing photon escape fraction $f_{\rm{esc}} \lesssim 0.1$ and $0.4 \lesssim \omega \lesssim 0.6$. 
Such observations reveal the presence of density-bounded channels (from which outflows originate) within an inhomogeneous ISM and highlight the anisotropic leakage of Lyman continuum (LyC) photons. 

This interpretation is supported by \cite{Hu_2023}, who report a significant fraction of CLASSY sample (9 out of 14 for our sample) have $\rm{Ly}\alpha$ emission in the bottom of the damped $\rm{Ly}\alpha$ absorber (DLA) profile. 
With the covering fraction of the DLA absorber, $f_{c, \text{DLA}}$, as measured in \cite{Hu_2023}, we estimate a median upper limit for $f_{\text{esc}} \sim 1 - f_{c, \text{DLA}} \leq 8\%$ for our sample exhibiting $f_{c, \text{DLA}}$, aligning with predictions from the two-zone photoionization model. 
Therefore, assuming both narrow and broad components are photoionized by the same stellar population (i.e., the same number of ionizing photons $Q$), a lower-density and more clumpy medium (smaller filling factor $\epsilon$) could account for broad components' reduced $\log U$ and enhanced [S\,{\footnotesize II}] and [O\,{\footnotesize I}] emissions \citep[see Equation (4) in][]{Ramambason_2020}. 

In summary, by integrating classical excitation diagnostic diagrams with both single and binary stellar population models, as well as the shock model across varying metallicities, $v_{\rm{sh}}$, and $B$ values, we have investigated the potential excitation mechanism of each velocity component. 
Our findings indicate that our sample is mainly dominated by stellar photoionization, but we cannot dismiss the possibility of an AGN in J1429+0643 and shock-related mechanisms in J1525+0757. 
Compared to the narrow components, the observed trend of broad component--higher low-ionization line ratios and lower [O\,{\footnotesize III}]/$\rm{H}\beta$--suggest the existence of an inhomogeneous ISM with low column density channels. 
Besides J1429+0643, the VB component eludes similar analysis due to its absence in most low-ionization lines. To elucidate the physical origin of this fast-outflowing cold gas, it is necessary to utilize the luminosities and velocity widths of these components to constrain the feasible mechanisms capable of producing such a faint yet high-velocity component (Section \ref{sec:broad_origins}).

\section{Origins of the Outflowing Components} \label{sec:broad_origins}
In this section, we intend to explore the physical origins of the outflowing components by examining two distinct galactic wind models under spherical symmetry: (1) a single-phase galactic wind model incorporating radiative cooling and gravity \citepalias{Thompson_2016}, and (2) a multiphase galactic wind model that accounts for mass, momentum, energy, and metallicity transfer between hot winds and cold clouds \citepalias{Fielding_2022}. 
To target the outflowing cold gas observable through UV absorption or optical emission lines, these analytical models are built upon the pioneering work of \citetalias{CC85}, which describes the steady-state radial structure of hot winds traced by X-ray emissions. 
Following \citetalias{CC85}, both models posit that supernova explosions deposit sources of mass and energy within a starburst region characterized by a radius $R_{\rm{SF}}$. This radius also corresponds to the sonic point (i.e., where the Mach number $\mathcal{M} = 1$) in the radial velocity structure of the hot wind. Here, the total mass, $\dot{M}_{\rm{{hot}}}$, and energy, $\dot{E}_{\rm{{hot}}}$, input rates are defined as
\begin{align}
    \label{eq:4.1} \dot{E}_{\rm{{hot}}} = 3 \times 10^{41} \ \rm{erg \ s^{-1}} \ \eta_{\rm{E}} \ \frac{\rm{SFR}}{M_{\odot} \ \rm{yr^{-1}}}
\end{align}
and 
\begin{align}
    \label{eq:4.2} \dot{M}_{\rm{{hot}}} = \eta_{\rm{M}} \ \rm{SFR},
\end{align}
respectively. $\eta_{\rm{E}}$ and $\eta_{\rm{M}}$ are the dimensionless thermalization efficiency and mass loading factors. The constant factor of the energy input rate $\dot{E}_{\rm{{hot}}}$ is based on the assumption that one supernova (SN) occurs for every 100 \(M_{\odot}\) formed and that each SN deposits \(E_{\rm{SN}} = 10^{51} \ \rm{erg}\) into the ISM. The validation of this SN rate is illustrated in Appendix \ref{sec:sn_rate}.
We refer readers to \citetalias{Thompson_2016} and \citetalias{Fielding_2022} for detailed descriptions of each model, and to Appendix \ref{sec:wind_model_solutions} for the initial conditions and cooling curves \citep[][hereafter PS20]{PS_2020} adopted in this work.

\subsection{Direct Radiative Cooling from Hot Winds} \label{subsec:hot_wind_cooling}
Previous studies \citepalias[e.g.,][]{Thompson_2016} have proposed a model suggesting that high-velocity cold clouds, traced by low-ionization UV absorption and optical emission lines, directly originate from the radiative cooling of hot winds. If these cold clouds, as identified by the outflowing components in strong emission lines, do indeed stem from hot winds, then the amount of energy released from radiative cooling would set the theoretical upper limit for the luminosities of these emission lines, as detailed in Appendix \ref{subsec:L_estimations_hot_wind}. 
In this work, our analysis primarily focuses on [O\,{\footnotesize III}] $\lambda 5007$, which, deblended with other strong lines, ensures accurate measurements of the outflowing components' luminosities. 

\defcitealias{PS_2020}{PS20}

One can also measure the velocity width of the outflowing components detected in strong lines to constrain the mass and energy input rates of the hot winds \citepalias[i.e., $v \propto \left( \eta_{\rm{E}} / \eta_{\rm{M}}  \right)^{1/2}$;][]{CC85}. 
We can gain insights into the velocity widths of these emission lines by simulating the galactic wind emission line profiles for a specified set of model parameters. 
Since the emissivities of both recombination and collisionally-excited lines are directly proportional to $n(r)^2$, the luminosity of the line radiated from a differential volume, given a specific geometry (spherical in this case), can be derived by integrating $n(r)^2$ over this volume. For a comprehensive breakdown of simulating galactic wind emission line profiles, please refer to Appendix \ref{sec:model}. 


\begin{figure*}[htb]
\centering
\includegraphics[width=0.49\textwidth]{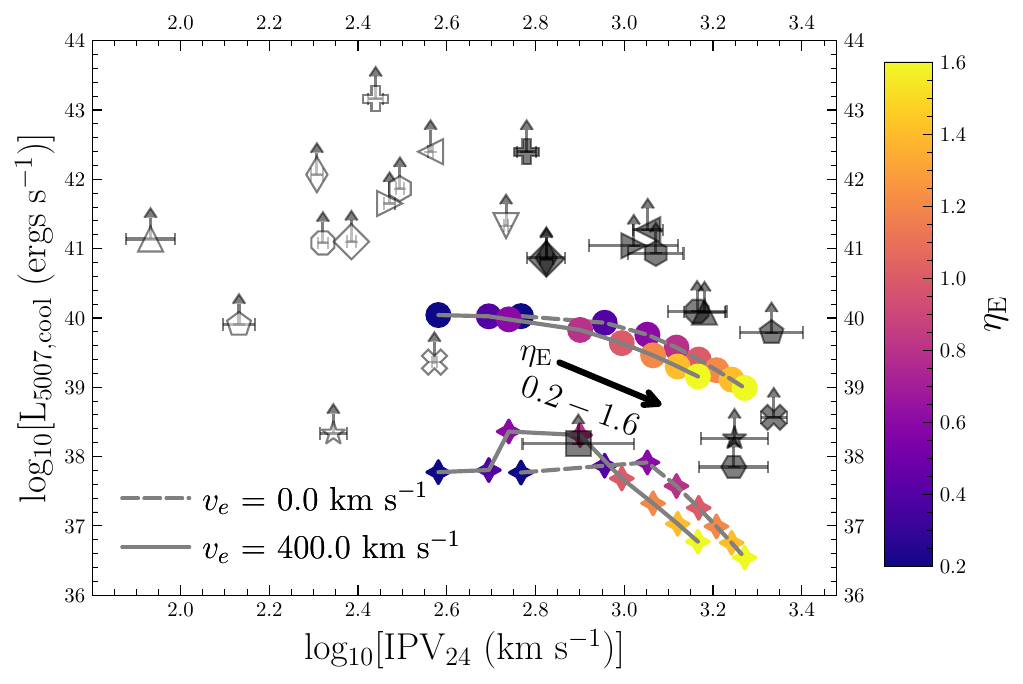}
\includegraphics[width=0.49\textwidth]{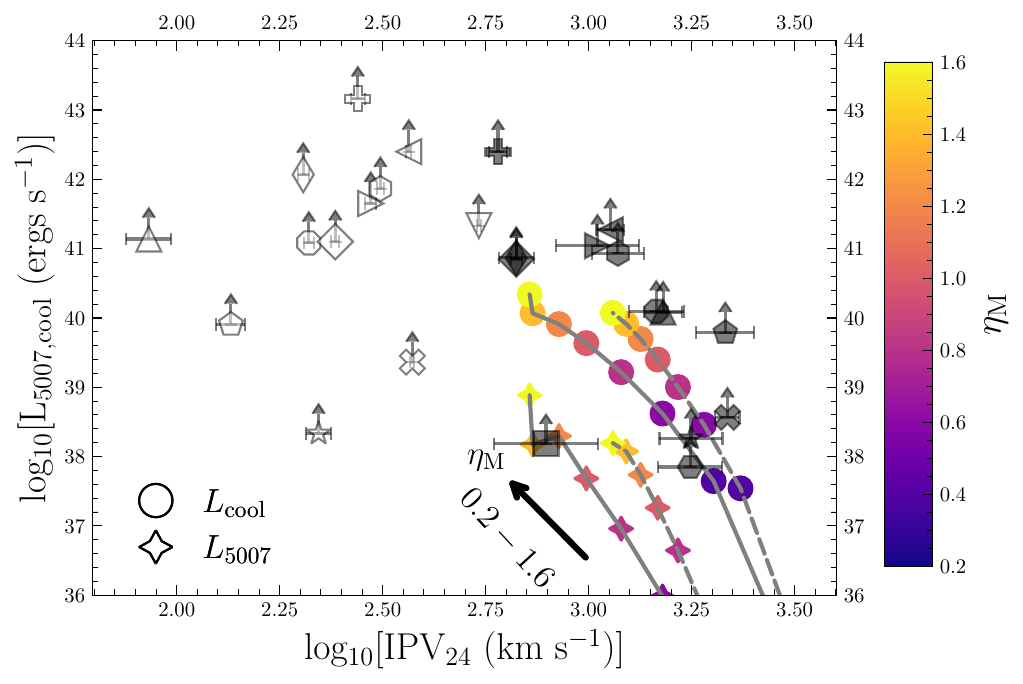}
\caption{Evaluating the role of direct condensation from hot winds in the physical origins of outflowing components. $L_{5007}$ for each galaxy is plotted against the measured IPV$_{24}$ of detected outflowing components. Based on \citetalias{Thompson_2016} galactic wind models with \(\eta_{\rm{M}} = 1.0\) (left panel) or \(\eta_{\rm{E}} = 1.0\) (right panel), 
the filled circles and four-pointed stars represent $L_{\rm{cool}}$ and $L_{5007}$, respectively, radiated within the temperature range from $10^4$ to $10^{5.5}$ K. 
Each model point is color-coded based on the \(\eta_{{\rm{E}}}\) (left panel) or \(\eta_{{\rm{M}}}\) (right panel) values ranging from 0.2 to 1.6, with the corresponding trend indicated by the direction of the black arrow (and will be used consistently in other figures of Section \ref{sec:broad_origins}). 
Model points connected by dashed (solid) lines represent models with \(v_e = 0\) (\(v_e = 400 \ \mathrm{km \ s^{-1}}\)).
Other key parameter values are summarized in Table \ref{table:model_parameters_t16}.
\label{fig:Lcool_SFR_vs_FWHM_plots}}
\end{figure*}

Combining the theoretical estimations of the cooling luminosity, $L_{\rm{cool}}$ (and the [O\,{\footnotesize III}] \(\lambda 5007\) luminosity, $L_{5007}$) with the velocity widths (i.e., IPV$_{24}$) derived from simulated emission line profiles, we can ascertain whether the detected outflowing components in our sample are a direct consequence of hot wind cooling. 

\subsubsection{Aperture Effects} \label{subsubsec:aperture_effects}
Before comparing the observed $L_{5007}$ to model estimations, it is important to note that most galaxies exhibit extended spatial profiles, indicating the ESI aperture cannot capture the entire [O\,{\footnotesize III}] nebulae. 
Therefore, in this study, we consider the measured luminosity of [O\,{\footnotesize III}] $\lambda 5007$ for each galaxy as a lower limit. 

Another important caveat in comparing these models to observations is the assumption that all model cooling luminosities are covered within the ESI aperture. This hypothesis holds only if the hot-phase mass-loading factor $\eta_{\rm{M}}$ is sufficiently high, causing the cooling radius (where radiative cooling becomes effective and the cooling time is less than the advection time) to collapse to the size of the ESI aperture. 
The expression of this critical value of $\eta_{\rm{M}}$ \citepalias[Equation 10 in][]{Thompson_2016} can be parameterized in terms of the median $\dot{\Sigma}_{\star}$ and the effective circular ESI aperture size ($R_{\mathrm{aper}} = \sqrt{1\farcs25 \times 20\farcs0 / \pi}$) as follows:
\begin{equation}
\label{eq:4.8} \eta_{\rm{M,crit}} (R_{\mathrm{aper}}) \simeq 1.78 \ \eta_{\rm{E}}^{0.73} \left(\frac{\Omega_{4 \pi}}{\xi_{M} \ \dot{\Sigma}_{\star,5.37} \ R_{\rm{aper,2.75}}} \right)^{0.27},
\end{equation}
where $R_{\rm{aper, 2.75}} = R_{\rm{aper}} \ / \  2.75 \ \rm{kpc}$, using the median pixel scale for our sample. Assuming solar metallicity ($\xi_{M} = 1$) and a spherical outflow geometry ($\Omega_{4 \pi} = 1$), our estimates for $\eta_{\rm{M,crit}}$ range approximately from 0.5 to 2.5, corresponding to $\eta_{\rm{E}}$ values varying between 0.2 and 1.6, considering the median $\dot{\Sigma}_{\star}$ and $R_{\mathrm{aper}}$. 
For the galaxies exhibiting the highest (J0808+3948) and lowest (J1444+4237) $\dot{\Sigma}_{\star}$ values, the $\eta_{\rm{M,crit}}$ values are approximately 0.6 and 102, respectively, assuming $\eta_{\rm{E}} = 1$. This substantial variation in $\eta_{\rm{M,crit}}$ suggests overestimations in the model $L_{\rm{cool}}$ for certain targets, particularly those with small $\dot{\Sigma}_{\star}$ and $R_{\mathrm{aper}}$ values. 
Consequently, the estimated $L_{\rm{cool}}$ from \citetalias{Thompson_2016} models should be regarded as upper limits. 
We carefully consider this aperture effect in Section \ref{subsec:trml_hot_and_cold}.

\subsubsection{Fiducial T16 Models} \label{subsubsec:fiducial_T16_models}
Assuming \citetalias{Thompson_2016} models with a median \(\dot{\Sigma}_{\star}\) of our targets (\(5.37 \ \rm{M_{\odot} \ yr^{-1} \ kpc^{-2}}\)), calculated using \citetalias{PS_2020} cooling curves at zero redshift and solar metallicity, Figure \ref{fig:Lcool_SFR_vs_FWHM_plots} shows how the estimated $L_{\rm{cool}}$ (circles) and $L_{5007}$ (four-pointed stars) depend on \(\eta_{{\rm{E}}}\), \(\eta_{{\rm{M}}}\), and gravitational deceleration. 
Since $v \propto \left( \eta_{\rm{E}} / \eta_{\rm{M}}  \right)^{1/2}$ and $n(r) \propto \eta_{\rm{M}}^{3/2} \eta_{\rm{E}}^{-1/2} $ (see Equation \ref{eq:4.4}), increasing $\eta_{\rm{M}}$ reduces the wind velocity but enhances radiative cooling, and vice versa for $\eta_{\rm{E}}$, given that $L_{\mathrm{cool}} \propto n^2$ (see the trends indicated by the black arrows in Figure \ref{fig:Lcool_SFR_vs_FWHM_plots}). 

The impact of an isothermal gravitational potential is illustrated by comparing models without gravity (the escape velocity \(v_e = 0\), shown as points connected by dashed lines) to models that include gravity (\(v_e = 400 \ \mathrm{km \ s^{-1}}\), shown as points connected by solid lines).
The models with \(v_e = 400 \ \rm{km \ s^{-1}}\) characterize the galaxy J1525+0757, which has the largest circular velocity \(v_{\rm{cir}} = 137.3^{+44.6}_{-23.5}\), and its corresponding estimated \(v_{\mathrm{e}} = 2 v_{\rm{cir}}^2 \left[1 + \log(r_{\rm{max}} / r) \right] \simeq 3 v_{\mathrm{cir}} \simeq 400 \ \rm{km \ s^{-1}}\) assuming a truncated singular isothermal potential \citep[for $r_{\rm{max}} / r = 10 - 100$;][]{Veilleux_2020}.
The differences in luminosities between models incorporating gravitational terms and those without them, typically remaining within one dex, are primarily attributable to the gravitational deceleration of hot winds. This deceleration increases the density of hot winds, which augments the cooling luminosity. 

The observed $L_{5007}$ of most outflowing components are too luminous to be reproduced by the \citetalias{Thompson_2016} models, given they are even orders of magnitudes larger than $L_{\rm{cool}}$ (i.e., the luminosity-deficit issue). 
Compared to broad components, the \citetalias{Thompson_2016} models can better reproduce the measured velocity widths of the VB components.
To assess whether the interpretations regarding the physical origins of the outflowing components are influenced by variations in the model configurations, we modify the model parameters including metallicity \(Z\), and \(\dot{\Sigma}_{\star}\), in the following analyses, respectively.
Since the redshift evolution of \citetalias{PS_2020} cooling curves is negligible within the redshift range \( z \sim 0 - 0.2 \) for our ESI sample (the model [O\,{\footnotesize III}] \(\lambda 5007\) at \( z = 0.2 \) are indistinguishable from those at \( z = 0 \)), we disregard this minimal redshift evolution in subsequent analyses and assume that the cooling curves are set at \( z = 0 \).
We summarize the parameter values used in \citetalias{Thompson_2016} models in Table \ref{table:model_parameters_t16}. 

\begin{figure*}[htb]
\centering
\includegraphics[width=0.49\textwidth]{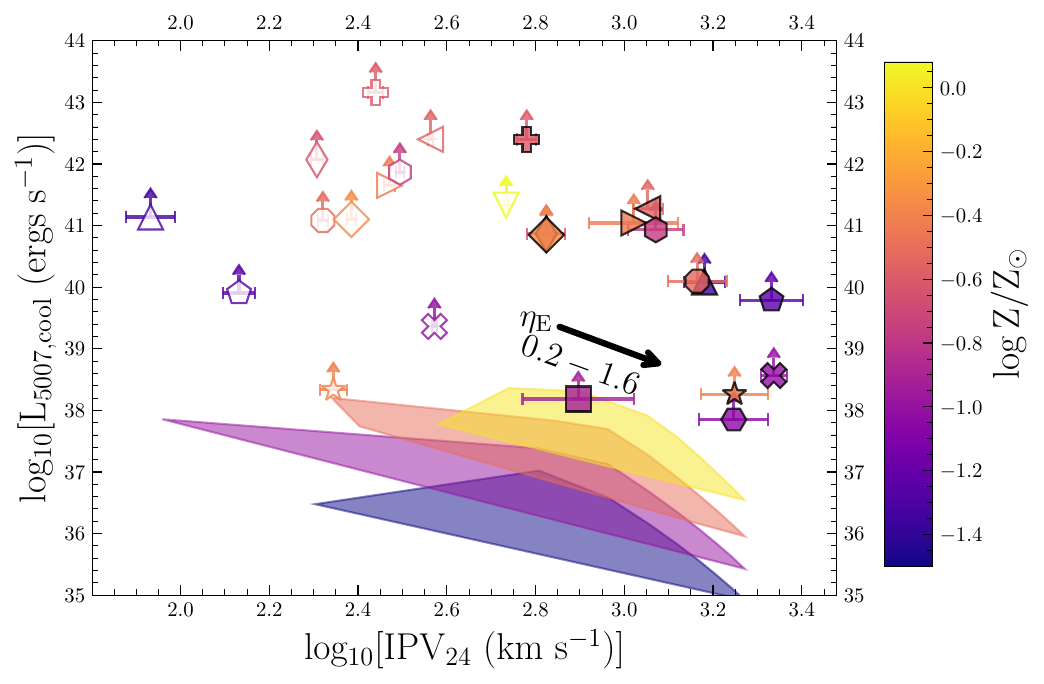}
\includegraphics[width=0.49\textwidth]{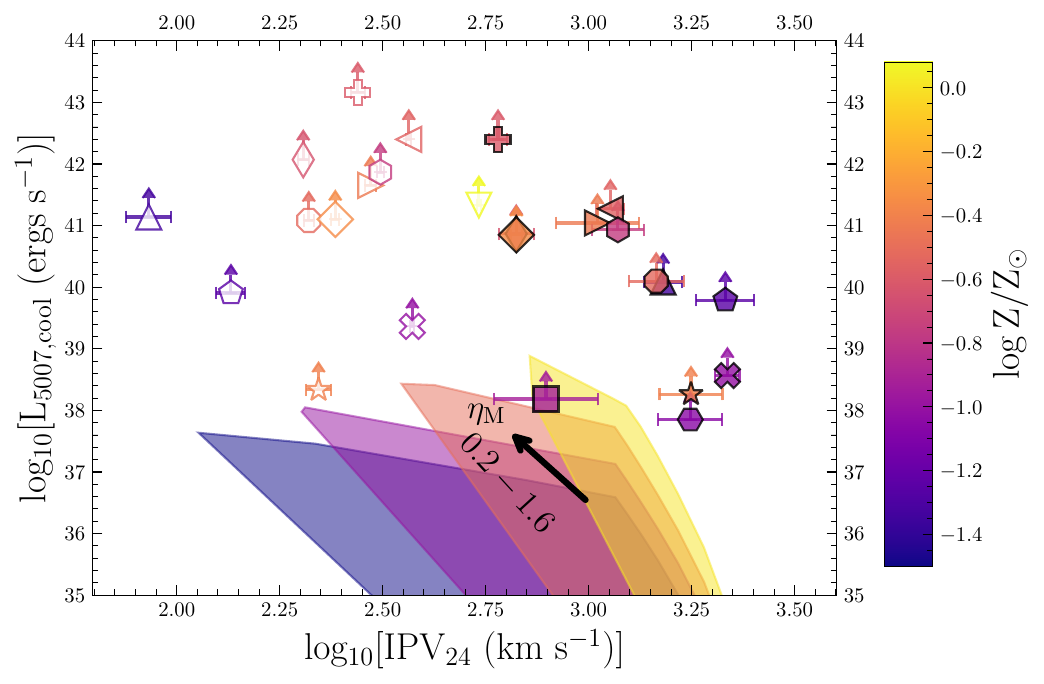}
\caption{Similar to Figure \ref{fig:Lcool_SFR_vs_FWHM_plots} but with varying $Z$ and only showing   $L_{5007}$. 
Each polygon-shaded region represents the parameter space that the \citetalias{Thompson_2016} model can reproduce. These polygon regions correspond to the ranges of $v_e$ from 0 to $400 \ \mathrm{km \ s^{-1}}$ and $\eta_{\rm{E}}$ from 0.2 to 1.6 in the left panel ($\eta_{{\rm{M}}} = 1.0$), or $\eta_{\rm{M}}$ from 0.2 to 1.6 in the right panel ($\eta_{{\rm{E}}} = 1.0$), like Figure \ref{fig:Lcool_SFR_vs_FWHM_plots}.
Each observation point or model polygon region (at $\log Z / Z_{\odot} = $ -1.5, -1.0, -0.5, and 0) is color-coded based on its $Z$ value.
Since [O\,{\footnotesize III}] $\lambda 4363$ of the outflowing components is usually not detected in our sample, we color-coded each galaxy by its gas-phase metallicity, derived from the narrow component, in Figure \ref{fig:Lcool_SFR_vs_FWHM_plots_vary_oh} \citep[galaxies without metallicity measurements are based on values extracted from][]{Berg_2022}. 
\label{fig:Lcool_SFR_vs_FWHM_plots_vary_oh}}
\end{figure*}

\subsubsection{\texorpdfstring{Varying $Z$ for T16 Models}{}} \label{subsubsec:varying_metallicity}
Both the forbidden and permitted heavy-element lines are primary coolants in the temperature ranges $T \sim 10^3 - 10^{6}$, as discussed above. Therefore, models with higher metallicity are expected to exhibit increased cooling luminosities. Figure \ref{fig:Lcool_SFR_vs_FWHM_plots_vary_oh} presents models from \citetalias{Thompson_2016} with the \citetalias{PS_2020} cooling curves at varying metallicity values, $\log Z/Z_{\odot} = -1.5$, -1.0, -0.5, and 0.0. 
Each model is depicted as a polygon-shaded region, illustrating the parameter space that the model can reproduce. 
Figure \ref{fig:Lcool_SFR_vs_FWHM_plots_vary_oh} shows that the solar-metallicity models can produce higher $L_{5007}$ compared to those with low metallicities.
The lower-metallicity models typically show lower velocity widths than higher-metallicity ones at the same $L_{5007}$. This behavior occurs because the radiative cooling rates for lower-metallicity, higher $\eta_{\rm{M}}$ models are similar to those for higher-metallicity, lower $\eta_{\rm{M}}$ ones \citepalias[i.e., $v$ is inversely proportional to $\eta_{\rm{M}}$;][]{Thompson_2016}. 
Although dwarf galaxies might exhibit metal-enriched outflows with outflow metallicities around $Z_{\odot}$ \citep{Chisholm_2018}, Figure \ref{fig:Lcool_SFR_vs_FWHM_plots_vary_oh} demonstrates that even the solar-metallicity models can hardly explain the least luminous outflowing components. 
Additionally, these \citetalias{Thompson_2016} models cannot account for the low IPV$_{24}$ values of most broad components, which is consistent with Figure \ref{fig:Lcool_SFR_vs_FWHM_plots}.


\subsubsection{\texorpdfstring{Varying $\dot{\Sigma}_{\star}$ for T16 Models}{}} \label{subsubsec:varying_sfrsd}
\begin{figure*}[htb]
\centering
\includegraphics[width=0.49\textwidth]{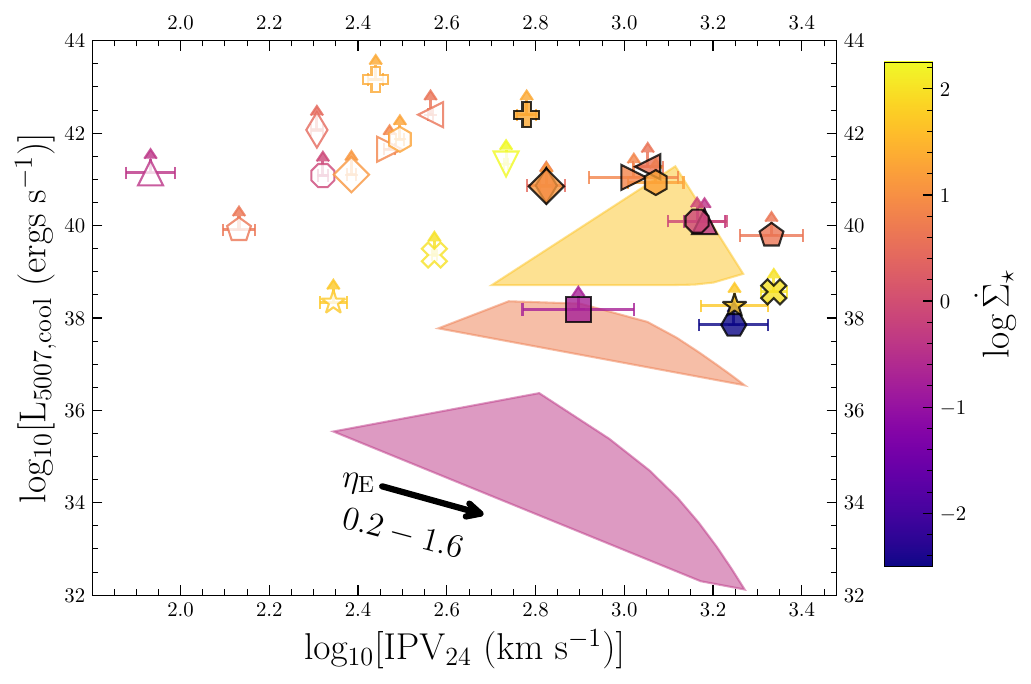}
\includegraphics[width=0.49\textwidth]{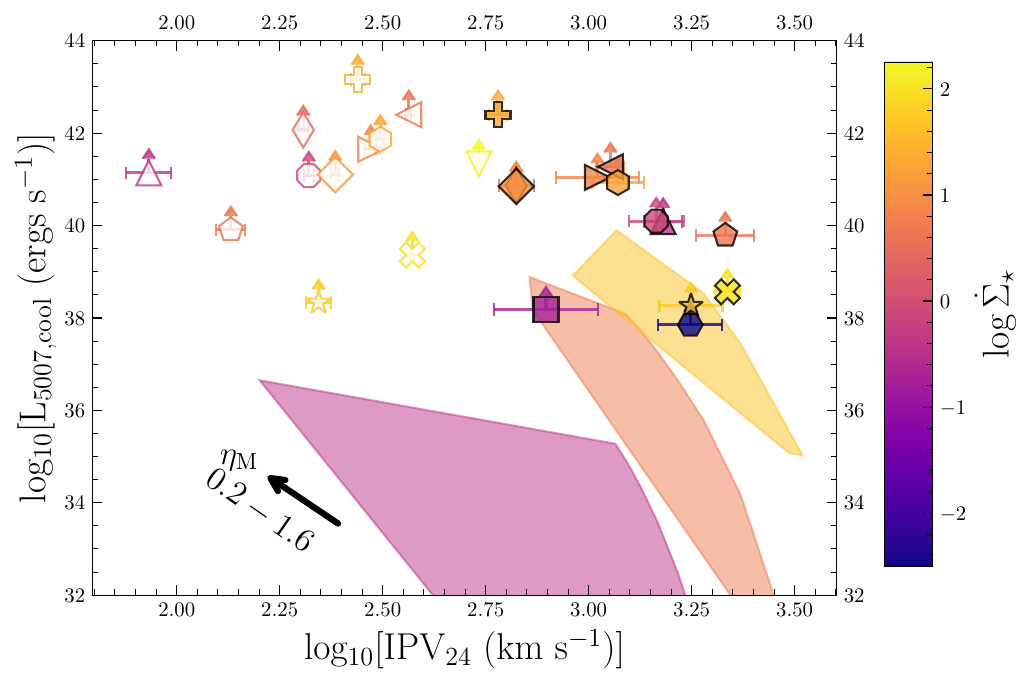}
\caption{Similar to Figures \ref{fig:Lcool_SFR_vs_FWHM_plots} and \ref{fig:Lcool_SFR_vs_FWHM_plots_vary_oh} but with varying $\dot{\Sigma}_{\star}$ and only showing $L_{5007}$. 
Each observation point or model polygon region (at the 16\textsuperscript{th}, 50\textsuperscript{th}, and  85\textsuperscript{th} percentile of \(\dot{\Sigma}_{\star}\) of our sample) is color-coded based on its $\dot{\Sigma}_{\star}$.
\label{fig:Lcool_SFR_vs_FWHM_plots_vary_sfrsd}}
\end{figure*}

\setlength{\tabcolsep}{6pt} 
\begin{deluxetable*}{ccccccc}
\tablecaption{\citetalias{Thompson_2016} Galactic Wind Model Parameters\label{table:model_parameters_t16}}
\tablehead{ 
\colhead{Figure} & \colhead{Panel} & \colhead{$\eta_{\rm{E}}$} & \colhead{$\eta_{\rm{M}}$} &
\colhead{log $\dot{\Sigma}_{\ast}$}  & \colhead{$v_{e}$} & \colhead{$\log Z / Z_{\odot}$} \\
\nocolhead{} & \nocolhead{} & \nocolhead{} & \nocolhead{} & \colhead{($\rm{M_{\odot} \  yr^{-1} \ kpc^{-2}}$)}  & \colhead{($\rm{km \ s^{-1}}$)} & \nocolhead{}
}
\decimalcolnumbers
\startdata
{Figure \ref{fig:Lcool_SFR_vs_FWHM_plots} or \ref{fig:thom16_radial_profiles}} & Left or Top & 0.2 -- 1.6 & 1.0 & 0.73 & 0, 400 & 0.0 \\
{} & Right or Bottom & 1.0 & 0.2 -- 1.6 & 0.73 & 0, 400 & 0.0 \\
{Figure \ref{fig:Lcool_SFR_vs_FWHM_plots_vary_oh}} & Left & 0.2 -- 1.6 & 1.0 & 0.73 & 0, 400 & -1.5, -1.0, -0.5, 0.0 \\
{} & Right & 1.0 & 0.2 -- 1.6 & 0.73  &  0, 400 & -1.5, -1.0, -0.5, 0.0 \\
{Figure \ref{fig:Lcool_SFR_vs_FWHM_plots_vary_sfrsd}} & Left & 0.2 -- 1.6 & 1.0 & -0.40, 0.73, 1.7  & 0, 400 & 0.0 \\
{} & Right & 1.0 & 0.2 -- 1.6 & -0.40, 0.73, 1.7 & 0, 400 & 0.0 \\
{Figure \ref{fig:correlation_plots_2}} & Lower Right & 1.0 & 0.2 -- 1.6 & -2.1, -0.40, 0.73, 1.7, 2.1 & 150 & 0.0 \\
\enddata
\tablecomments{Columns (3), $\eta_{\rm{E}}$, and (4), $\eta_{\rm{M}}$, represent the thermalization and mass-loading efficiency factors of hot winds. 
Column (5), log $\dot{\Sigma}_{\ast}$, is the input SFR surface density of the star-forming region. 
The selected size of the star-forming region, $R_{\rm{SF}}$, matches the median half-light radii $r_{\rm{50}} = 0.41 \ \rm{kpc}$ of our sample. 
Column (6), $v_e$, is the escape velocity assuming a singular isothermal potential. Column (7), $\log Z / Z_{\odot}$, is the gas-phase metallicity of the \citetalias{PS_2020} `UVB\_dust1\_CR1\_G1\_shield1' cooling tables. All cooling tables are assumed at redshift $z = 0$.}
\end{deluxetable*}

The luminosities of collisionally excited lines (also total cooling luminosities) are proportional to  \( n(r)^2 \). Given that \( n(r) \propto \dot{M}_{\rm{hot}}^{3/2} \ \dot{E}_{\rm{hot}}^{-1/2} \ R^{-2} \propto \rm{SFR} \ R^{-2} \propto \dot{\Sigma}_{\star} \), a greater value of \( \dot{\Sigma}_{\star} \) suggests elevated $L_{\rm{cool}}$ and $L_{5007}$. 
Lower-$\dot{\Sigma}_{\star}$ models generally show smaller IPV$_{24}$ values than higher-$\dot{\Sigma}_{\star}$ ones because the former require higher $\eta_{\rm{M}}$ to achieve comparable $L_{5007}$, resulting in smaller velocity widths (similar to the effect of varying $Z$).
These effects are elucidated in Figure \ref{fig:Lcool_SFR_vs_FWHM_plots_vary_sfrsd}, where each galaxy marker or model (polygon-shaded region) is color-coded by its \( \dot{\Sigma}_{\star} \) value. 
Similar to our argument in Section \ref{subsubsec:varying_metallicity}, even the \citetalias{Thompson_2016} models with the highest \( \dot{\Sigma}_{\star} \) values struggle to reproduce the luminosities of most outflowing components, and the predicted line widths of these models are generally larger than those of broad components but agree with VB components.

\subsection{Turbulent Radiative Mixing Between Hot and Cold Phases} \label{subsec:trml_hot_and_cold}
\begin{figure*}[htb]
\centering
\includegraphics[width=0.49\textwidth]{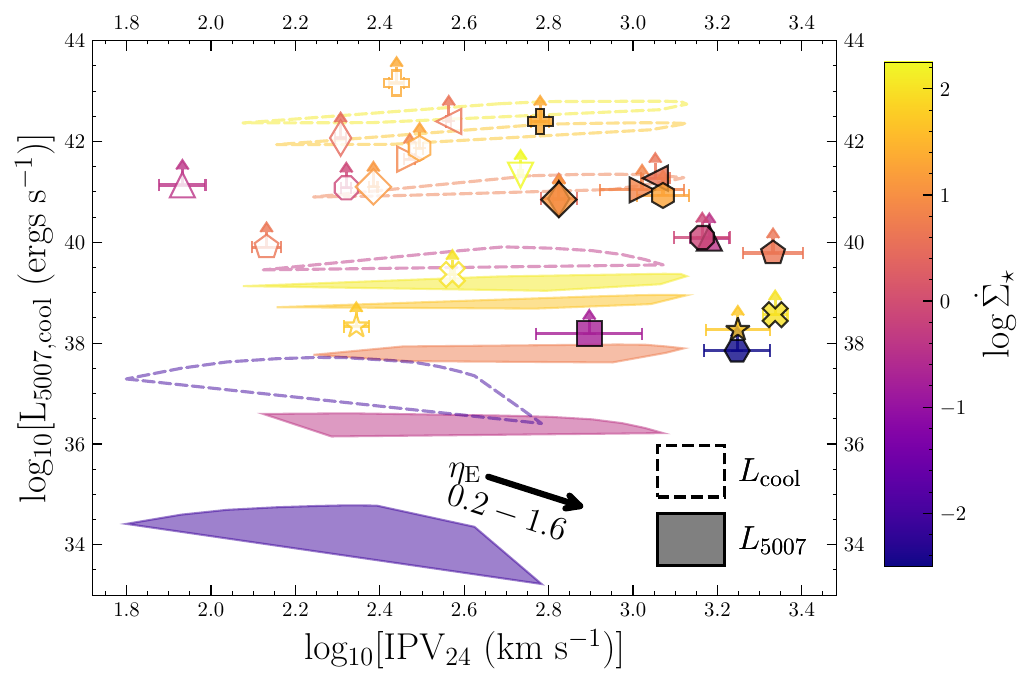}
\includegraphics[width=0.49\textwidth]{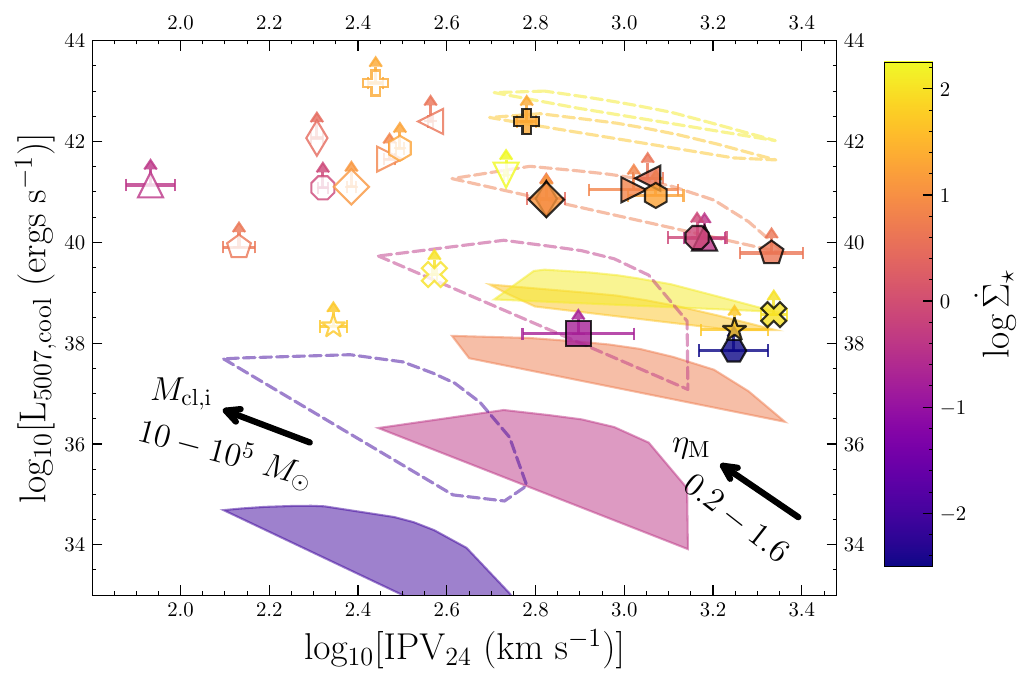}
\caption{Evaluating the role of TRMLs in the physical origins of outflowing components. $L_{5007}$ for each galaxy is plotted against the measured IPV$_{24}$ of outflowing velocity components. 
Based on \citetalias{Fielding_2022} galactic wind models with \(\eta_{\rm{M}} = 1.0\) (left panel) or \(\eta_{\rm{E}} = 1.0\) (right panel), 
polygon-shaded regions show the model predictions of $L_{5007}$ (within the effective ESI aperture radius $R_{\rm{aper}} = 2.75 \ {\rm{kpc}}$). 
Each shaded region, color-coded by its $\dot{\Sigma}_{\star}$ (at the 
5\textsuperscript{th},
16\textsuperscript{th},
50\textsuperscript{th}, 
84\textsuperscript{th}, and 95\textsuperscript{th}
percentile of our sample's \(\dot{\Sigma}_{\star}\)), represents the ranges of $M_{\rm{cl,i}}$ from $10 \ M_{\odot}$ to $10^{5.0} \ M_{\odot}$ ($\eta_{\rm{M,cl,i}}$ fixed at 0.1) and $\eta_{\rm{E}}$ from 0.2 to 1.6 in the left panel, or $\eta_{\rm{M}}$ from 0.2 to 1.6 in the right panel. 
The corresponding $L_{\rm{cool}}$, including the radiative cooling from the hot wind and the TRMLs, are shown as open-polygon shapes marked by dashed lines. 
The model $M_{\rm{cl,i}}$ value increases in the direction of the black arrow for each polygon region. 
Other key parameter values are summarized in Table \ref{table:model_parameters_fb22}.
\label{fig:Lcool_SFR_vs_FWHM_plots_fb22}}
\end{figure*}

\begin{figure*}[htb]
\centering
\includegraphics[width=0.49\textwidth]{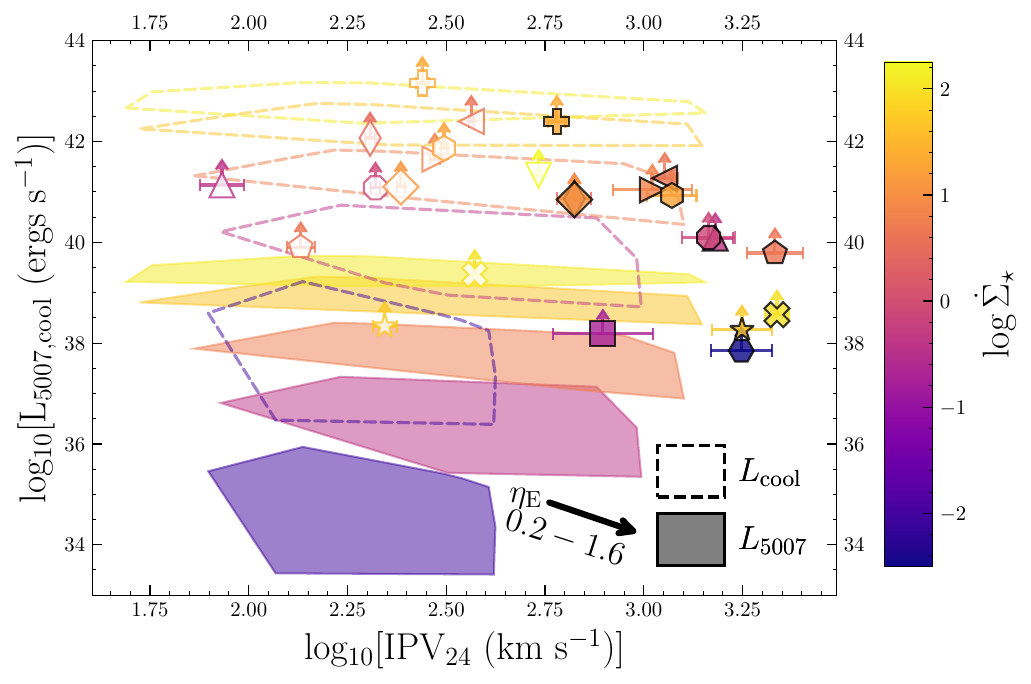}
\includegraphics[width=0.49\textwidth]{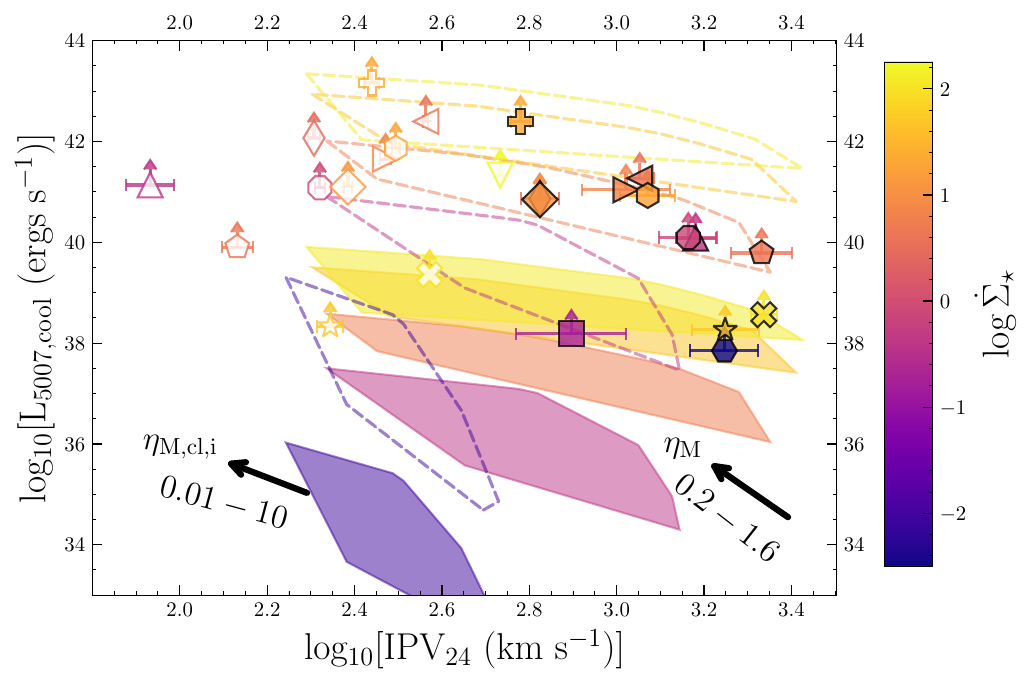}
\caption{Similar to Figure \ref{fig:Lcool_SFR_vs_FWHM_plots_fb22} but with varying $\eta_{\rm{M,cl,i}}$ (from 0.01 to 10) instead of $M_{\rm{cl,i}}$ values ($M_{\rm{cl,i}}$ fixed at $10^{2.5} \ M_{\odot}$). For each polygon region, the model $\eta_{\rm{M,cl,i}}$ value increases in the direction of the black arrow.
\label{fig:Lcool_SFR_vs_FWHM_plots_fb22_2}}
\end{figure*}

Recent studies on the turbulent radiative mixing between the hot and cold phases \citep[e.g.,][]{Fielding_2020, Fielding_2022, Chen_2023} suggest that mixing primarily drives the acceleration of cold clouds. 
The temperature and shear velocity differences between these two phases drive the advection of enthalpy and kinetic energy flux into the intermediate temperature regime ($T_{\rm{mix}} \sim \sqrt{T T_{\rm{cl}}} \sim 10^5 \ \mathrm{K}$) where the radiative cooling function peaks (see Figure \ref{fig:cooling_curves}), thereby forming TRMLs. 
Notably, the fraction of cooling luminosities contributing to a specific line within the hot winds and the TRMLs are not identical. 
Applying the 1.5-dimensional TRML model proposed in \cite{Chen_2023}, we generate a grid of pressure, \(P\), and the relative Mach number, \(\mathcal{M}_{\mathrm{rel}} = v_{\mathrm{rel}} / c_s\), to determine the flux fraction grid of [O\,{\footnotesize III}] $\lambda 5007$ within the mixing layers (i.e., the ratio between the line flux and the total cooling flux; see Appendix \ref{subsec:flux_frac_5007_trml} for details). 
We investigate whether the TRML model could solve the luminosity-deficit issue of most outflowing components discussed in Section \ref{subsec:hot_wind_cooling}.


As \citetalias{Fielding_2022} consider the metallicity transfer between hot and cold phases as a function of radius, we need to interpolate the cooling curves as a function of metallicity (besides temperature and density) to solve for the radial evolution of each physical quantity (i.e., metallicity is not a constant value as in \citetalias{Thompson_2016}). Therefore, we focus on examining how variations in the $\dot{\Sigma}_{\star}$ models impact our arguments. 
We are also interested in exploring how the velocity widths and cooling luminosities change with the initial cloud mass, $M_{\rm{cl,i}}$, considering that the measured $M_{\rm{cl}}$ from \citet{Xu_2023} display a wide range for our sample (with a median of $10^{2.85} \ M_{\odot}$ and a maximum of $10^{4.81} \ M_{\odot}$, although only 7 out of 14 galaxies have $M_{\rm{cl}}$ measurements). 
Besides $M_{\rm{cl,i}}$, the initial mass-loading factor of cold clouds, $\eta_{\rm{M,cl, i}}$, also effectively influences the velocity width of \citetalias{Fielding_2022} models, e.g., the $\eta_{\rm{M,cl, i}} = 1.0$ (or 0.1) model could yield a $v_{\rm{max}} \sim 500 \ \rm{km \ s^{-1}}$(or $1000 \ \rm{km \ s^{-1}}$). To quantify the impact of each parameter, we traverse in the $M_{\mathrm{cl,i}}$ and $\eta_{\rm{M,cl, i}}$ parameter spaces relevant to our sample in Figures \ref{fig:Lcool_SFR_vs_FWHM_plots_fb22} and \ref{fig:Lcool_SFR_vs_FWHM_plots_fb22_2}, respectively.

\subsubsection{\texorpdfstring{Varying $M_{\mathrm{cl,i}}$ and $\eta_{\rm{M,cl, i}}$ for FB22}{} Models}  \label{subsubsec:varying_beta_cl_and_M_cl}
Figure \ref{fig:Lcool_SFR_vs_FWHM_plots_fb22} illustrates the estimated $L_{5007}$ from both the hot winds and the TRMLs in the \citetalias{Fielding_2022} models, represented as polygon-shaded regions color-coded by their $\dot{\Sigma}_{\ast}$. 
The upper and lower bounds of each polygon correspond to models with initial cloud masses of $M_{\rm{cl,i}} = 10^{5.0} \ M_{\odot}$ and $10 \ M_{\odot}$, respectively. 
This distinction is due to the fact that more massive clouds are harder to accelerate. Consequently, the relative velocity, $v_{\rm{rel}} = v - v_{\rm{cl}}$, and the kinetic energy thermalization term \citepalias[$\propto v_{\rm{rel}}$; see Equation (26) in][]{Fielding_2022} between these phases will increase. As a result, models with more massive clouds exhibit higher cooling luminosities within the TRMLs but have smaller cloud velocities, reflected in smaller IPV$_{24}$ values (the trend of increasing model $M_{\rm{cl,i}}$ values is shown as a black arrow in the right panel). 

Each shaded polygon region in Figure \ref{fig:Lcool_SFR_vs_FWHM_plots_fb22_2} represents the range of \(\eta_{\rm{M, cl, i}}\) values from 0.01 to 10.0. 
Models with higher \(\eta_{\rm{M, cl, i}}\) exhibit a higher number flux of clouds or TRMLs at each radius (i.e., \(\dot{N}_{\mathrm{cl}} \propto \eta_{\mathrm{M,cl, i}}\)). 
Hence, these models demonstrate increased cooling luminosity and a reduced velocity width, illustrated with a black arrow in the right panel (similar to the effects of \(M_{\mathrm{cl,i}}\)). 
Since the total cooling luminosity of \citetalias{Fielding_2022} models is the sum of the luminosity radiated from both the hot winds and the TRMLs, models with higher \(\eta_{\rm{M, cl, i}}\) not only present increased luminosity radiated from the TRMLs but also a decrease from the hot winds. 
This compensation implies that the overall increase in total cooling luminosity with rising \(\eta_{\rm{M, cl, i}}\) is not substantial (i.e., \(\lesssim 1.0\) dex for a tenfold increase in \(\eta_{\rm{M, cl, i}}\)).

For Figures \ref{fig:Lcool_SFR_vs_FWHM_plots_fb22} and \ref{fig:Lcool_SFR_vs_FWHM_plots_fb22_2}, we incorporate the aperture effect, as detailed in Section \ref{subsec:hot_wind_cooling}, by integrating cooling emissivities only within the median effective ESI aperture radius of our sample, $R_{\rm{aper}} = 2.75 \ \rm{kpc}$. 
Although $R_{\rm{aper}}$ can reach up to approximately 6 kpc for the most redshift target (J0926+4427; extending only about half of the ESI slit), the majority of cooling luminosities are radiated within $\sim 3 \ \rm{kpc}$. 
Hence, our arguments are not qualitatively sensitive to our choice of $R_{\rm{aper}}$.

\begin{figure*}[htb]
\centering
\includegraphics[width=0.99\textwidth]{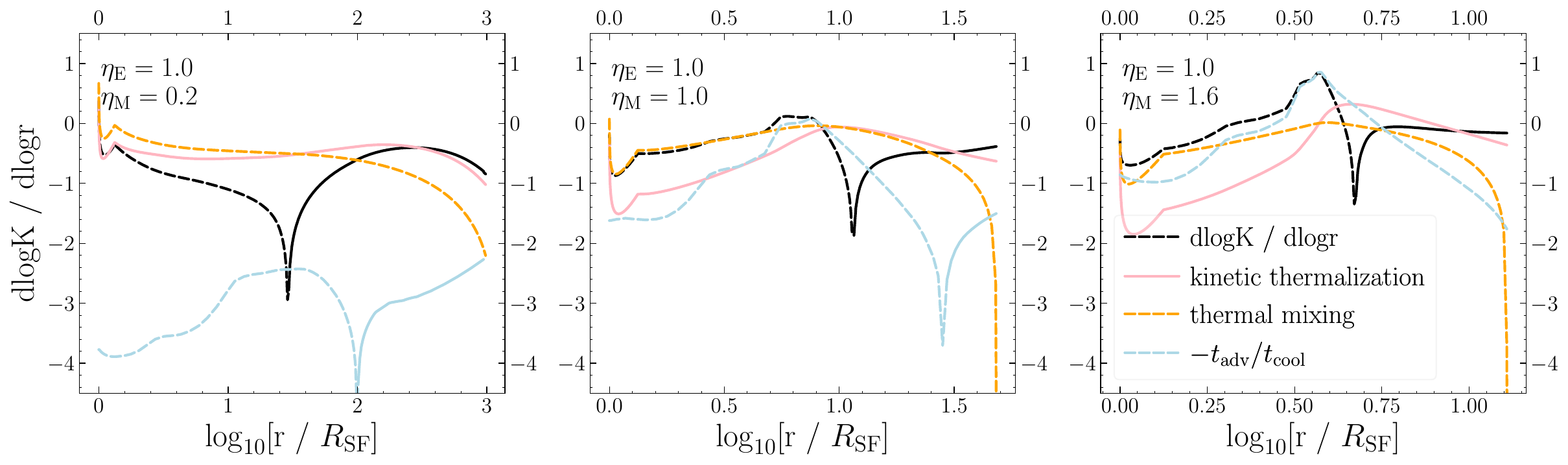}
\includegraphics[width=0.99\textwidth]{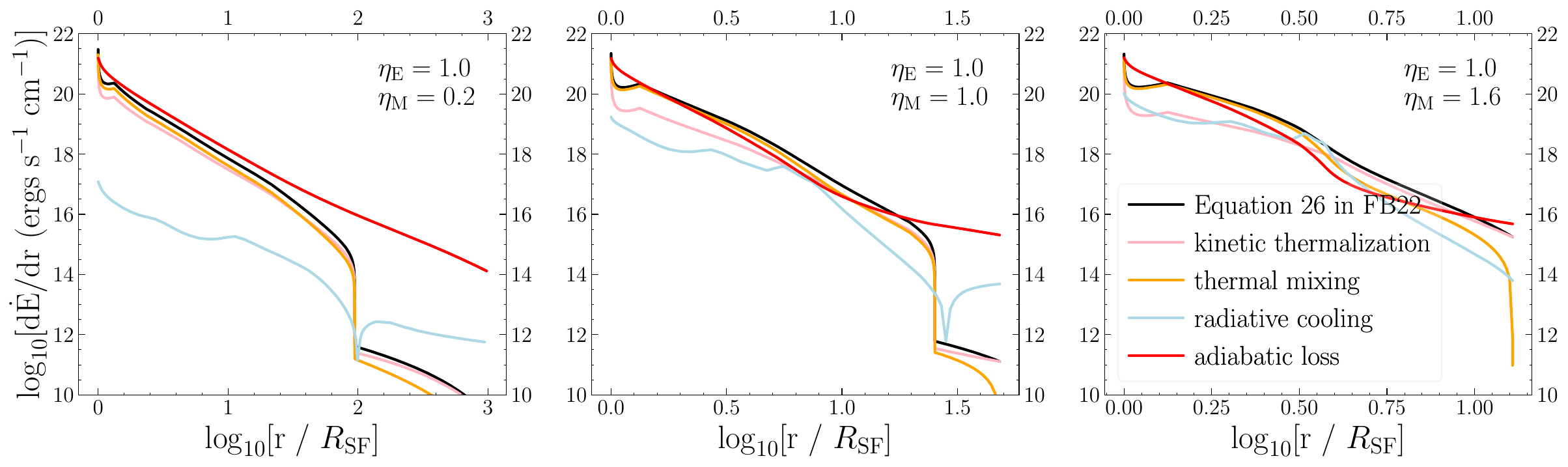}
\caption{\textit{Top}: Evolutions of entropy $K$ with radius for the
\citetalias{Fielding_2022} models at different $\eta_{\rm{M}}$. 
The expression of $\dd \log K / \dd \log r$ in the \citetalias{Fielding_2022} model is shown in Equation \ref{eq:7.9}, which includes four terms: (1) radiative cooling (i.e., $- t_{\rm{adv}} / t_{\rm{cool}}$; blue), (2) kinetic energy thermalization (pink), (3) thermal energy mixing (orange), and (4) drag work (not shown in the figure). 
For each top panel, solid lines indicate positive values and dashed lines indicate negative values. 
\textit{Bottom}:
Evolutions of the radiative cooling (blue), kinetic energy thermalization (pink), thermal energy mixing (orange), and adiabatic loss (red) terms of advection enthalpy with radius for the \citetalias{Fielding_2022} models at different $\eta_{\rm{M}}$. 
The summation of kinetic energy thermalization and thermal energy mixing terms \citepalias[Equation (26) in][]{Fielding_2022} is shown as the solid black line.
Each term of advection enthalpy is shown in Section \ref{sec:enthalpy_model}.
\label{fig:entropy_plots}}
\end{figure*}

Compared to the \citetalias{Thompson_2016} models, which fail to reproduce the velocity widths of broad components, the \citetalias{Fielding_2022} models can cover the range of IPV$_{24}$ values for both broad (at high $M_{\rm{cl,i}}$ or $\eta_{\rm{M,cl,i}}$) and VB components (at low $M_{\rm{cl,i}}$ or $\eta_{\rm{M,cl,i}}$).
Assuming both broad and VB components originate from the TRMLs, they should represent different cloud populations with distinct $M_{\rm{cl,i}}$ and $\eta_{\rm{M,cl,i}}$ values.
Nevertheless, we argue that multiple cloud populations cannot explain the difference in IPV$_{24}$ for broad and VB components (ranging from a factor of $\sim 2 - 10$ with a median of $\sim 4$). 
At the $\dot{\Sigma}_{\ast}$ values shown in 
Figures \ref{fig:Lcool_SFR_vs_FWHM_plots_fb22} and \ref{fig:Lcool_SFR_vs_FWHM_plots_fb22_2},
IPV$_{24}$ would increase by a factor of $\sim 1.2 - 2.4$ for $M_{\rm{cl,i}}$ varying from $10^5$ to $10 \ M_{\odot}$ (fixed $\eta_{\rm{M,cl,i}}$), and by a factor of $\sim 1.8 - 4.1$ for $\eta_{\rm{M,cl,i}}$ varying from 10 to 0.01 (fixed $M_{\rm{cl,i}}$), 
which are insufficient to explain the velocity difference between broad and VB components for most targets.

Moreover, since $\eta_{\rm{M,cl,i}}$ is inversely proportional to $M_{\rm{cl,i}}$ if multiple cloud populations exist \citep[i.e., populations with more massive clouds have lower $\eta_{\rm{M,cl,i}}$ than those with less massive clouds;][]{NG_2024}, multiphase models incorporating distinct cloud populations should show relatively small inter-population velocity dispersions compared to the observed velocity difference between broad and VB components \citep[see Figure 12 in][]{NG_2024}.
Consequently, if the VB component originates from the TRMLs of hot winds, the broad component cannot be generated within the same hot winds; therefore, the broad and VB components should have distinct physical origins. This conclusion is qualitatively consistent with the \citetalias{Thompson_2016} models, which can only explain the velocity widths of the VB components. 

\citetalias{Fielding_2022} models' estimated $L_{5007}$ are smaller than those of most outflowing components by at least one dex. 
Most outflowing components' observed $L_{5007}$ is similar to, or even exceeds, $L_{\rm{cool}}$ (open-polygon regions marked by dashed lines), which defines the total radiative cooling from the hot winds and TRMLs within the temperature range from $10^4$ to $10^{5.5} \ \rm{K}$, emitted within $R_{\rm{aper}}$. 

\setlength{\tabcolsep}{6pt} 
\begin{deluxetable*}{ccccccc}
\tablecaption{\citetalias{Fielding_2022} Galactic Wind Model Parameters\label{table:model_parameters_fb22}}
\tablehead{ 
\colhead{Figure} & \colhead{Panel} & \colhead{$\eta_{\rm{E}}$} & \colhead{$\eta_{\rm{M}}$} & \colhead{$\eta_{\rm{M,cl,i}}$} &
\colhead{log $\dot{\Sigma}_{\ast}$} &  \colhead{$M_{\rm{cl,i}}$}  \\
\nocolhead{} & \nocolhead{} & \nocolhead{} & \nocolhead{} & \nocolhead{} & \colhead{($\rm{M_{\odot} \  yr^{-1} \ kpc^{-2}}$)} & \colhead{($M_{\odot}$)} 
}
\decimalcolnumbers
\startdata
{Figure \ref{fig:Lcool_SFR_vs_FWHM_plots_fb22}} & Left & 0.2 -- 1.6 & 1.0 & 0.1 & -2.1, -0.40, 0.73, 1.7, 2.1  & $10$, $10^{2.5}$, $10^{5.0}$ \\
{} & Right & 1.0 & 0.2 -- 1.6 & 0.1 & -2.1, -0.40, 0.73, 1.7, 2.1 & $10$, $10^{2.5}$, $10^{5.0}$ \\
{Figure \ref{fig:Lcool_SFR_vs_FWHM_plots_fb22_2}} & Left & 0.2 -- 1.6 & 1.0 & 0.01, 0.1, 1.0, 10.0 & -2.1, -0.40, 0.73, 1.7, 2.1 &  $10^{2.5}$ \\
{} & Right & 1.0 & 0.2 -- 1.6 & 0.01, 0.1, 1.0, 10.0 & -2.1, -0.40, 0.73, 1.7, 2.1 & $10^{2.5}$ \\
{Figure \ref{fig:entropy_plots}} & All & 1.0 & 0.2, 1.0, 1.6 & 0.1 & 0.73 & $10^{2.5}$ \\
{Figure \ref{fig:correlation_plots_2}} & Lower Right & 1.0 & 0.2 -- 1.6 & 0.1 & -2.1, -0.40, 0.73, 1.7, 2.1 & $10^{2.5}$ \\
{Figure \ref{fig:fb22_radial_profiles}} & Top & 0.2 -- 1.6 & 1.0 & 0.1 & 0.73 &  $10^{2.5}$ \\
{} & Bottom & 1.0 & 0.2 -- 1.6 & 0.1 & 0.73 & $10^{2.5}$  
\enddata
\tablecomments{Columns (3), $\eta_{\rm{E}}$, and (4), $\eta_{\rm{M}}$, represent the thermalization-efficiency and mass-loading factors of hot winds. Column (5), $\eta_{\rm{M,cl,i}}$, indicate the initial mass-loading factor of cold clouds. Column (6), log $\dot{\Sigma}_{\ast}$, is the input SFR surface density. 
The selected size of the star-forming region, $R_{\rm{SF}}$, matches the median half-light radii $r_{\rm{50}} = 0.41 \ \rm{kpc}$ of our sample. 
Columns (7), $M_{\rm{cl,i}}$, is the initial cloud mass. 
Other \citetalias{Fielding_2022} parameters are set as constants: $v_{\rm{cir}} = 50 \ \rm{km \ s^{-1}}$, the circular velocity assuming a singular isothermal potential, $Z_{\rm{cl,i}} = 0.25 \ Z_{\odot}$, the initial cloud metallicity, $Z_{\rm{i}} = 2.0 \ Z_{\odot}$, the initial wind metallicity, and $v_{\rm{cl,i}} = 30 \ \rm{km \ s^{-1}}$, the initial cloud velocity.
All \citetalias{PS_2020} `UVB\_dust1\_CR1\_G1\_shield1' cooling tables are assumed at redshift $z = 0$.}
\end{deluxetable*}

\subsubsection{Why Does the Multiphase Model Fail to Generate Significantly More \texorpdfstring{$L_{5007}$ than the Single-phase Model?}{}}  \label{subsubsec:T16_FB22_comparisons}

Intuitively, we might think that turbulent radiative mixing between hot winds and cold clouds is a more plausible physical origin for these outflowing components than the single-phase theory. 
This is because TRMLs can emit more cooling luminosities in the intermediate temperature regime, where cooling curves peak, than in the hot-wind phase. 
However, this statement does not hold (1) in scenarios where the mass-loading factor of the hot phase is high (i.e., $\eta_{\rm{M}} \gtrsim 1.0$), and (2) considering that the flux fraction of [O\,{\footnotesize III}] $\lambda 5007$ in TRMLs ($\sim 0.1 \%$) is nearly an order of magnitude lower than in hot winds ($\sim 1 \%$).

To illustrate the first point, we show the evolutions of the radiative cooling (blue), kinetic energy thermalization (pink), and thermal energy mixing (orange) terms of entropy, $K = P / \rho^{\gamma}$, and advection enthalpy in radial space in Figure \ref{fig:entropy_plots} (see Appendix \ref{sec:enthalpy_model} or \ref{sec:entropy_model} for the derivation of each enthalpy or entropy term). 
The top panels of Figure \ref{fig:entropy_plots} illustrate that the contribution of radiative cooling to the entropy gradient, denoted as \(\dd \log K / \dd \log r\), is represented by the ratio of the wind advection time, \(t_{\rm{adv}} =r/v\), to the wind cooling time, \(t_{\rm{cool}} = P/(\gamma - 1) \dot{\epsilon}\), for both \citetalias{Thompson_2016} and \citetalias{Fielding_2022} models. 
When \(\eta_{{\rm{M}}}\) is small (\(\eta_{\rm{M}} = 0.2\)), $t_{\rm{adv}}$ is considerably shorter than $t_{\rm{cool}}$, indicating that most of the supernovae-injected energy is lost to adiabatic losses rather than radiative cooling (i.e., by 2$-$4 orders of magnitude; see the leftmost bottom panel in Figure \ref{fig:entropy_plots}). 
In this scenario, the entropy gradient is predominantly influenced by the interplay between kinetic energy thermalization and thermal energy mixing ($\propto c_s^2 - c_{s,\rm{cl}}^2$) across all radii. Conversely, for \(\eta_{\rm{M}} \geq 1.0\), the radiative cooling term becomes comparable to or even exceeds the other two terms, indicating efficient radiative cooling where \(t_{\rm{adv}} / t_{\rm{cool}} \gtrsim 1.0\)—reaching as high as 6.0 at \(\log[r / R_{\mathrm{SF}}] \approx 0.6\) for \(\eta_{\rm{M}} = 1.6\). 


The second point is evident from Figure \ref{fig:flux_fraction_5007_trml_grid}, especially when $\log\left[P / k_{\rm{B}} \ / \ \rm{(K \ cm^{-3})} \right] \gtrsim 1.5$. 
The disparity between the [O\,{\footnotesize III}] $\lambda 5007$ flux fraction in TRMLs and hot winds can be attributed to a relatively steep temperature gradient, $\dd T / \dd z$, within TRMLs at the temperature where the [O\,{\footnotesize III}] $\lambda 5007$ emissivity peaks (T $\sim 10^5$ K, see Figure \ref{fig:o3_5007_emissivities}). In general, the temperature gradient in a TRML is controlled by a complex interplay between physical processes like radiative cooling, conduction, and viscous heating. \cite{Chen_2023} find that near the cold phase (T $\sim 10^4$ K), viscous heating balances radiative cooling to produce a very shallow temperature gradient, resulting in a region of viscous heat dissipation that takes up a tiny fraction of the temperature space but a significant portion of the thickness of the mixing layer \citep[see Figure 7 of][]{Chen_2023}. 
However, since the emissivity of [O\,{\footnotesize III}] $\lambda 5007$ is negligible at $\sim 10^4$ K, this region of viscous dissipation, which typically takes up more than half of the thickness of a TRML, contributes little to the observed flux of [O\,{\footnotesize III}] $\lambda 5007$. 
In the remainder of a TRML, where the bulk of the temperature transition happens, conductive and radiative cooling dominate to drive a steep temperature gradient. This steep temperature gradient means that the temperature range where the [O\,{\footnotesize III}] $\lambda 5007$ emissivity peaks (T $\sim 10^5$ K) correspond to a very thin slice in TRMLs that contains few ions available for the observed [O\,{\footnotesize III}] $\lambda 5007$ flux.

\section{Discussion} \label{sec:discussion}
We investigate the excitation mechanisms (from strong emission line ratios) and the physical origins (from $L_{5007}$ and velocity widths) of each velocity component in Sections \ref{sec:bpt_diagram} and \ref{sec:broad_origins}. 
From the optical excitation diagnostics, we argue that most targets' narrow and broad components are dominated by stellar photoionization with little contribution from shocks or AGN activities. 
Since the VB component is only detected in [O\,{\footnotesize III}] $\lambda 5007$ and $\rm{H}\alpha$ for most targets, we are not able to employ similar excitation diagnostics on this faint velocity component. 
Hence, we compare the observed $L_{5007}$ of outflowing components to model predictions based on \citetalias{Thompson_2016} and \citetalias{Fielding_2022}. 
These comparative analyses suggest that broad and VB components should have distinct physical origins, and both single-phase and multiphase galactic wind models better reproduce the velocity widths of VB components.
Nonetheless, both wind models underestimate $L_{5007}$ of outflowing components by orders of magnitude (at least one dex for \citetalias{Fielding_2022} models). 
Therefore, other physical mechanisms (e.g., stellar photoionization) should be dominant in explaining the luminosities of outflowing components.
To further illuminate the physical origins of these outflowing components, we compare the outflows traced by UV absorption lines and optical emission lines in Section \ref{subsec:em_vs_ab}. 
Moreover, we investigate whether merger-induced shocks can provide an additional energy budget for the outflowing components in Section \ref{subsec:merger_flows}.
Last, we discuss the implications of the physical origins of broad and VB components for studies on high-redshift galaxies and AGNs in Section \ref{subsec:connect_high_z}.

\subsection{Emission-line Based Versus Absorption-line Based Outflows}\label{subsec:em_vs_ab}
Our arguments above are solely based on optical emission lines, which scale differently with gas density compared to UV absorption lines (i.e., the column density $N(v) \equiv \int n(v) \ d l$ and the emission measure $EM(v) \equiv \int n(v)^2 \ d l$). Consequently, it is crucial to explore if optical emission and UV absorption lines trace the same outflowing gas. 

\begin{figure}[htb]
\centering
\includegraphics[width=0.45\textwidth]{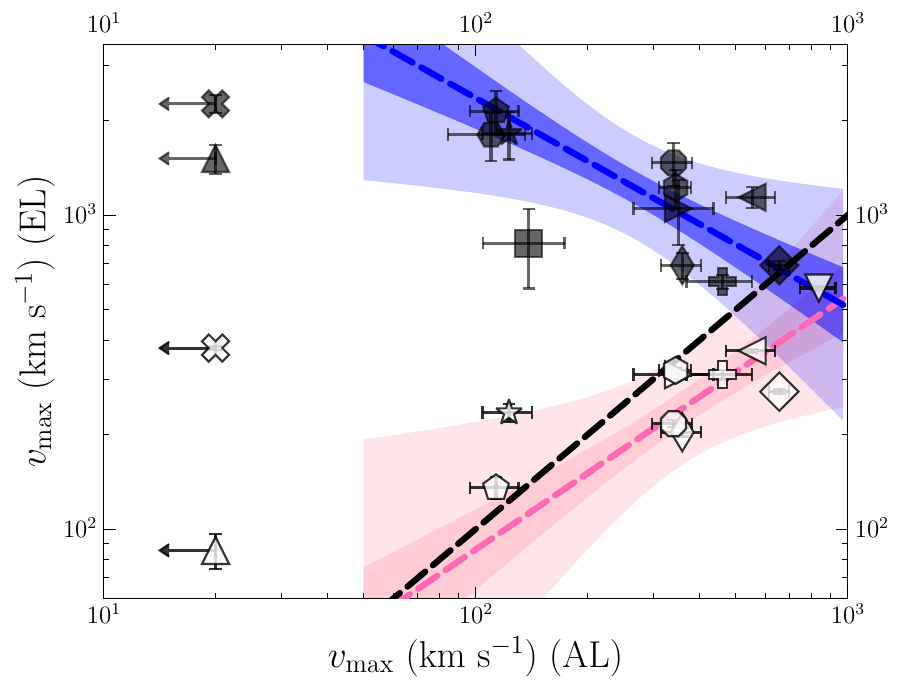}
\caption{
$v_{\rm{max}}$ derived from the outflowing components of absorption lines versus those from emission lines. 
The black dashed line denotes the line of equality. Galaxies labeled ``no outflows'' in \citetalias{Xu_2022} are assigned an upper limit of $v_{\rm{max}} \ (\rm{AL}) = 20 \ \rm{km \ s^{-1}}$ to include them in the plot. 
Using orthogonal distance regression, the best-fit linear relation for the broad (or VB) components is shown as the pink (or blue) dashed line (best-fit slopes and intercepts in log-log space is summarized in Table \ref{table:scaling_relations}). 
The pink or blue shaded areas represent these linear relations' $68 \%$ (dark) and $99 \%$ (light) confidence bands.
\label{fig:correlation_plots}}
\end{figure}

\begin{figure*}[htb]
\centering
\includegraphics[width=1.0\textwidth]{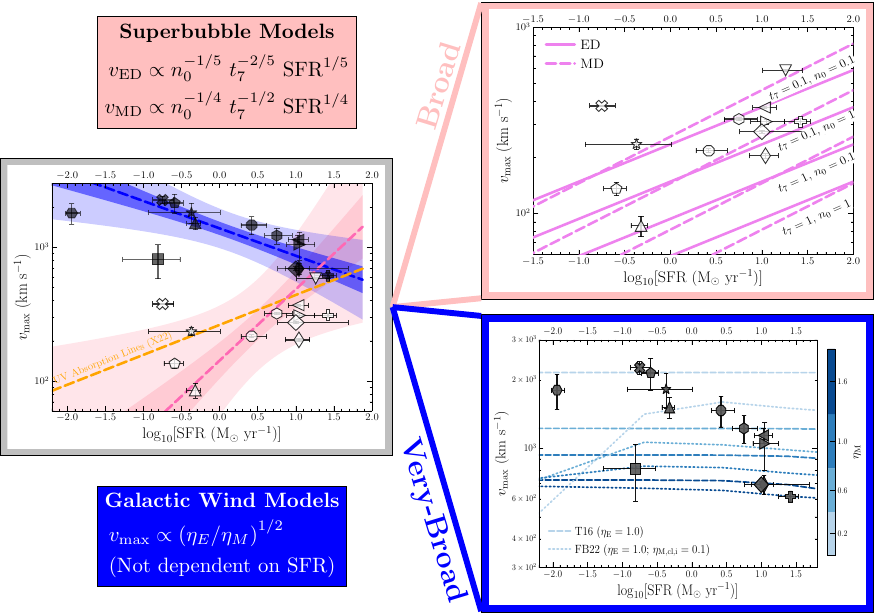}
\caption{
$v_{\rm{max}}$--SFR correlation derived from the outflowing components of absorption lines versus those from emission lines.  
\textit{Left}: Best-fitting linear relations for the broad and VB components compared to UV absorption lines from \citetalias{Xu_2022}, with the same plotting style as Figure \ref{fig:correlation_plots}.
The orange dashed line represents the best-fit linear relation determined by \citetalias{Xu_2022} using UV absorption lines.
\textit{Top Right:} The violet solid and dashed lines represent the predictions of energy-driven and momentum-driven superbubble models, respectively, for bubble expansion times of 1 Myr ($t_7 = 0.1$) and 10 Myr ($t_7 = 1.0$), and ISM ambient densities $n_0$ of 0.1 and 1 $\rm{cm^{-3}}$. 
Equations \ref{eq:5.2} and \ref{eq:5.4} show the expressions of shell velocity for both models.
The scatter of broad components in the $v_{\rm{max}}$–SFR space can be attributed to variations in bubble expansion times and ISM ambient densities across galaxies.
\textit{Bottom Right:} The dashed and dotted lines (color-coded by $\eta_{\rm{M}}$) illustrate the \citetalias{Thompson_2016} and \citetalias{Fielding_2022} galactic wind models, respectively. Both wind models align with the VB components of $\sim 1000$ km s$^{-1}$ reasonably well. 
\label{fig:correlation_plots_2}}
\end{figure*}

First, we test whether the absorption and emission lines share a similar $v_{\rm{max}}$. 
Figure \ref{fig:correlation_plots} shows that the ``no outflows'' galaxies categorized in \citetalias{Xu_2022}, such as J0127-0619 and J0337-0502, exhibit the existence of high-velocity cold gas ($v_{\rm{max}} \ \rm{(VB)} \gtrsim 1000 \ \rm{km \ s^{-1}}$). 
We argue that the substantial velocity discrepancy derived from emission and absorption lines in these ``no outflows'' galaxies might suggest the presence of spatially extended outflows, which are captured by the ESI slit but not by the HST/COS aperture. 
For example, a $\sim 15 \ \rm{kpc}$ outflow cone has been found around the starburst region of J0337-0502 \citep{Herenz_2023}.

Excluding these ``no outflows'' galaxies, $v_{\rm{max}}$ of broad components roughly align with $v_{\rm{max}}$ (AL) (see the caption of Table \ref{table:velocity_info}) from UV absorption lines for most galaxies. 
Using the orthogonal distance regression that considers the errors on both independent and dependent variables, the best-fit linear relation is shown as a pink dashed line (slope = $0.81 \pm 0.27$). 
Consistent with our results, \citet{Xu_2024} compare their $v_{\rm{max}}$ (the 95\textsuperscript{th} percentile velocity of the blueshifted, broad component) of H$\alpha$ and UV absorption lines in a larger sample and find $v_{\rm{max}} \ (\rm{H}\alpha) = (0.56 \pm 0.10) \ v_{\rm{max}} \ (\mathrm{AL}) - (40 \pm 30) \ \mathrm{km \ s^{-1}}$. 
\citet{Xu_2024} attribute the lower outflow velocities observed in optical emission lines compared to UV absorption lines to the different density dependencies of emission and absorption lines.
They also argue that the same radial outflow density and velocity fields can simultaneously account for the observed outflow features in both emission and absorption lines.

In contrast, $v_{\rm{max}}$ of VB components show a relatively negative correlation with $v_{\rm{max}}$ (AL) (slope = $-0.67 \pm 0.15$), but primarily due to the fact that our sample does not include galaxies that have valid $v_{\rm{max}}$ (AL) measurements of $\lesssim 100 \ \rm{km \ s^{-1}}$. 
If we include the ``no outflows'' galaxies in deriving the best-fit linear regression, the slope would appear flatter, indicating weak dependence on $v_{\rm{max}}$ (AL). 
Nevertheless, these markedly different correlations indicate distinct physical origins for the broad and VB components, consistent with the conclusion based on galactic wind models in Section \ref{sec:broad_origins}.
 
It is important to note that we do not directly compare $\mathrm{FWHM_{out}}$ (AL) and $\mathrm{FWHM_{out}}$ (EL) because absorption lines are generated only in the material situated between the source and the observer, whereas emission lines result from photons emitted in all directions, producing both positive and negative velocity components along the observer's line of sight. 
Consequently, $\mathrm{FWHM_{out}}$ (AL) and $\mathrm{FWHM_{out}}$ (EL) do not exhibit a one-to-one correspondence.

\setlength{\tabcolsep}{6pt} 
\begin{deluxetable}{ccc}
\tablecaption{Fitted Slopes and Intercepts for Scaling Relations\label{table:scaling_relations}}
\tablehead{
\colhead{Correlations} & \colhead{Slope} & \colhead{Intercept}  \\
\colhead{(1)} & \colhead{(2)} & \colhead{(3)}}
\startdata
\colhead{$v_{\rm{max}}$ (AL) vs. $v_{\rm{max}}$ (Broad)} & $0.81 \pm 0.27$ & $0.32 \pm 0.71$  \\
\colhead{$v_{\rm{max}}$ (AL) vs. $v_{\rm{max}}$ (VB)} & $-0.67 \pm 0.15$  & $4.71 \pm 0.39$ \\
\colhead{SFR vs. $v_{\rm{max}}$ (Broad)} & $0.53 \pm 0.26$ & $2.15 \pm 0.16$  \\
\colhead{SFR vs. $v_{\rm{max}}$ (VB)} & $-0.20 \pm 0.03$ & $3.14 \pm 0.03$  
\enddata
\tablecomments{Fitted slopes and intercepts of linear relations are in log–log scale.}
\end{deluxetable}

\subsubsection{Broad Velocity Components: \texorpdfstring{$v_{\rm{max}}$ vs. SFR}{}}\label{subsubsec:B_Vmax_vs_SFR}
Drawing from recent absorption line studies on the CLASSY sample \citepalias{Xu_2022}, we can examine whether the gas traced by emissions and absorptions has different scaling relations with the galaxy's global physical properties like SFR. 
\citetalias{Xu_2022} discuss how the momentum-driven clouds model proposed by \cite{Heckman_2015} offers a comprehensive understanding of the potential origins of outflowing gas as traced by absorption lines. 
\cite{Heckman_2015} derive a critical momentum flux, denoted as $\dot{p}_{\rm{crit}}$, at which the outward force from both core-collapsed supernovae (CCSN) and radiation pressure supplied by the starburst balances the inward gravitational force. Furthermore, \citetalias{Xu_2022} observe that the majority of the CLASSY galaxies fit within the strong-outflow regime, characterized by $\dot{p}_{\rm{out}} / \dot{p}_{\rm{crit}} > 10$. Within this regime, Equation (5) in \cite{Heckman_2015} anticipates that the maximum outflow velocity, $v_{\rm{max}}$, scales as $(\dot{p}_{\rm{out}} / \dot{p}_{\rm{crit}})^{0.5} \propto \rm{SFR}^{0.5}$. 
This scaling relation is roughly consistent with our best-fit linear relation for the broad components (slope = $0.53 \pm 0.26$). 
If these broad components trace the momentum-driven clouds, their scatter can be due to the extensive ranges of the half-light radii $r_{50}$, the hydrogen column density $N_{\rm{H}}$, and $v_{\rm{cir}}$ ($\dot{p}_{\rm{crit}} \propto r_{50} N_{\rm{H}} v_{\rm{cir}}^2$) of our sample.


\citetalias{Xu_2022} also claim that the momentum-driven bubble model \citep{Dyson_1989}, which suggests $v \propto \rm{SFR^{0.25}}$ can qualitatively explain the scaling relation between $v_{\rm{max}}$ and SFR from UV absorption lines (represented by the orange dashed line in Figure \ref{fig:correlation_plots_2}). This shallower slope is within one standard deviation of our best-fit slope.
Here we quantitatively compare the energy and momentum-driven bubble models and determine if they can reproduce the broad components' positions in $v_{\rm{max}}$-SFR space. 

The well-known energy-driven bubble model proposed by \cite{Weaver_1977} for stellar winds
can be extended to supernova-driven superbubbles \citep[e.g.,][]{Mac_Low_1988, El-Badry_2019} if the mechanical energy released remains roughly constant over time and successive supernovae fully deposit their mechanical energy within the hot bubble interior (i.e., remain subsonic), rather than in the cold shell. 
To be consistent with our analysis in Section \ref{sec:broad_origins}, we substitute the supernova rate with $\dot{E}_{\rm{hot}}$, as given in Equation \ref{eq:4.1}.
Following \cite{El-Badry_2019}, we can then rewrite the shell radius from the energy-driven superbubble model (when the shocked interstellar gas has collapsed into a thin, dense shell) as
\begin{align}
    \label{eq:5.1} R_{\rm{ED}} &= \left( \frac{125}{154 \pi} \ \right)^{1/5} \ \left(\dot{E}_{\rm{hot}} \ (1 - \theta_c) \right)^{1/5} \ \rho_0^{-1/5} \ t^{3/5}.
\end{align}
Compared to the original \cite{Weaver_1977} model, \cite{El-Badry_2019} introduce a cooling term $(1 - \theta_c)$ (shown in Equation \ref{eq:5.1}) to account for the fraction of energy not lost to radiative cooling at the shell-bubble interface \citep[$\theta_c \sim 0.6-0.8$ if the ISM ambient density $n_0 = 1 \ \rm{cm^{-3}}$; Figure 7 in][]{El-Badry_2019}. 
The corresponding shell velocity is defined as $\dd R_{\rm{ED}} / \dd t$, 
\begin{align}
    \label{eq:5.2} &v_{\rm{ED}} = \frac{3}{5}\left( \frac{125}{154 \pi} \ \right)^{1/5} \ \left(\dot{E}_{\rm{hot}} \ (1 - \theta_c) \right)^{1/5} \ \rho_0^{-1/5} \ t^{-2/5} \nonumber \\
    &= 82 \ \rm{km \ s^{-1}} \ n_0^{-1/5} \ t_7^{-2/5} \ \rm{SFR_1}^{1/5} \ \eta_{\rm{E}}^{1/5} \ (1 - \theta_c)^{1/5}.
\end{align}
Here $t_7$ is the bubble expansion time in units of $10^7$ years and $\rm{SFR}_1$ is in units of $1 \ \rm{M_{\odot} \ yr^{-1}}$. 
It is important to mention that $v_{\rm{ED}} \propto \rm{SFR}^{1/5}$, which is similar to the scaling relation of the momentum-driven model (shown below).

If cooling within the bubble is so intense that the majority of the input supernova energy is radiated away, the superbubble becomes momentum-driven rather than energy-driven. The shell radius in the momentum-driven model \citep[e.g.,][]{Koo_1992_I, Koo_1992_II, Lancaster_2021} is given by
\begin{align}
    \label{eq:5.3} R_{\rm{MD}} &= \left( \frac{3}{2 \pi} \ \right)^{1/4} \ \dot{p}_{\rm{hot}}^{1/4} \ \rho_0^{-1/4} \ t^{1/2},
\end{align}
where $\dot{p}_{\rm{hot}} = v_{\ast} (\dot{E}_{\rm{hot}} \dot{M}_{\rm{hot}})^{1/2}$ is the momentum input rate of supernova and $v_{\ast}$ is the dimensionless hot-wind velocity shown in Equation \ref{eq:4.3}. The corresponding shell velocity is 
\begin{align}
    \label{eq:5.4} &v_{\rm{MD}} = \frac{1}{2} \left( \frac{3}{2 \pi} \ \right)^{1/4}\ \dot{p}_{\rm{hot}}^{1/4} \ \rho_0^{-1/4} \ t^{-1/2} \nonumber \\
    &= 48 \ \rm{km \ s^{-1}} \ n_0^{-1/4} \ t_7^{-1/2} \ \rm{SFR_1}^{1/4} \ (\eta_{\rm{E}} \eta_{\rm{M}})^{1/8} \ v_{\ast,1}^{1/4}.
\end{align}
Here we normalize $v_{\rm{MD}}$ based on $v_{\ast}$ at the starburst radius (the sonic point) where $r_{\ast} = 1$. At this radius, $v_{\ast} (r_{\ast} = 1) = 0.707$. As $v_{\ast}$ approaches the asymptotic value, i.e., $v_{\ast} (r_{\ast} \gg 1) = 1.414$, we have $\dot{p}_{\rm{hot}} =(2\dot{E}_{\rm{hot}}\dot{M}_{\rm{hot}})^{1/2}$ \citep{Lancaster_2021}. 

Shell velocities of energy-driven (Equation \ref{eq:5.2}) and momentum-driven models (Equation \ref{eq:5.4}) are shown as violet solid and dashed lines in the upper right panel of Figure \ref{fig:correlation_plots_2}, respectively. These models are normalized based on the fiducial parameter values $\eta_{\rm{E}} = 0.65$, $\eta_{\rm{M}} = 1$, $v_{\ast,1} = 0.707$, and $\theta_c = 0.7$. 
$\eta_{\rm{E}} = 0.65$ accounts for approximately $65 \%$ of the total energy input left in the expanding bubbles \citep{El-Badry_2019}.
$t_7$ and $n_0$ (in units of $\rm{cm^{-3}}$) are set to 0.1 and 1.0 to reflect the scatters of stellar population ages (the median stellar population age at 51 Myr; see Table \ref{table:physical_properties}) and ISM ambient densities for our sample.
If the broad components indeed trace the expanding superbubble shells, their scatter should result from variations in bubble expansion times and ambient densities of different galaxies.
Consistent with this argument, \citet*{Martin_2024} demonstrate that the broad component in J1044+0353 (with a bubble expansion time of approximately 2 Myr) traces expanding superbubble shells from its starburst region.
This interpretation of broad components is also supported by the success of two-zone photoionization models with density-bounded channels in reproducing broad components' [S\,{\footnotesize II}]/$\rm{H}\alpha$ and [O\,{\footnotesize I}]/$\rm{H}\alpha$ ratios near the maximal starburst line. 
These density-bounded channels may result from anisotropic superbubble blowouts driven by stellar populations with $\gtrsim 10 \ \rm{Myr}$ in an inhomogeneous ISM and are subsequently ionized by young massive stars ($\lesssim 5 \ \rm{Myr}$), enabling anisotropic LyC photon leakage through these low-density channels \citep[the two-stage burst scenario for LyC escape;][]{Hogarth_2020, Martin_2024, Flury_2024}.

In summary, while parameters like $\eta_{\rm{E}}$, $\eta_{\rm{M}}$, $v_{\ast}$, and $\theta_c$ might not significantly influence our arguments within the parameter ranges of interest, the ambient ISM density $n_0$ and the bubble expansion time $t$ can vary substantially across different cases. 
Accurate measurements of these variables are crucial for more in-depth analysis. For instance, to determine $n_0$, a proper estimate of the filling factor $\epsilon$ may be required, as $n_0 \approx \epsilon n_e$. 
The expansion of superbubbles within the starburst region can be resolved using HST $\rm{H}\alpha$ imaging, providing insights into the bubble expansion time \citep*{Martin_2024}. 
Spatially resolved studies using HST images can explore whether relatively young stellar populations (i.e., $\lesssim 10$ Myr) exist in most of our sample.
These young stellar populations can be formed through the hierarchical triggering of star formation by expanding superbubbles \citep{Oey_2005}.
This spatial information will also help determine whether the aperture effect, characterized by substantial variation in physical scale (from $\sim 0.05 \ \rm{kpc \ \arcsec^{-1}}$ to $3.0 \ \rm{kpc \ \arcsec^{-1}}$) of our sample, influences our interpretation of the physical origin of broad components.
Such variation in physical scale results in vastly different fractional coverage within the same ESI aperture, ranging from single stellar clumps to nearly the entire galaxy.



\subsubsection{VB Velocity Components: \texorpdfstring{$v_{\rm{max}}$ vs. SFR}{}}\label{subsubsec:VB_Vmax_vs_SFR}
Analysis of galactic wind models suggests VB components should originate from galactic winds.
If this argument holds, one would expect no correlation (i.e., slope $\sim 0$) between the $v_{\rm{max}}$ of these VB components and SFR since the hot and cold phase outflow velocity are mainly proportional to the ratio of $\eta_{\rm{E}}$ and $\eta_{\rm{M}}$ (see Figures \ref{fig:thom16_radial_profiles} and \ref{fig:fb22_radial_profiles}). 
Contrarily, we find a shallow negative slope ($-0.20 \pm 0.03$) between $v_{\rm{max}}$ and SFR for VB components (left panel of Figure \ref{fig:correlation_plots_2}).
Nevertheless, this negative slope could result from different $\eta_{\rm{M}}$ values of our targets. Both the hot wind velocities from the single-phase \citetalias{Thompson_2016} model (dashed lines) and the cold cloud velocities from the multiphase \citetalias{Fielding_2022} model (dotted lines), which are insensitive to SFR, can reproduce the $v_{\rm{max}}$ of VB components with $\eta_{\rm{M}}$ ranging from 0.2 to 1.6 (with a fiducal $\eta_{\rm{E}}$ value at 1.0). 
Further investigations incorporating more targets at the lower SFR end could further prove this argument.
Despite their luminosity-deficit issue (Section \ref{sec:broad_origins}), both \citetalias{Thompson_2016} and \citetalias{Fielding_2022} models can explain the $v_{\rm{max}}$ and velocity widths (i.e., $\rm{IPV_{24,out}}$) of these VB components; therefore, other dynamic processes should dominantly contribute to the luminosity budget of these components.

\subsection{Merger-induced Shocks}\label{subsec:merger_flows}

Given that a significant fraction of our galaxy sample exhibits irregular morphologies, such as multiple nuclei (see Figure \ref{fig:esi_aperture}), it is plausible that merger-induced shocks could contribute to the luminosities of outflowing components through the thermalization of merger progenitors' orbital energy \citep{Cox_2004, Martin_2006}.

To determine which galaxies' outflowing components are significantly influenced by merger-induced shocks, we must identify which galaxies are mergers and whether their outflowing components are predominantly governed by shock-related mechanisms (as indicated by the excitation diagnostic diagrams in Section \ref{sec:bpt_diagram}). By measuring the merger morphology indicators \citep[$G$, $M_{20}$, $C$, and $A$;][]{Conselice_2003,Lotz_2004,Lotz_2008} in the DECalS g-band images, we classify our targets into two categories: (1) seven merger-like and (2) seven non-merger candidates (see Appendix \ref{sec:merger_indicators} for details). Among the seven merger-like candidates, only two, J0808+3948 and J1525+0757, display shock-like strong line ratios in their broad components.

To estimate the thermal energy from the merger-induced shocks, we follow \cite{Cox_2004} to assume the gas in the two merger progenitors will experience strong shocks when they follow nearly radial orbits (i.e., pass through rather than revolve around each other) during the final coalesced stage. The Rankine-Hugoniot jump conditions for a strong shock indicate a nearly 50$\%$ transfer efficiency from the kinetic energy prior to the final merger into thermal energy \citep[Equations (2) and (3) in][]{Cox_2004}: 
\begin{align}
      k T_{\rm{sh}} \sim \frac{3}{16} m_p v_{r}^2 \sim \frac{3}{16} m_p \frac{e^2}{1 + e} \frac{G M_{\rm{vir}}}{R_{\rm{peri}}},
     \label{eq:5.2.1}
\end{align} 
where $T_{\rm{sh}}$ is the gas temperature after the shock, $v_{r}$ is the radial velocity of the merger progenitor, and $M_{\rm{vir}}$ is the virial mass. 
Based on Table 2 in \cite{Cox_2004}, the initial pericentric distance and the orbit eccentricity are assumed to be $R_{\rm{peri}} \sim 1 - 10 \ \rm{kpc}$ and $e \sim 1$ (i.e., a parabolic trajectory), respectively. 

According to Equation \ref{eq:5.2.1}, the total gas thermal energy via the merger-induced strong shocks can be approximated as $\frac{3}{16} M_{\rm{g}} v_{r}^2 \sim \frac{3}{16} G M_{\rm{g}} M_{\rm{vir}} / R_{\rm{peri}}$, where $M_{\rm{g}}$ is the gas mass of the system. 
To determine the gas mass, we can first convert  $\dot{\Sigma}_{\ast}$ to the gas surface densities, $\Sigma_{g}$, by inverting the Kennicutt-Schmidt law \citep{Kennicutt_1998}. 
The gas fraction can be calculated as $\mu = \Sigma_{g} / (\Sigma_{g} + \dot{\Sigma}_{\ast})$, and then $M_{\rm{g}} = \mu M_{\ast} / (1 - \mu)$.
Utilizing the reported stellar mass from \cite{Berg_2022} and the stellar-to-halo mass relation from \cite{Girelli_2020}, we approximate the total gas thermal energy to be 
$10^{56} - 10^{57} \ \rm{ergs}$ for J0808+3948 and J1525+0757, when $R_{\rm{peri}}$ ranges from $1 - 10 \ \rm{kpc}$. 

We then estimate the duration of the starburst period for CCSN energy deposition based on the stellar population ages derived from \texttt{BEAGLE} SED fitting (see Table \ref{table:physical_properties}).
Assuming a starburst period of 
$10 - 100$ Myr for J0808+3948 or J1525+0757, the CCSN provide a mechanical energy of roughly 
$10^{55} - 10^{57}$ ergs for a SN energy injection rate of a constant star formation history (CSFH) based on Figure \ref{fig:sn_rate}. 
Since the merger-induced thermal energy from strong shocks is roughly comparable to the mechanical energy supplied by CCSN, we argue that the contribution of merger-induced shocks to the observed luminosity of broad components is not negligible in these two merger-like galaxies. 
Nevertheless, this merger process cannot explain the orders of magnitude luminosity-deficit issue discussed in Section \ref{sec:broad_origins} for most targets.
Because excitation diagnostic diagrams (Section \ref{sec:bpt_diagram}) suggest that stellar photoionization contributes to most of the luminosity of outflowing components, future investigations into multiphase galactic wind models should incorporate ionizing flux from the galaxy to validate this argument.

In conclusion, our previous assumption that a single physical mechanism is responsible for both the velocity widths and the luminosities of the outflowing components is not technically accurate.
Quantitatively disentangling these dynamic processes is nearly impossible because no self-consistent physical model in the literature accounts for all these processes simultaneously. 
From the observers' perspective, it is imperative to extract more spatial information, including determining the spatial variation of spectral line surface brightness and shape on a large galactic scale, along the ESI slit or by leveraging IFU (integral field unit) data, such as from KCWI/KCRM, to gain more insights into the physical origins of the outflowing components.

\subsection{Implications for High-redshift Galaxies and AGNs} \label{subsec:connect_high_z}

High-velocity galactic winds (i.e., FWHM$_{\rm{out}}$ $\gtrsim 1000 \ \rm{km \ s^{-1}}$) in star-forming galaxies are extremely rare in both the local and high-redshift universe \citep{Heckman_2015, Carniani_2024}.
The threshold of $\gtrsim 1000 \ \rm{km \ s^{-1}}$ is often used to define broad lines originating from an AGN \citep[e.g.,][]{Harikane_2023}, as most star-forming galaxies report FWHM$_{\rm{out}}$ values below $500 \ \rm{km \ s^{-1}}$. 
Nevertheless, our study systematically reports extreme outflows in local star-forming dwarf ($M_{\ast} \lesssim 10^{10} \ M_{\odot}$) galaxies, with 9 out of 14 exhibiting VB components with FWHM$_{\rm{out}} \gtrsim 1000 \ \rm{km \ s^{-1}}$.
We argue that some ``winds'' identified in the literature (FWHM$_{\rm{out}}$ $\lesssim 500 \ \rm{km \ s^{-1}}$) may actually be expanding superbubble shells, as indicated by the broad components ($150 \ \rm{km \ s^{-1}} \lesssim FWHM_{\rm{out}} \lesssim 500 \ \rm{km \ s^{-1}}$), rather than galactic winds escaping from the gravitational potential well, which are instead traced by the VB components ($\gtrsim 500 \ \rm{km \ s^{-1}}$). 

In addition to this work, a few relatively massive ($M_{\ast} \gtrsim 10^{10} \ M_{\odot}$) local galaxies exhibiting extreme outflows ($v_{\rm{max}} \gtrsim 1000 \ \rm{km \ s^{-1}}$) have been reported in the literature \citep{Diamond-Stanic_2012, Heckman_2016, Perrotta_2023}. 
Their high-redshift counterparts are predicted by the Feedback-Free Starbursts model proposed in \citet{Li_2024} when the star formation efficiency is high (i.e., $\gtrsim 0.35$; see their Equation 22).
As shown in Figure \ref{fig:correlation_plots_2}, since $v_{\rm{max}}$ is not sensitive to global galaxy properties like $M_{\ast}$ and SFR, we anticipate that future spectroscopic observations with SNR of [O\,{\footnotesize III}] $\lambda 5007$ or H$\alpha$ $\gtrsim 200$ to reveal more dwarf star-forming galaxies exhibiting extreme outflows.
Consequently, future studies should consider the possibility of extreme outflows in star-forming galaxies with wind velocities comparable to AGN-driven winds, which could blur the boundary between star-forming galaxies and AGNs solely based on the FWHM of outflowing components in optical emission and UV absorption lines. 

\begin{figure}[htb]
    \includegraphics[width=0.5\textwidth]{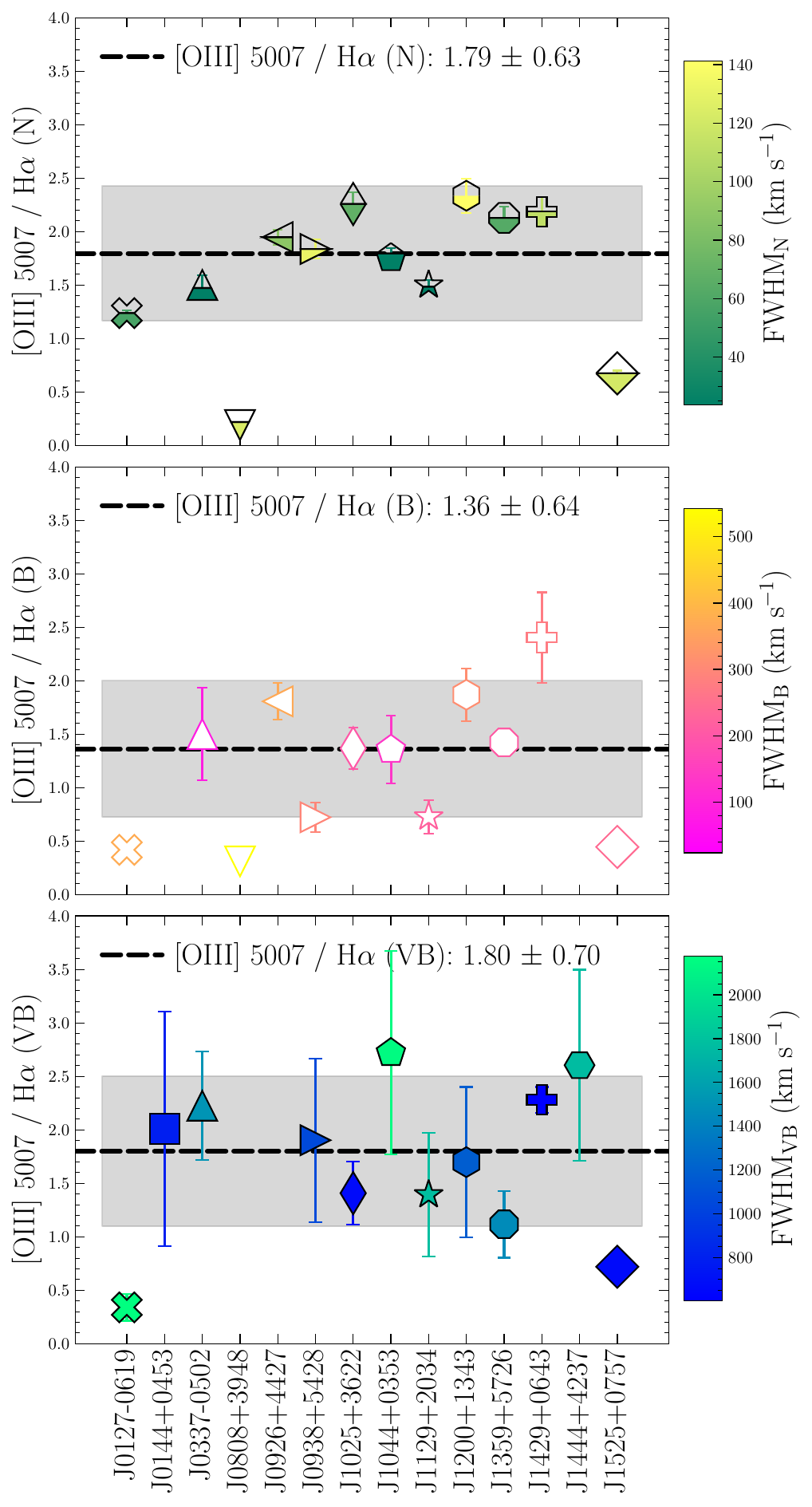}
\caption{[O\,{\footnotesize III}] $\lambda 5007$ / H$\alpha$ flux ratio for the narrow (\textit{Top}), broad (\textit{Middle}), and VB (\textit{Bottom}) components. Each velocity component's observed flux ratio is color-coded by its corresponding FWHM. 
The median flux ratio for each component and its scatter are indicated by the black dashed line and the grey shaded area, respectively, with their values reported in the upper-left corner of each panel.
\label{fig:o3_over_halpha}}
\end{figure}

Our work also suggests the inappropriate assumption of a consistent O3H$\alpha$ ratio across all velocity components has been used in the literature to justify the absence of the high-velocity component in the forbidden line [O\,{\footnotesize III}] $\lambda 5007$ in potential high-redshift Type-1 AGNs \citep[e.g., Figure 4 in][]{Xu_JWST_2023}.
Figure \ref{fig:o3_over_halpha} shows the distribution of observed [O\,{\footnotesize III}] $\lambda 5007$ / H$\alpha$ (O3H$\alpha$) flux ratio of each velocity component. 
In general, narrow components have a median O3H$\alpha$ ratio ($1.79 \pm 0.63$) similar to that of VB components ($1.80 \pm 0.70$) and consistent with that of broad components ($1.36 \pm 0.64$) within one standard deviation.
However, since individual galaxies typically do not necessarily exhibit similar O3H$\alpha$ across different components, it is not valid to fix the same flux ratio between each velocity component of [O\,{\footnotesize III}] $\lambda 5007$ and H$\alpha$ when determining the best-fitting three-component model.
For example, we find that the outflowing components of J0127-0619 (also known as Mrk 996) have significantly lower O3H$\alpha$ ratios than its narrow component.
This could be due to the extremely high number of dense clouds ($n_e$ of $\sim 10^7 \ \rm{cm^{-3}}$) in its core region \citep{James_2009} for the broad component, where the $n_e$ is high enough to collisionally de-excite [O\,{\footnotesize III}] $\lambda 5007$, with a critical density of $\sim 10^6 \ \rm{cm^{-3}}$.

\section{Conclusion} \label{sec:conclusion}
We have studied the physical origins of outflowing cold gas, traced by broad (FWHM $\sim 260 \ \rm{km  \ s^{-1}}$) and VB (FWHM $\sim 1200 \ \rm{km \ s^{-1}}$) velocity components of strong optical emission lines, in a sample of 14 low-redshift star-forming dwarf galaxies drawn from CLASSY \citep{Berg_2022} using Keck/ESI data. We summarize our main results from this study as follows:

\begin{enumerate}[leftmargin=*]
    \item We have detected up to three velocity components in strong emission lines like $\rm{H}\alpha$ and [O\,{\footnotesize III}] $\lambda 5007$. 
    The detected first, second, and third velocity components are defined as the narrow, broad, and VB components for galaxies without the double-narrow component feature.
    The approximate boundaries between these velocity components are based on the median and one-sigma scatter of the FWHM
    and the component/total flux ratio of [O\,{\footnotesize III}] $\lambda 5007$: (1) ``\textit{narrow}'': $\rm{FWHM} \lesssim 150 \ \rm{km \ s^{-1}}$ and $F_{\rm{comp}} / F_{\rm{total}} \gtrsim 0.40$ (2)``\textit{broad}'': $150 \ \rm{km \ s^{-1}} \lesssim \rm{FWHM} \lesssim 500 \ \rm{km \ s^{-1}}$ and $ 0.10 \lesssim F_{\rm{comp}} / F_{\rm{total}} \lesssim 0.40$ 
    (3) ``\textit{very-broad (VB)}'': $\rm{FWHM} \gtrsim 500 \ \rm{km \ s^{-1}}$ and $F_{\rm{comp}} / F_{\rm{total}} \lesssim 0.10$. 
    \item We have quantitatively modeled the cooling luminosities and velocity widths of [O\,{\footnotesize III}] $\lambda 5007$ using both single-phase \citepalias[i.e., direct condensation of hot winds;][]{Thompson_2016} and multiphase \citepalias[i.e., the TRML model;][]{Fielding_2022} supernova-driven galactic wind models to explore whether they align with those of the broad and VB components. 
    Our results indicate that single-phase models can explain the velocity widths of VB components but not those of broad components.
    Although multiphase models can cover the range of velocity widths for both broad (at high $M_{\rm{cl,i}}$ or $\eta_{\rm{M,cl,i}}$) and VB (at low $M_{\rm{cl,i}}$ or $\eta_{\rm{M,cl,i}}$) components, we argue that multiphase models incorporating multiple cloud populations (with distinct $M_{\rm{cl,i}}$ and $\eta_{\rm{M,cl,i}}$) should show relatively small inter-population velocity dispersions compared to the observed velocity differences between these two components. 
    Therefore, both models suggest that the VB components, but not the broad components, should originate from galactic winds.
    This argument is further supported by the fact that the scatter of VB components in the $v_{\rm{max}}$--$\rm{SFR}$ space can be explained by single-phase and multiphase galactic wind models with different $\eta_{\rm{M}}$.
    \item Single-phase galactic wind models significantly underestimate the luminosities of both broad and VB components by orders of magnitude. 
    Multiphase wind models reduce this shortfall to at least one dex for most VB components (i.e., the luminosity-deficit issue).
    The observed $L_{5007}$ of a few galaxies even exceed $L_{\rm{cool}}$ that includes all radiative cooling from the hot winds and TRMLs radiated within the temperature range from $10^4$ to $10^{5.5} \ \rm{K}$. 
    We find that the multiphase model can generate more radiative cooling within the TRMLs than within the hot winds only when the cold-phase mass-loading factor is small (i.e., $\eta_{\rm{M}} \lesssim 1.0$), and only a considerably small fraction ($\sim 0.1 \%$) of the TRML luminosities directly contribute to [O\,{\footnotesize III}] $\lambda 5007$.
    Consequently, most [O\,{\footnotesize III}] $\lambda 5007$ luminosity should stem from physical mechanisms other than mechanical energy from CCSN. 
    \item 
    We estimate the merger-induced thermal energy from strong shocks to be roughly $10^{56} - 10^{57} \ \rm{ergs}$ for J0808+3948 and J1525+0757 (the only two merger-like galaxies with shock-like strong line ratios in their broad components), which are comparable to their mechanical energy supplied by CCSN. 
    Therefore, the contribution of merger-induced shocks to the observed luminosity of these two merger-like galaxies' broad components should not be neglected.
    Nonetheless, we argue that merger-induced shocks cannot resolve the orders of magnitude luminosity-deficit issue observed for most targets.
    \item We propose that either classical energy-driven or momentum-driven superbubble models can account for the scatter of broad components in the $v_{\rm{max}}$--$\rm{SFR}$ space with different bubble expansion times ($1- 10$ Myr) and ISM ambient densities ($0.1 - 1 \ \rm{cm^{-3}}$).
    This result indicates that broad components trace expanding superbubble shells rather than galactic winds that can escape from the gravitational potential well.
    Consistent with \cite{Xu_2024}, we find that broad components exhibit lower $v_{\rm{max}}$ compared to UV absorptions, likely due to the different density dependencies of emission and absorption lines.
    \item
    We explore the excitation mechanism of each velocity component using excitation diagnostics diagrams and find that most of our targets are dominated by stellar photoionization, unlike in ULIRGs where broad emissions often trace shocked gas from galactic winds \citep{SM_2012_2}.  
    This result aligns with the findings of \cite{mingozzi_classy_2024} and suggests that stellar photoionization contributes to most of the luminosity of the broad and VB components.
    Nevertheless, the potential presence of an AGN in J1429+0643 and shocked gas in J1525+0757 cannot be ruled out.
    \item We also find broad components generally have higher low-ionization line ratios, including [N\,{\footnotesize II}]/$\rm{H}\alpha$, [S\,{\footnotesize II}]/$\rm{H}\alpha$, and [O\,{\footnotesize I}]/$\rm{H}\alpha$, and lower [O\,{\footnotesize III}]/$\rm{H}\beta$ ratio than those of narrow components. 
    One-zone (single $\log U$) photoionization models fail to explain broad components' [S\,{\footnotesize II}] and [O\,{\footnotesize I}] emissions near the maximal starburst line, which 
    two-zone photoionization models (two $\log U$ with differing covering fractions) with density-bounded channels instead reproduce.
    The success of two-zone photoionization models supports the anisotropic leakage of LyC photons through low-density channels formed by expanding superbubbles, which are traced by broad components.
    \item Our study highlights that low-mass ($M_{\ast} \lesssim 10^{10} \ M_{\odot}$) star-forming galaxies can power high-velocity galactic winds ($v_{\rm{max}} \gtrsim 1000 \ \rm{km \ s^{-1}}$) that are comparable to winds driven by Type-1 AGN.
    While extreme outflows have primarily been identified in relatively massive star-forming galaxies ($M_{\ast} \gtrsim 10^{10} \ M_{\odot}$), 
    we expect that future spectroscopic observations with SNR of [O\,{\footnotesize III}] $\lambda 5007$ or H$\alpha$ $\gtrsim 200$ to uncover more dwarf star-forming galaxies exhibiting these extreme outflows, as the $v_{\rm{max}}$ of galactic winds appears insensitive to global galaxy properties such as $M_{\ast}$ and SFR.
    Furthermore, we find that the [O\,{\footnotesize III}] $\lambda 5007$/H$\alpha$ flux ratios for different velocity components are not necessarily identical for individual galaxies. 
    This result suggests the misclassification of high-redshift Type-1 AGNs in studies that assume a consistent [O\,{\footnotesize III}] $\lambda 5007$/H$\alpha$ ratio across all velocity components to rule out the high-velocity component ($v_{\rm{max}} \gtrsim 1000 \ \rm{km \ s^{-1}}$) in the forbidden line [O\,{\footnotesize III}] $\lambda 5007$ \citep[e.g.,][]{Xu_JWST_2023}.
\end{enumerate}
Our study quantitatively demonstrates the diverse physical origins of outflowing cold clouds in local star-forming galaxies and highlights that multiple physical mechanisms are responsible for explaining the velocity widths and luminosities of the broad and VB components.

Future work can leverage spatial resolution along the ESI slit and new IFU observations with KCWI/KCRM to test our expectation that VB components (originating from galactic winds) are more spatially extended than broad components (tracing expanding superbubble shells) by examining their spatial variations of surface brightness. 
IFU observations will also clarify the impact of aperture effects on our interpretations of the physical origins of outflowing components$-$for instance, whether superbubble models would still explain broad components when the aperture size is limited to the scale of these shells.
Additionally, HST H$\alpha$ imaging can resolve expanding ionized gas shells, offering valuable insights into the bubble expansion time.
Furthermore, since VB components in most targets are not observable in low-ionization lines and thus not available for excitation diagnostic analysis, we emphasize the necessity of incorporating ionizing flux from central continuum sources, such as massive stars, in a multiphase galactic wind model. 
This approach will help determine whether stellar photoionization is the primary source of the VB components' luminosity.




\facilities{Keck:II (ESI)}

\software{{\tt\string Astropy} \citep{astropy:2022},
{\tt\string BEAGLE} \citep{chevallard_modelling_2016},
{\tt\string LMFIT} \citep{lmfit},
{\tt\string Matplotlib} \citep{Matplotlib:2007},
{\tt\string NumPy} \citep{numpy},
{\tt\string pandas} \citep{reback2020pandas},
{\tt\string Photutils} \citep{larry_bradley_2022_6825092},
{\tt\string PyNeb} \citep[][]{pyneb,pyneb_2015}, 
{\tt\string PypeIt} \citep{pypeit:zenodo},
{\tt\string SciPy} \citep{2020SciPy-NMeth},
{\tt\string statmorph} \citep{statmorph}, 
{\tt\string VerEmisFitting} \citep{veremisfit}}

\begin{acknowledgments}
Z.P. would like to acknowledge Joseph Hennawi and J. Xavier Prochaska for their guidance through the ESI data reduction procedures. Z.P. also appreciates the insightful discussions with Dalya Baron, Chad Bustard, Greg Bryan, Christophe Morisset, Mordecai Mac Low, Matilde Mingozzi, Norman Murray, Aaron Smith, and Yongda Zhu. 
Z.P. has gained valuable insights from the KITP Program ``Cosmic Origins: The First Billion Years''.
The authors are grateful to the anonymous referee for constructive suggestions. 
Z.P. sincerely acknowledges support for this work from NSF grant AST-1817125, NASA FINESST (Future Investigators in NASA Earth and Space Science and Technology) grant 80NSSC23K1450, 
HST-GO-15840.008-A, 
and HST-GO-15866.002-A.
L.R. gratefully acknowledges funding from the Deutsche Forschungsgemeinschaft (DFG, German Research Foundation) through an Emmy Noether Research Group (grant number CH2137/1-1).
\end{acknowledgments}

\appendix
\section{Line-fitting Results} \label{sec:line_fitting_results}
The best line-fitting result of each galaxy is shown in Figure \ref{fig:line_fittings}. 
The line fluxes and derived physical properties of each velocity component are summarized in Tables \ref{table:line_Flux} and \ref{table:derived_physical_properties}, respectively.


\begin{figure*}[htbp]
\gridline{\fig{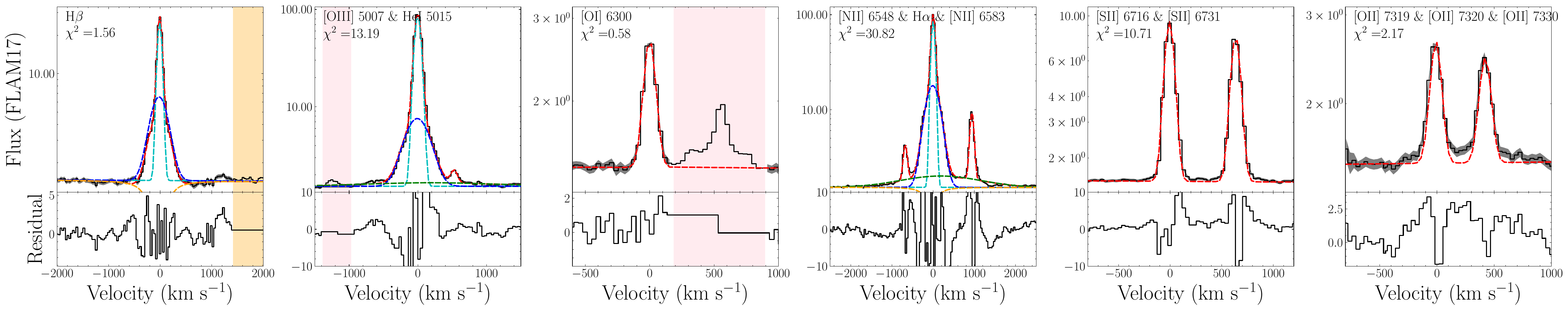}
{1.0\textwidth}{\textbf{{\large (J0127-0619)}}}}
\gridline{\fig{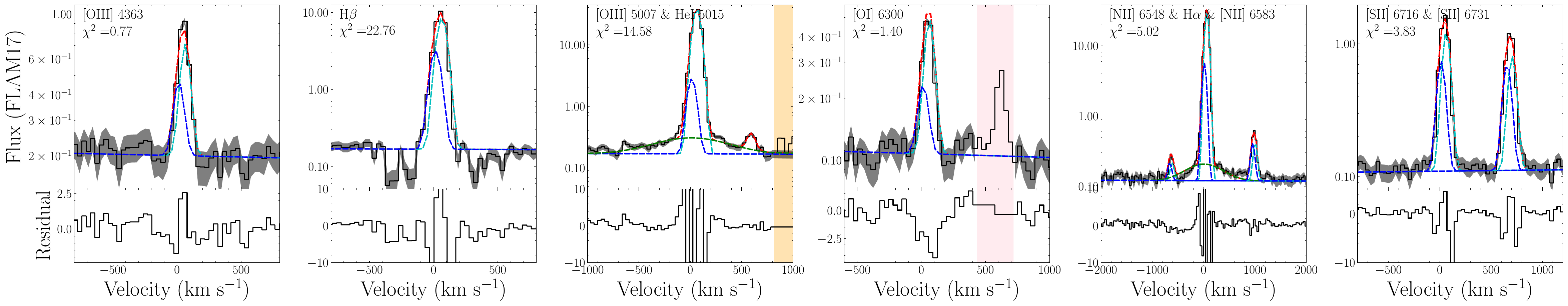}
{1.0\textwidth}{\textbf{{\large (J0144+0453)}}}}
\gridline{\fig{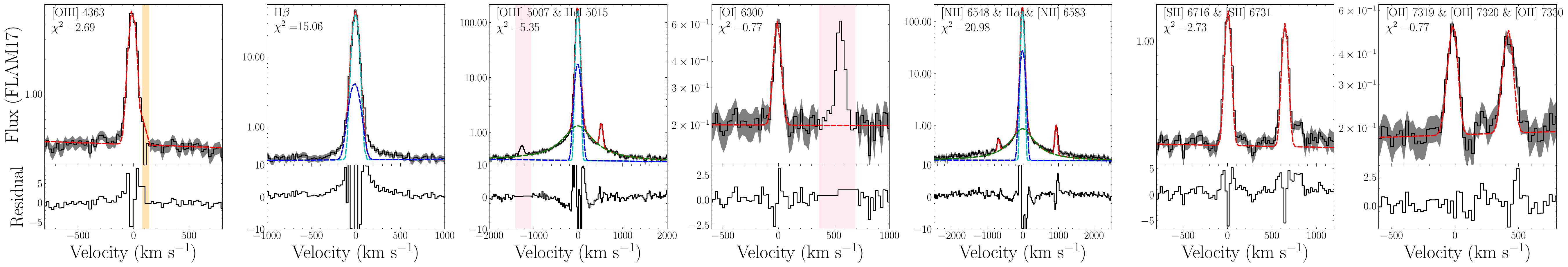}
{1.0\textwidth}{\textbf{{\large (J0337-0502)}}}}
\gridline{\fig{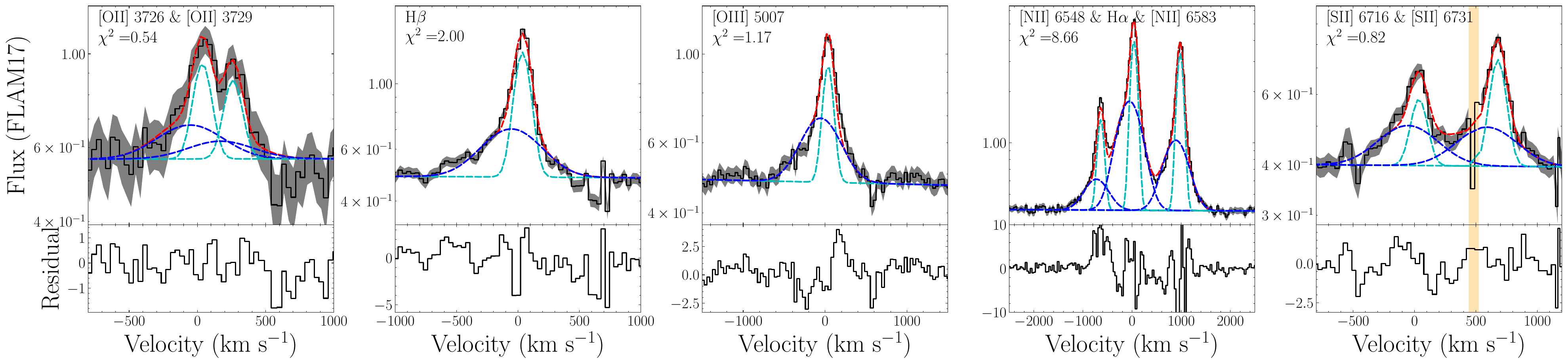}
{1.0\textwidth}{\textbf{{\large (J0808+3948)}}}}
\caption{Line-fitting results of each galaxy. 
The best-fitting model (red dashed line) is plotted alongside the observed line profile (black solid line) in the first row, with flux units labeled as ``FLAM17'' ($10^{-17} \ \rm{erg \ s^{-1} \ cm^{-2} \ (km \ s^{-1})^{-1}}$) in velocity space.
The narrow, broad, and VB emission velocity components are shown as cyan, blue, and green-dashed lines for multi-component fittings. 
For DNC galaxies (J0144+0453 and J1444+4237), the weaker and stronger narrow components are represented by blue and cyan dashed lines, respectively.
The orange dashed lines represent the Lorentzian profiles that fit the Balmer stellar absorption lines.
The grey areas represent the 1-$\sigma$ uncertainties of raw data. 
The residual plot (data - model / error) is displayed in the second row, with certain strong lines (e.g., [O\,{\footnotesize III}] $\lambda 5007$ and H$\alpha$) zoomed in to emphasize residuals in the broad wings.
Strong emission lines (or bad sky-subtraction features) next to the intended lines are masked during line fittings and are shaded in pink (or yellow). 
Please see Section \ref{subsec:line_fitting} for details. \label{fig:line_fittings}}
\end{figure*}

\begin{figure*}[htbp]
\addtocounter{figure}{-1}
\gridline{\fig{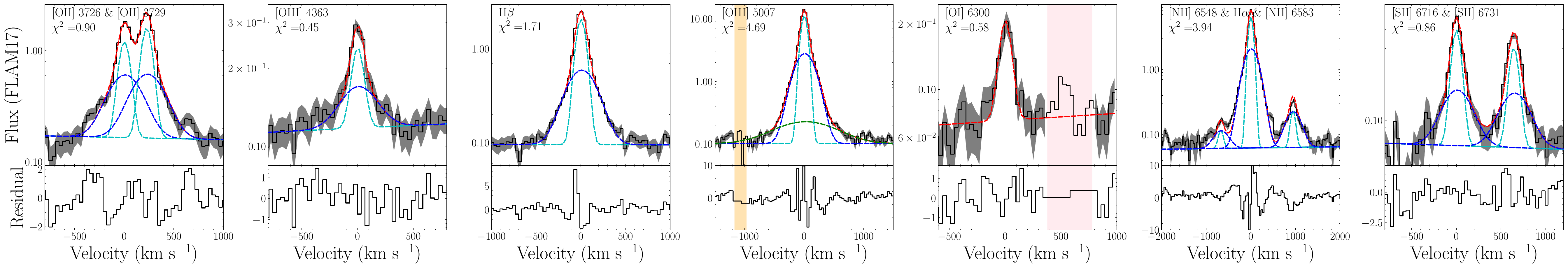}
{1.00\textwidth}{\textbf{{\large (J0926+4427)}}}}
\gridline{\fig{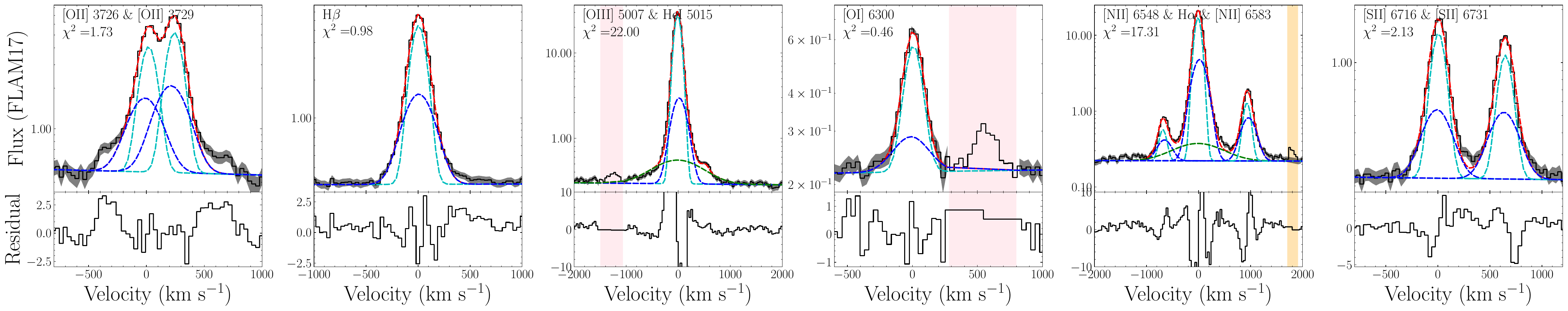}
{1.00\textwidth}{\textbf{{\large (J0938+5428)}}}}
\gridline{\fig{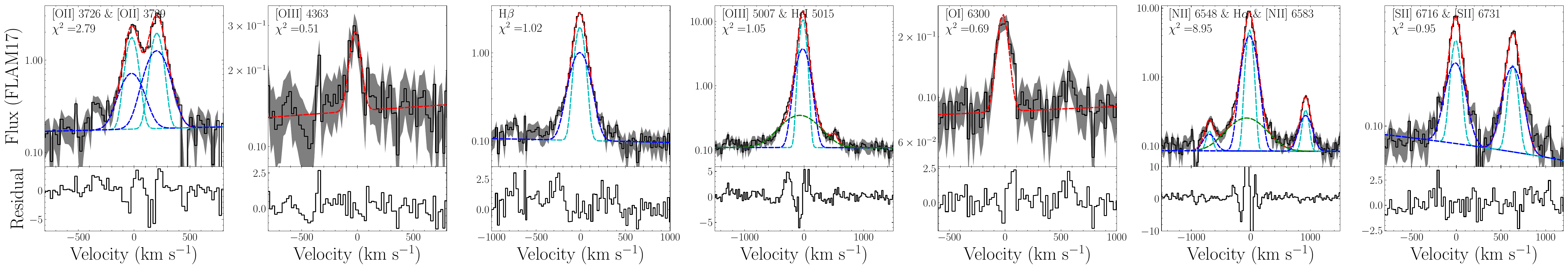}
{1.00\textwidth}{\textbf{{\large (J1025+3622)}}}}
\gridline{\fig{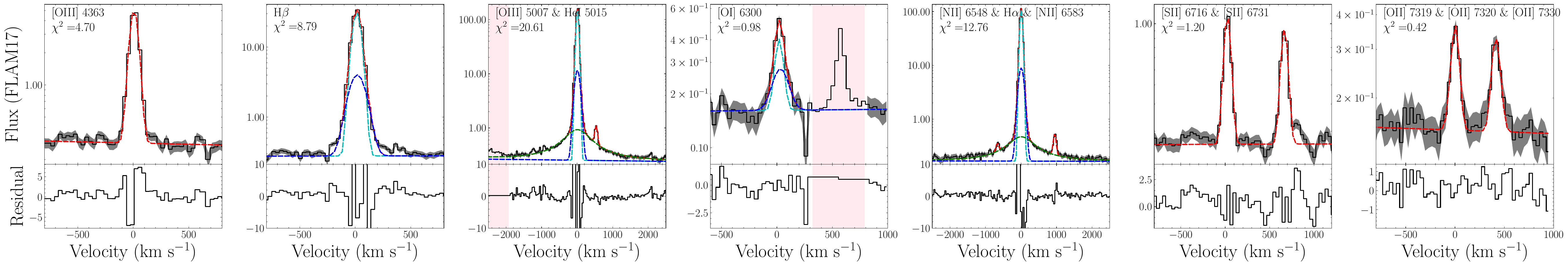}
{1.00\textwidth}{\textbf{{\large (J1044+0353)}}}}
\gridline{\fig{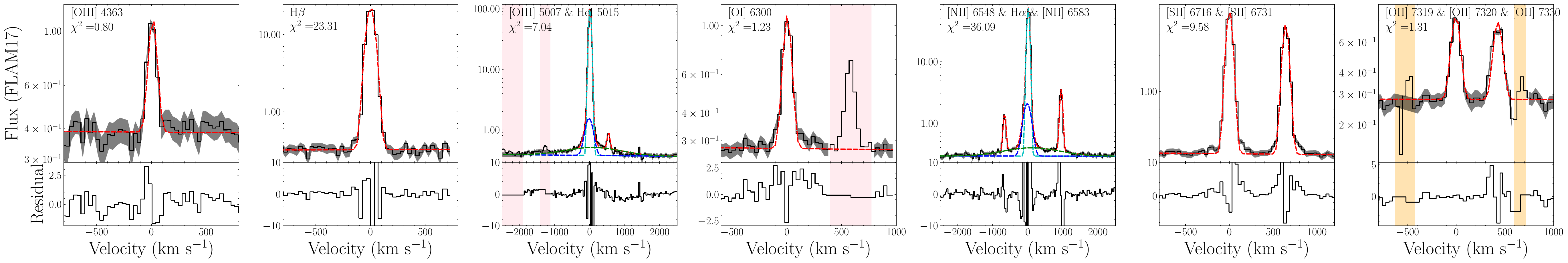}
{1.00\textwidth}{\textbf{{\large (J1129+2034)}}}}
\caption{(\textit{continued}). \label{fig:line_fittings_2}}
\end{figure*}

\begin{figure*}[htbp]
\addtocounter{figure}{-1}
\gridline{\fig{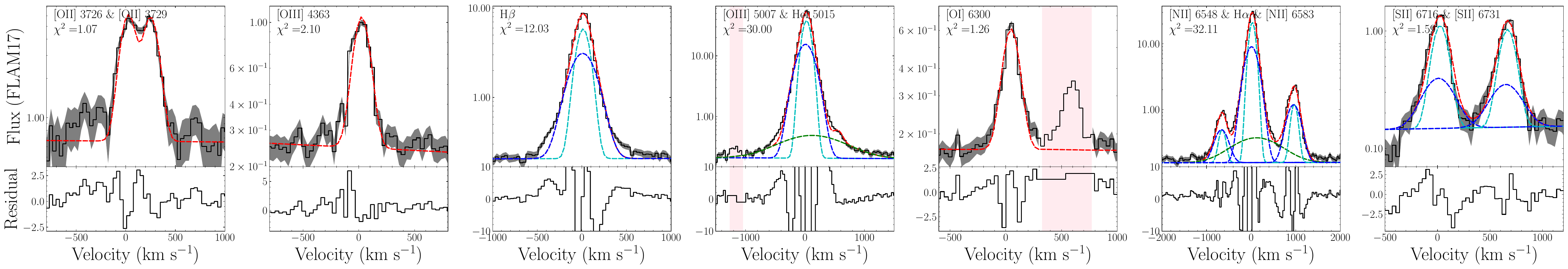}
{1.00\textwidth}{\textbf{{\large (J1200+1343)}}}}
\gridline{\fig{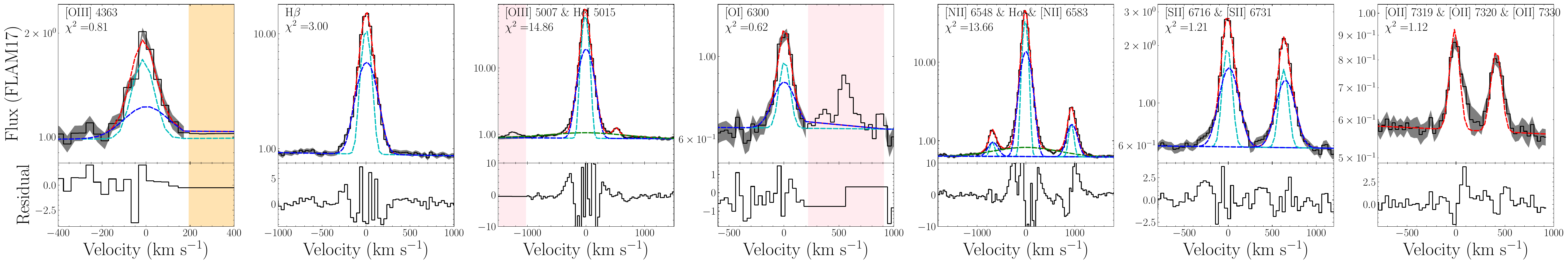}
{1.00\textwidth}{\textbf{{\large (J1359+5726)}}}}
\gridline{\fig{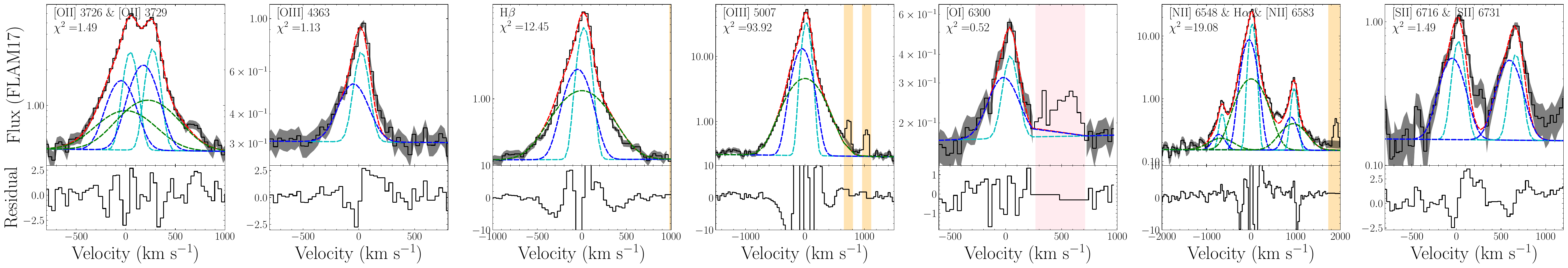}
{1.00\textwidth}{\textbf{{\large (J1429+0643)}}}}
\gridline{\fig{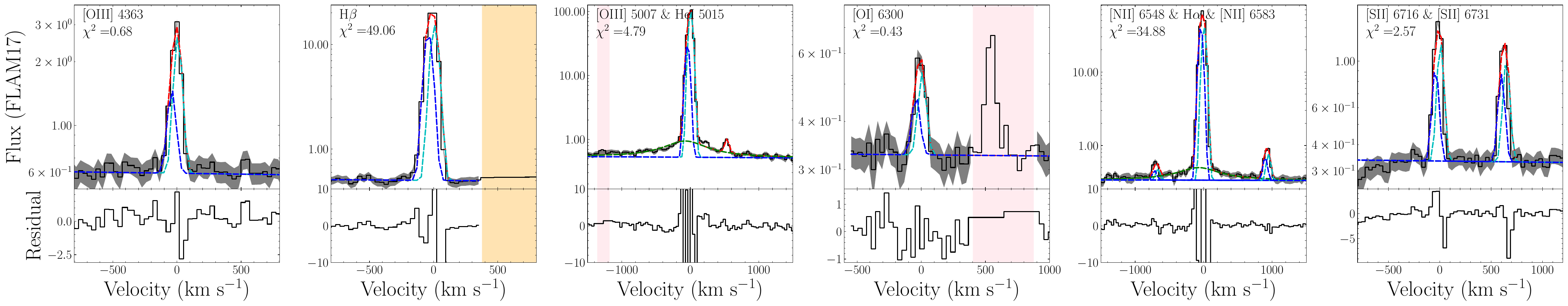}
{1.00\textwidth}{\textbf{{\large (J1444+4237)}}}}
\gridline{\fig{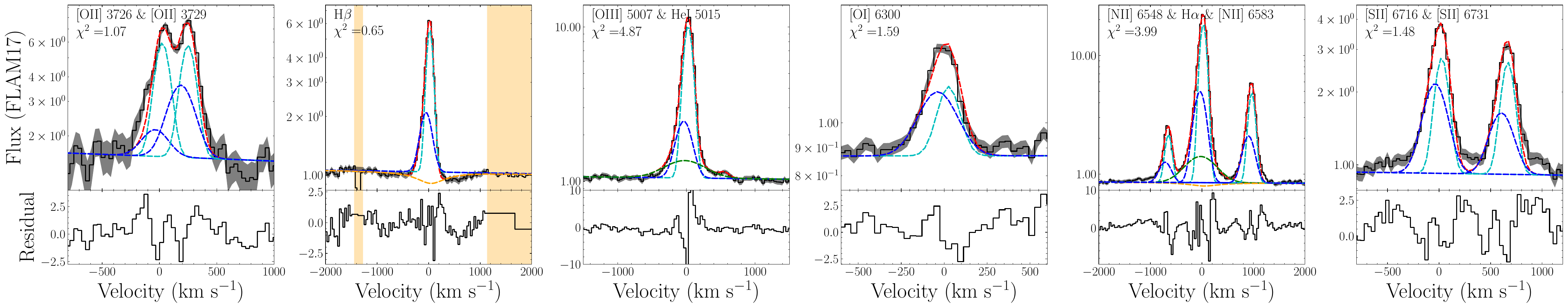}
{1.00\textwidth}{\textbf{{\large (J1525+0757)}}}}
\caption{(\textit{continued}). \label{fig:line_fittings_3}}
\end{figure*}

\setlength{\tabcolsep}{3pt} 
\renewcommand{\arraystretch}{1.05} 
\begin{longrotatetable}
\begin{splitdeluxetable*}{lcccccccBcccccccc}
\tabletypesize{\footnotesize}
\tablecaption{Line Fluxes from Best-fitting Results \label{table:line_Flux}}
\tablehead{
\colhead{Emission line} & 
\colhead{J0127--0619} & 
\colhead{J0144+0453} & 
\colhead{J0337--0502} & 
\colhead{J0808+3948} &
\colhead{J0926+4427} &
\colhead{J0938+5428} & 
\colhead{J1025+3622} &
\colhead{J1044+0353} &
\colhead{J1129+2034} &
\colhead{J1200+1343} &
\colhead{J1359+5726} &
\colhead{J1429+0643} & 
\colhead{J1444+4237} &
\colhead{J1525+0757}
}
\startdata
$\rm{[O\,{\footnotesize II}] \ \lambda 3726}$ & 
\nodata & 
\nodata & 
\nodata & 
\makecell{76.56$\pm$28.98, 284.1$\pm$219.9} & 
\makecell{58.94$\pm$4.54, 183.1$\pm$26.3} & 
\makecell{79.51$\pm$5.68, 80.77$\pm$16.67} & 
\makecell{85.92$\pm$8.98, 295.7$\pm$79.6} & 
\nodata & 
\nodata & 
130.0$\pm$20.0 & 
\nodata & 
\makecell{46.38$\pm$4.55, 338.1$\pm$76.5, \\ 39.03$\pm$6.80} & 
\nodata & 
129.2$\pm$17.2 \\
$\rm{[O\,{\footnotesize II}] \ \lambda 3729}$ & 
\nodata & 
\nodata & 
\nodata & 
\makecell{60.28$\pm$31.69, 141.8$\pm$316.0} & 
\makecell{81.04$\pm$5.39, 187.8$\pm$26.8} & 
\makecell{104.4$\pm$7.3, 107.4$\pm$20.7} & 
\makecell{97.45$\pm$9.42, 590.0$\pm$107.9} & 
\nodata & 
\nodata & 
124.5$\pm$19.3 & 
\nodata & 
\makecell{49.61$\pm$4.82, 499.6$\pm$108.4, \\ 55.74$\pm$5.28} & 
\nodata & 
125.4$\pm$16.0 \\
$\rm{[O\,{\footnotesize III}] \ \lambda 4363}$ & 
\nodata & 
\makecell{6.85$\pm$1.36, 1.71$\pm$0.63} & 
15.24$\pm$1.19 & 
\nodata & 
\makecell{6.53$\pm$1.29, 17.67$\pm$4.97} & 
\nodata & 
\nodata & 
17.13$\pm$1.15 & 
3.43$\pm$0.63 & 
16.32$\pm$2.07 & 
\makecell{7.50$\pm$3.28, 3.05$\pm$2.69} & 
\makecell{8.00$\pm$2.97, 45.89$\pm$23.13} & 
14.14$\pm$0.78, 9.42$\pm$0.74 & 
\nodata \\
$\rm{H}\beta$ & 
\makecell{100.0$\pm$1.7, 60.16$\pm$1.70} & 
\makecell{100.0$\pm$2.4, 19.28$\pm$0.70} & 
\makecell{100.0$\pm$4.8, 83.35$\pm$19.08} & 
\makecell{100.0$\pm$8.1, 247.0$\pm$44.4} & 
\makecell{100.0$\pm$3.4, 146.6$\pm$13.6} & 
\makecell{100.0$\pm$3.8, 65.21$\pm$8.57} & 
\makecell{100.0$\pm$1.8, 317.7$\pm$32.8} & 
\makecell{100.0$\pm$3.9, 13.63$\pm$0.76} & 
\makecell{100.0$\pm$2.7} & 
\makecell{100.0$\pm$5.6, 56.47$\pm$6.49} & 
\makecell{100.0$\pm$4.2, 61.14$\pm$2.76} & 
\makecell{100.0$\pm$5.9, 293.2$\pm$52.2, \\ 78.19$\pm$7.64} & 
\makecell{100.0$\pm$3.4, 126.3$\pm$6.3} & 
\makecell{100.0$\pm$3.5, 41.41$\pm$4.07} \\
$\rm{[O\,{\footnotesize III}] \ \lambda 5007}$ & 
\makecell{345.1$\pm$8.5, 70.04$\pm$2.71, \\ 11.16$\pm$3.93} & 
\makecell{438.5$\pm$14.3, 16.62$\pm$0.81, \\ 14.56$\pm$5.51} & 
\makecell{412.2$\pm$28.0, 343.7$\pm$82.0, \\ 30.26$\pm$4.94} & 
\makecell{54.63$\pm$6.91, 218.9$\pm$38.9} & 
\makecell{557.7$\pm$25.1, 758.4$\pm$63.9, \\ 56.69$\pm$9.75} & 
\makecell{525.9$\pm$28.5, 134.9$\pm$22.3, \\ 33.14$\pm$8.94} & 
\makecell{566.1$\pm$17.3, 1210.9$\pm$141.8, \\ 75.85$\pm$10.27} & 
\makecell{481.4$\pm$27.2, 41.43$\pm$6.99, \\ 31.57$\pm$7.42} & 
\makecell{429.8$\pm$15.5, 14.30$\pm$2.54, \\ 12.18$\pm$3.63} & 
\makecell{651.0$\pm$50.9, 294.2$\pm$35.2, \\ 34.13$\pm$11.09} & 
\makecell{593.3$\pm$34.2, 236.2$\pm$10.4, \\ 24.36$\pm$5.01} & 
\makecell{609.4$\pm$47.5, 1965.5$\pm$306.7, \\ 338.4$\pm$21.9} & 
\makecell{684.3$\pm$46.9, 296.2$\pm$53.7, \\ 41.48$\pm$9.87} & 
\makecell{193.1$\pm$9.0, 52.78$\pm$4.73, \\ 29.49$\pm$2.76} \\
$\rm{[O\,{\footnotesize I}] \ \lambda 6300}$ & 
4.35$\pm$0.30 & 
\makecell{3.69$\pm$0.42, 0.78$\pm$0.25} & 
0.72$\pm$0.17 & 
\nodata & 
6.24$\pm$0.92 & 
\makecell{6.14$\pm$0.36, 2.84$\pm$0.46} & 
9.24$\pm$1.10 & 
\makecell{0.64$\pm$0.10, 0.40$\pm$0.09} & 
3.34$\pm$0.13 & 
6.03$\pm$0.42 & 
\makecell{2.84$\pm$0.37, 2.70$\pm$0.36} & 
\makecell{4.09$\pm$0.66, 14.75$\pm$2.72} & 
\makecell{1.39$\pm$0.13, 1.20$\pm$0.16} & 
\makecell{5.03$\pm$0.65, 8.08$\pm$1.03} \\
$\rm{[N\,{\footnotesize II}] \ \lambda 6548}$ & 
8.36$\pm$0.21 & 
\makecell{0.89$\pm$0.16, 0.58$\pm$0.10} & 
0.41$\pm$0.05 & 
\makecell{77.52$\pm$7.28, 109.1$\pm$13.9} & 
\makecell{2.49$\pm$0.26, 11.14$\pm$1.17} & 
\makecell{5.96$\pm$0.55, 8.04$\pm$1.28} & 
\makecell{4.31$\pm$1.00, 14.47$\pm$3.22} & 
0.27$\pm$0.08 & 
4.17$\pm$0.16 & 
\makecell{4.59$\pm$0.57, 5.95$\pm$0.75} & 
\makecell{2.84$\pm$0.34, 4.11$\pm$0.41} & 
\makecell{7.18$\pm$1.14, 15.43$\pm$5.95 \\ 8.57$\pm$2.87} & 
\makecell{0.96$\pm$0.29, 1.01$\pm$0.39} & 
\makecell{21.29$\pm$0.91, 11.83$\pm$0.97} \\
$\rm{H}\alpha$ & 
\makecell{278.8$\pm$5.9, 167.7$\pm$4.0, \\ 32.94$\pm$3.48} & 
\makecell{278.8$\pm$8.0, 33.94$\pm$1.03, \\ 7.24$\pm$2.85} & 
\makecell{274.7$\pm$16.8, 228.9$\pm$40.5, \\ 13.60$\pm$2.34} & 
\makecell{285.7$\pm$26.8, 705.6$\pm$88.2} & 
\makecell{286.3$\pm$11.5, 419.8$\pm$27.2} & 
\makecell{286.3$\pm$14.1, 186.7$\pm$20.5, \\ 17.42$\pm$5.27} & 
\makecell{250.7$\pm$11.6, 885.5$\pm$73.1, \\ 53.90$\pm$8.71} & 
\makecell{274.7$\pm$13.7, 30.56$\pm$5.17, \\ 11.61$\pm$3.08} & 
\makecell{286.3$\pm$9.2, 19.69$\pm$2.54, \\ 8.74$\pm$2.54} & 
\makecell{278.8$\pm$19.3, 157.4$\pm$15.4, \\ 20.10$\pm$5.40} & 
\makecell{278.8$\pm$14.7, 166.0$\pm$9.8, \\ 21.81$\pm$4.34} & 
\makecell{278.8$\pm$19.4, 817.4$\pm$94.5, \\ 148.4$\pm$11.1} & 
\makecell{278.8$\pm$11.2, 352.1$\pm$14.4, \\ 15.93$\pm$4.01} & 
\makecell{286.3$\pm$11.8, 118.6$\pm$8.2, \\ 41.01$\pm$2.15} \\
$\rm{[N\,{\footnotesize II}] \ \lambda 6583}$ & 
25.42$\pm$0.64 & 
\makecell{2.71$\pm$0.47, 1.78$\pm$0.31} & 
1.26$\pm$0.14 & 
\makecell{235.1$\pm$22.0, 329.6$\pm$41.8} & 
\makecell{7.60$\pm$0.80, 33.85$\pm$3.56} & 
\makecell{18.16$\pm$1.67, 24.44$\pm$3.88} & 
\makecell{13.15$\pm$3.05, 43.89$\pm$9.78} & 
0.82$\pm$0.23 & 
12.71$\pm$0.49 & 
\makecell{13.94$\pm$1.72, 18.13$\pm$2.27} & 
\makecell{8.66$\pm$1.04, 12.54$\pm$1.25} & 
\makecell{21.89$\pm$3.49, 46.75$\pm$18.01, \\ 26.12$\pm$8.76} & 
\makecell{2.92$\pm$0.87, 3.07$\pm$1.17} & 
\makecell{64.69$\pm$2.75, 35.95$\pm$2.94} \\
$\rm{[S\,{\footnotesize II}] \ \lambda 6717}$ & 
25.84$\pm$0.56 & 
\makecell{10.18$\pm$0.51, 3.85$\pm$0.29} & 
2.24$\pm$0.13 & 
\makecell{14.50$\pm$3.10, 51.35$\pm$10.43} & 
\makecell{14.95$\pm$1.33, 18.71$\pm$3.30} & 
\makecell{21.18$\pm$1.20, 12.91$\pm$1.56} & 
\makecell{19.16$\pm$1.15, 49.16$\pm$4.57} & 
2.44$\pm$0.12 & 
18.04$\pm$0.59 & 
\makecell{12.67$\pm$2.94, 4.38$\pm$2.75} & 
\makecell{11.27$\pm$0.68, 11.93$\pm$0.65} & 
\makecell{10.64$\pm$5.33, 37.57$\pm$19.38} & 
\makecell{5.90$\pm$0.84, 5.07$\pm$1.08} & 
\makecell{29.74$\pm$1.83, 35.10$\pm$3.11} \\
$\rm{[S\,{\footnotesize II}] \ \lambda 6731}$ & 
21.01$\pm$0.47 & 
\makecell{6.32$\pm$0.45, 3.70$\pm$0.29} & 
1.78$\pm$0.11 & 
\makecell{26.14$\pm$3.58, 49.68$\pm$11.75} & 
\makecell{10.81$\pm$1.20, 17.63$\pm$3.30} & 
\makecell{14.95$\pm$0.90, 12.26$\pm$1.48} & 
\makecell{11.65$\pm$1.12, 47.03$\pm$4.47} & 
1.95$\pm$0.10 & 
13.31$\pm$0.45 & 
\makecell{11.49$\pm$2.78, 3.52$\pm$2.59} & 
\makecell{7.61$\pm$0.53, 9.29$\pm$0.55} & 
\makecell{8.06$\pm$4.98, 35.89$\pm$18.13} & 
\makecell{4.51$\pm$0.78, 4.50$\pm$1.02} & 
\makecell{28.00$\pm$2.04, 20.51$\pm$2.59} \\
$\rm{[O\,{\footnotesize II}] \ \lambda 7319+}$ & 
3.85$\pm$1.04 & 
\nodata & 
0.58$\pm$0.20 & 
\nodata & 
\nodata & 
\nodata & 
\nodata & 
0.57$\pm$0.37 & 
2.58$\pm$0.57 & 
\nodata & 
2.84$\pm$1.00 & 
\nodata & 
\nodata & 
\nodata \\
$\rm{[O\,{\footnotesize II}] \ \lambda 7330+}$ & 
3.31$\pm$1.10 & 
\nodata & 
0.55$\pm$0.23 & 
\nodata & 
\nodata & 
\nodata & 
\nodata & 
0.50$\pm$0.36 & 
2.31$\pm$0.50 & 
\nodata & 
2.36$\pm$1.22 & 
\nodata & 
\nodata & 
\nodata \\
\hline 
$F_{\rm{H}\beta}$ (narrow) & 
5196.8$\pm$86.8 & 
1656.9$\pm$40.5 & 
9955.3$\pm$478.2 & 
471.9$\pm$38.1 & 
360.4$\pm$12.2 & 
1257.4$\pm$48.0 & 
230.9$\pm$4.2 & 
5367.4$\pm$209.5 & 
2852.3$\pm$75.8 & 
2319.1$\pm$129.2 & 
1953.5$\pm$82.2 & 
882.2$\pm$51.9 & 
1178.4$\pm$39.8 & 
1697.7$\pm$58.9 \\
\enddata
\tablecomments{Extinction-corrected (see Table \ref{table:physical_properties} for $E(B-V)_{\rm{MV}}$ and Table \ref{table:derived_physical_properties} for $A_V$) emission-line fluxes are listed relative to the narrow component (or the cyan narrow component in Figure \ref{fig:line_fittings} for DNC galaxies) of H$\beta \times 100$ for each velocity component of our sample. The last row below the emission line shows the extinction-corrected flux for the narrow component of H$\beta$, in units of $10^{-17} \ \rm{ergs}\ \rm{s^{-1}}\ \rm{cm^{-2}}$.
Fluxes for the narrow, broad, and VB components of each line are listed in order, separated by commas. If only one or two entries are shown, it indicates that only the narrow and/or broad components are fitted for that emission line.}
\end{splitdeluxetable*}
\end{longrotatetable}

\setlength{\tabcolsep}{6pt} 
\begin{deluxetable*}{cccccc}
\centerwidetable
\tablecaption{Spectroscopically Derived Physical Properties\label{table:derived_physical_properties}}
\tablehead{
\colhead{Object} & 
\colhead{\(n_e\) ([O\,{\footnotesize II}])} & 
\colhead{\(n_e\) ([S\,{\footnotesize II}])} & 
\colhead{\(T_e\) ([O\,{\footnotesize III}])} & 
\colhead{12 + log(O/H)} &
\colhead{\(A_V\)} \\
\nocolhead{} & \colhead{($\rm{cm^{-3}}$)} & \colhead{($\rm{cm^{-3}}$)} & \colhead{($\rm{K}$)} & \nocolhead{} & \nocolhead{} \\
\colhead{(1)} & \colhead{(2)} & \colhead{(3)} & \colhead{(4)} & \colhead{(5)} & \colhead{(6)}
}
\startdata
J0127--0619 & \nodata & $280_{-50}^{+50}$ & \nodata & \nodata & $0.48_{-0.01}^{+0.01}$, $0.25_{-0.02}^{+0.02}$ \\
J0144+0453 & \nodata & $40_{-40}^{+40}$, $430_{-390}^{+390}$ & $12000_{-2300}^{+2300}$, $20000_{-3700}^{+3700}$ & \nodata & $0.59_{-0.02}^{+0.02}$, $0$ \\
J0337--0502 & \nodata & $250_{-110}^{+110}$ & $21000_{-1100}^{+1100}$ & $7.36_{-0.06}^{+0.06}$ & $0.94_{-0.04}^{+0.04}$, $2.39_{-0.20}^{+0.18}$  \\
J0808+3948 & $790_{-790}^{+790}$, $680_{-680}^{+680}$ & $4200_{-3200}^{+3200}$, $640_{-550}^{+540}$ & \nodata & \nodata & $1.03_{-0.06}^{+0.07}$, $1.85_{-0.13}^{+0.13}$ \\
J0926+4427 & $160_{-60}^{+60}$, $550_{-290}^{+290}$ & $230_{-160}^{+160}$, $740_{-550}^{+550}$ & $12000_{-1000}^{+1000}$, $16000_{-2000}^{+2000}$ & $8.13_{-0.10}^{+0.10}$, $7.91_{-0.16}^{+0.16}$ & $0.23_{-0.03}^{+0.03}$, $0.82_{-0.07}^{+0.07}$ \\
J0938+5428 & $140_{-90}^{+90}$, $230_{-220}^{+210}$ & $70_{-60}^{+60}$, $560_{-380}^{+370}$ & \nodata & \nodata & $0.28_{-0.03}^{+0.03}$, $0.57_{-0.12}^{+0.11}$ \\
J1025+3622 & $370_{-210}^{+210}$, $140_{-130}^{+130}$ & $130_{-80}^{+80}$, $680_{-390}^{+390}$ & $13000_{-1200}^{+1200}$ & $8.05_{-0.21}^{+0.21}$ & $0$, $1.07_{-0.09}^{+0.08}$  \\
J1044+0353 & \nodata & $260_{-90}^{+90}$ & $21000_{-900}^{+900}$ & $7.44_{-0.06}^{+0.06}$ & $0.47_{-0.03}^{+0.03}$, $0$ \\
J1129+2034 & \nodata & $90_{-40}^{+40}$ & $11000_{-660}^{+660}$ & $8.28_{-0.10}^{+0.10}$ & $0.25_{-0.02}^{+0.02}$  \\
J1200+1343 & $710_{-420}^{+420}$ & $750_{-600}^{+600}$, $850_{-850}^{+850}$ & $17000_{-1200}^{+1200}$ & $7.93_{-0.12}^{+0.12}$ & $0.67_{-0.05}^{+0.05}$, $0.16_{-0.09}^{+0.08}$ \\
J1359+5726 & \nodata & $140_{-50}^{+50}$, $200_{-80}^{+80}$ & $13000_{-2200}^{+2200}$, $13000_{-3200}^{+3200}$ & $8.16_{-0.23}^{+0.23}$ & $0.44_{-0.04}^{+0.04}$, $0$ \\
J1429+0643 & $430_{-170}^{+170}$, $300_{-200}^{+200}$, $150_{-130}^{+130}$ & $850_{-850}^{+850}$, $1000_{-1000}^{+1000}$ & $14000_{-2200}^{+2200}$, $13000_{-3800}^{+3800}$ & $8.08_{-0.24}^{+0.24}$, $8.09_{-0.33}^{+0.33}$ & $0$, $1.30_{-0.13}^{+0.13}$ \\
J1444+4237 & \nodata & $390_{-290}^{+290}$, $730_{-560}^{+560}$ & $16000_{-540}^{+540}$, $19000_{-2100}^{+2100}$ & \nodata & $0.03_{-0.03}^{+0.03}$, $0.36_{-0.03}^{+0.03}$ \\
J1525+0757 & $570_{-430}^{+420}$, $90_{-90}^{+90}$ & $550_{-250}^{+220}$, $80_{-70}^{+60}$ & \nodata & \nodata & $0.64_{-0.03}^{+0.03}$, $0.68_{-0.08}^{+0.07}$ \\
\enddata
\tablecomments{Columns: (1) Object name; (2) Electron density \(n_e\) from [O\,{\footnotesize II}] $\lambda\lambda 3726, 3729$; (3) Electron density \(n_e\) from [S\,{\footnotesize II}] $\lambda\lambda 6716, 6731$; (4) Electron temperature \(T_e\) based on the direct method using [O\,{\footnotesize III}] $\lambda 4363$ and [O\,{\footnotesize III}] $\lambda 5007$; (5) Oxygen abundance 12 + log(O/H); (6) Internal extinction \(A_V\) from the Balmer decrement. 
Values for each physical property's narrow, broad, and VB components are listed as the first, second, and third entries, respectively. If only one or two entries are shown, it indicates that only the narrow and broad components could be determined for that galaxy.
Please check Section \ref{subsubsec:derived_phys_prop} for details in deriving each physical property.}
\end{deluxetable*}

\section{Supernova Rate}\label{sec:sn_rate}
\begin{figure}[htbp]
\centering
\includegraphics[width=0.45\textwidth]{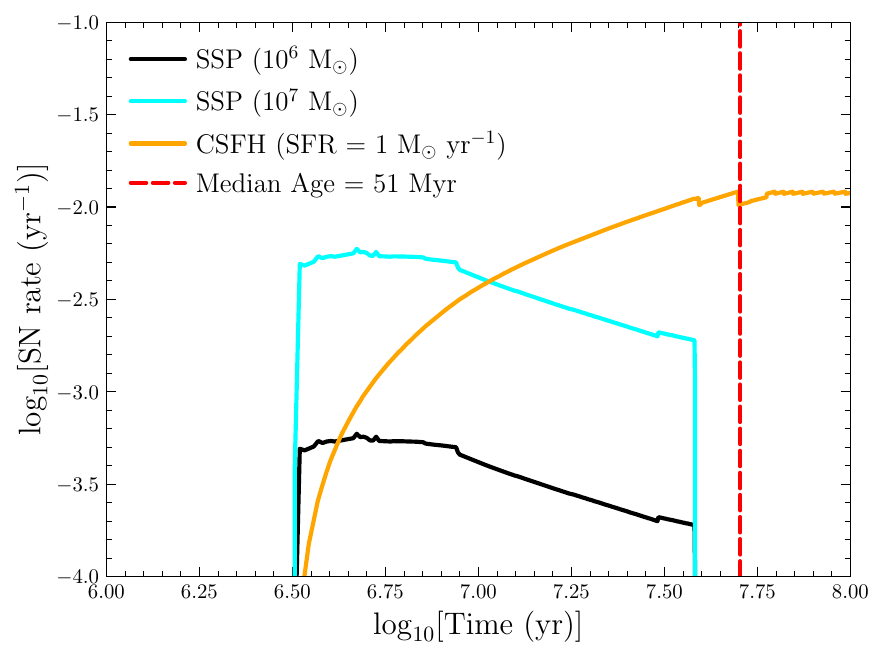}
\caption{SN rate predicted by \texttt{STARBURST99} models. The black solid (cyan) line represents the SSP model with the total stellar mass of $10^6 M_{\odot}$ ($10^7 M_{\odot}$). The orange solid line shows the CSFH model with SFR = $1 \ \rm{M_{\odot} \ yr^{-1}}$. For the CSFH model, the SN rate is around $10^{-2} \ \rm{yr^{-1}}$ at the median age of our sample (51 Myr).}   \label{fig:sn_rate}
\end{figure}

The following rationale validates the SN rate used in this study. A series of \texttt{STARBURST99} simulations are conducted, following a \cite{Kroupa_2001} initial mass function (IMF) spanning 0.1 to 100 $M_{\odot}$: (1) assume an SSP (simple stellar population) with a total stellar mass of $10^6 \ M_{\odot}$ or $10^7 \ M_{\odot}$ (2) CSFH with SFR = $1 \ \rm{M_{\odot} \ yr^{-1}}$ (Figure \ref{fig:sn_rate}). We then measured the stellar population age of each galaxy within the HST/COS aperture using a CSFH \texttt{BEAGLE} model with a \cite{chabrier_galactic_2003} IMF between 0.1 and 100 $M_{\odot}$ \citep[consistent with the SED model setups in][]{Berg_2022}. The systematic difference between these two IMFs is negligible \citep[$\sim 0.03$ dex;][]{Moustakas_2013}. At the median age of our targets (51 Myr), shown as a red dash line in Figure \ref{fig:sn_rate}, the SN rate of the CSFH model is around $10^{-2} \ \rm{yr^{-1}}$, corresponding to the SN energy input rate given in Equation \ref{eq:4.1}. 

\begin{figure}[htb]
    \includegraphics[width=0.45\textwidth]{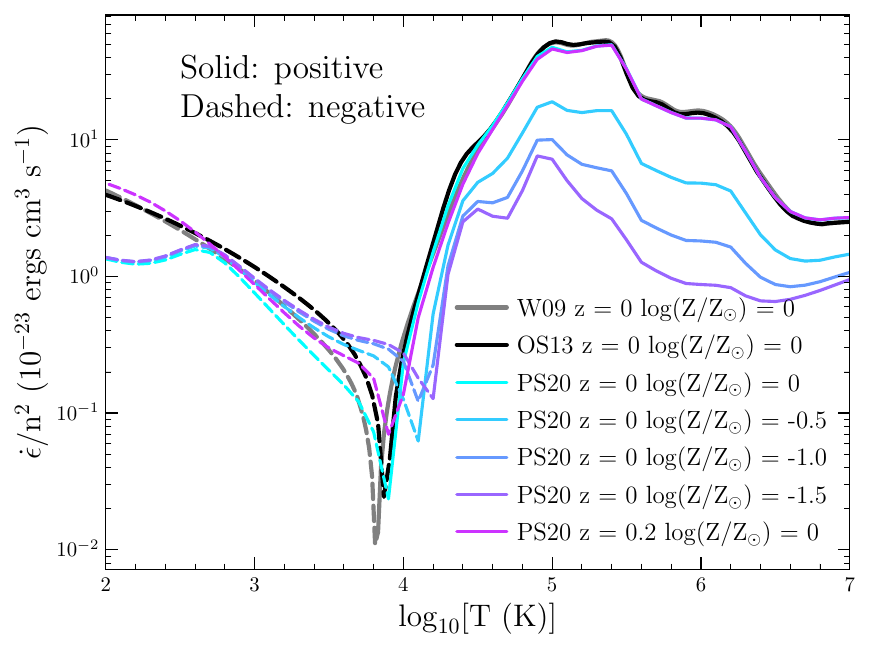}
\caption{Examples of the normalized net cooling functions $\dot{\epsilon} / n^2$ of \cite{Wiersma_2009}, \cite{OS_13}, and \citetalias{PS_2020}. \cite{Wiersma_2009} solar-metallicity cooling curve (the grey curve) includes heating from the $z = 0$ UV background \citep{HM01}. 
The net cooling function of \cite{OS_13}, depicted as a black curve, is in photoionization equilibrium with a metagalactic UV background, set at a redshift $z = 0$ and with solar metallicity.  Cooling functions from the \citetalias{PS_2020} models at distinct redshifts and metallicities are shown in different colors. 
The hydrogen densities of all cooling functions are set at $n =  10^{-3} \ \mathrm{cm^{-3}}$. 
The `spike' at $\sim 10^4$ K occurs when the heating rate exceeds the cooling rate (i.e., solid lines indicate positive values and dashed lines indicate negative values). 
At the same redshift and metallicity, all three cooling tables have negligible differences in the temperature range from $10^4$ to $10^7$ K. \label{fig:cooling_curves}}
\end{figure}

\begin{figure}[htb]
    \includegraphics[width=0.5\textwidth]{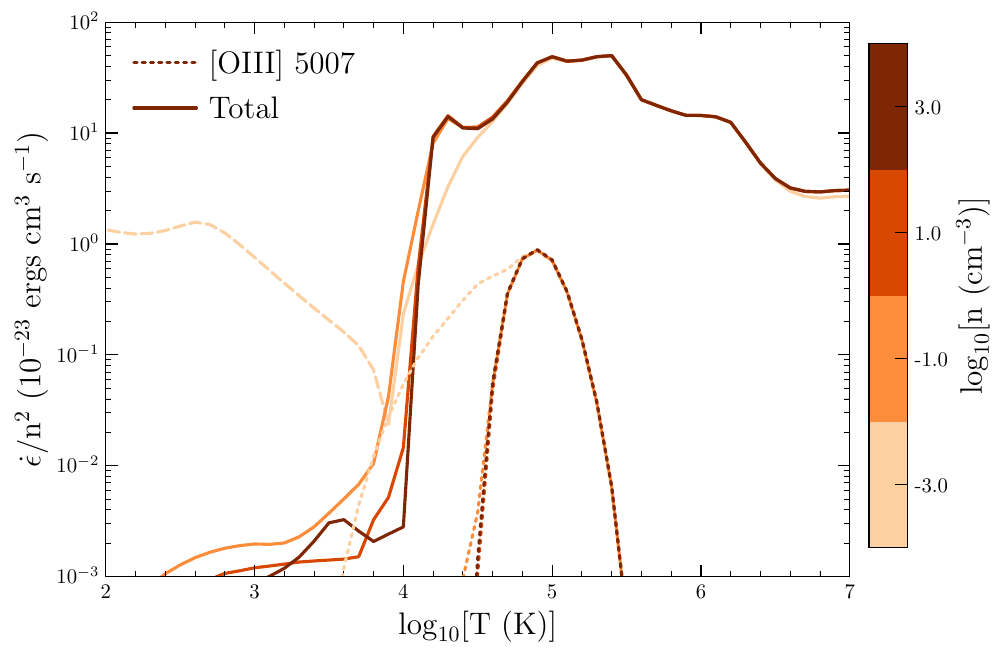}
\caption{Normalized emergent emissivities of the [O\,{\footnotesize III}] $\lambda 5007$ line, shown as dotted lines as functions of temperature, are juxtaposed with the normalized net cooling functions from the \citetalias{PS_2020} model at redshift $z = 0$ and solar metallicity, consistent with the style in Figure \ref{fig:cooling_curves}.
[O\,{\footnotesize III}] $\lambda 5007$ emissivities peak in the temperate range $T \sim 10^4 - 10^{5.5} \ \rm{K}$.
\label{fig:o3_5007_emissivities}}
\end{figure}

\section{Galactic Wind Model Solutions} \label{sec:wind_model_solutions}
Here we summarize the adopted initial conditions and cooling curves for the \citetalias{Thompson_2016} and \citetalias{Fielding_2022} models, and we refer readers to both papers for detailed descriptions of these models. 

\subsection{Initial Conditions} \label{subsec:wind_models_ic} 
With the \citetalias{CC85} model determining the radial structure of hot winds inside the starburst region (i.e., $\mathcal{M} \leq 1$), both \citetalias{Thompson_2016} and \citetalias{Fielding_2022} models integrate from the sonic point (serving as initial conditions) to derive the radial velocity, density, pressure, and temperature solutions of hot winds: \( v(r_{\ast}) \), \( \rho(r_{\ast}) \), \( P(r_{\ast}) \), and \( T(r_{\ast}) \) at larger radii \( r_{\ast} = r / R_{\mathrm{SF}} \). 
The dimensionless form (with a subscript *) of each physical quantity, as described by \citetalias{CC85}, is given by:
\begin{equation}
   \label{eq:4.3} v(r_{\ast}) = v_{\ast}(r_{\ast}) \  \dot{M}_{\rm{hot}}^{-1/2} \ \dot{E}_{\rm{hot}}^{1/2}
\end{equation}
\begin{equation}
    \label{eq:4.4} \rho(r_{\ast}) = \rho_{\ast}(r_{\ast}) \ \dot{M}_{\rm{hot}}^{3/2} \ \dot{E}_{\rm{hot}}^{-1/2} \ R_{\rm{SF}}^{-2}
\end{equation}
\begin{equation}
    \label{eq:4.5} P(r_{\ast}) = P_{\ast}(r_{\ast}) \ \dot{M}_{\rm{hot}}^{1/2} \  \dot{E}_{\rm{hot}}^{1/2} \ R_{\rm{SF}}^{-2}
\end{equation}
and 
\begin{equation}
    \label{eq:4.6} T(r_{\ast}) = \frac{\mu m_p}{k_\mathrm{B}} \ \frac{P(r_{\ast})}{\rho(r_{\ast})}  = \frac{\mu m_p}{k_\mathrm{B}} \ \frac{\dot{E}_{\rm{hot}}}{\dot{M}_{\rm{hot}}} \ T_{\ast}(r_{\ast}).
\end{equation}
The initial conditions at the sonic point are defined as \( v_{\ast}(r_{\ast} = 1) = 0.707 \), \( \rho_{\ast}(r_{\ast} = 1) = 0.113 \), \( P_{\ast}(r_{\ast} = 1) = 0.0338 \), and \( T_{\ast}(r_{\ast} = 1) = 0.0338 / 0.113 \).

Besides the initial conditions for hot winds shared by both the \citetalias{Thompson_2016} and \citetalias{Fielding_2022} models, there are additional initial conditions for cold clouds introduced in \citetalias{Fielding_2022}. 
These new parameters include the initial mass-loading factor of the cold cloud $\eta_{\rm{M,cl,i}}$, the initial cloud mass \(M_\mathrm{cl, i}\), the initial cloud metallicity \(Z_{\mathrm{cl, i}}\), the initial wind metallicity \(Z_{\rm{i}}\), and the initial cloud velocity \(v_{\mathrm{cl, i}}\). These model parameter values are summarized in Table \ref{table:model_parameters_fb22}, and their radial evolutions are plotted in Figure \ref{fig:fb22_radial_profiles}.

\subsection{Cooling Curves} \label{subsec:wind_models_cc} 
We utilize state-of-the-art cooling functions from \citetalias{PS_2020} (colored curves in Figure \ref{fig:cooling_curves}) instead of the \cite{Wiersma_2009} and \cite{OS_13} cooling curves used in \citetalias{Fielding_2022} and \citetalias{Thompson_2016}, respectively, because the former enables us to explore the cooling curves at the redshift and metallicity ranges relevant to our sample. 
\citetalias{PS_2020} cooling functions tabulate gas cooling and heating rates, considering influences from dust, cosmic rays, and a redshift-dependent UV/X-ray background, as well as an interstellar radiation field over an extensive range in redshift ($z = $ 0 -- 9), temperature (log $T$(K) = 1 -- 9.5), metallicity (log $Z/Z_{\odot}$ = -4 -- 0.5), and density (log $\mathrm{n_{H} (cm^{-3})}$ = -8 -- 6). 
The \cite{Wiersma_2009}, \cite{OS_13} and \citetalias{PS_2020} cooling tables are indistinguishable at solar metallicity within the temperature range of $T \sim 10^4 - 10^7 \ \mathrm{K}$. 


Figures \ref{fig:thom16_radial_profiles} and \ref{fig:fb22_radial_profiles} present the illustrative examples of radial profiles of physical properties for \citetalias{Thompson_2016} and \citetalias{Fielding_2022} models\footnote{The publicly available code provided by \citetalias{Fielding_2022} can be accessed at \url{https://github.com/dfielding14/MultiphaseGalacticWind}.}, respectively. 
These radial profiles are crucial for estimating the cooling luminosities radiated within a temperature range and simulating the corresponding outflowing line profiles, as described in Appendices \ref{sec:L_estimations} and \ref{sec:model}. 

\section{Cooling and [O\,{\footnotesize III}] $\lambda 5007$ Luminosities from Galactic Wind Models}\label{sec:L_estimations}
\subsection{Hot winds}\label{subsec:L_estimations_hot_wind}
Since the radiative cooling luminosity, $L_{\rm{cool}}$, from  hot winds is sensitive to the radial profiles of physical properties (i.e., $T(r)$, $n(r)$, and $v(r)$), we first integrate \citetalias{PS_2020} cooling functions 
within the \citetalias{Thompson_2016} or \citetalias{Fielding_2022} galactic wind model and solve for these radial profiles. 
The $L_{\rm{cool}}$ of hot winds within a specific temperature range can be determined as follows:
\begin{align}
\label{eq:4.7}
L_{\rm{cool}} 
&= \int_{T_{\rm{min}}}^{T_{\rm{max}}} \frac{\dd L}{\dd r} \frac{\dd r}{\dd T} \ \dd T \nonumber \\
&= \int_{T_{\rm{min}}}^{T_{\rm{max}}} 
\frac{\gamma - \mathcal{M}^{-2}}{1 - \mathcal{M}^{-2}}
\left(\dot{\epsilon} \ 4 \pi r^2\right) \left(\frac{\dd T}{\dd r} \right)^{-1} \dd T ,
\end{align}
In Equation \ref{eq:4.7}, $\dot{\epsilon}$ is the net cooling rate shown in Figures \ref{fig:cooling_curves} and \ref{fig:o3_5007_emissivities}, $\gamma = 5 / 3$ is the adiabatic index\footnote{At the isothermal limit, $\gamma$ would approach 1, and Equation \ref{eq:4.7} would become $L_{\rm{cool}} =  \int_{T_{\rm{min}}}^{T_{\rm{max}}} 
\left(\dot{\epsilon} \ 4 \pi r^2\right) \left(\frac{\dd T}{\dd r} \right)^{-1} \dd T$.}, $\mathcal{M} = v / c_s$ is the Mach number of hot winds, and $T_{\rm{min}}$ and $T_{\rm{max}}$ denote the lower and upper bounds for temperature integration.
To calculate the corresponding luminosity of a specific line within the same temperature range, we can substitute the line emissivity, $\dot{e}$, for  $\dot{\epsilon}$ in Equation \ref{eq:4.7}.
We refer the reader to Appendix \ref{sec:enthalpy_model} for details on deriving Equation \ref{eq:4.7}. 

\begin{figure}[t]
\centering
\includegraphics[width=0.5\textwidth]{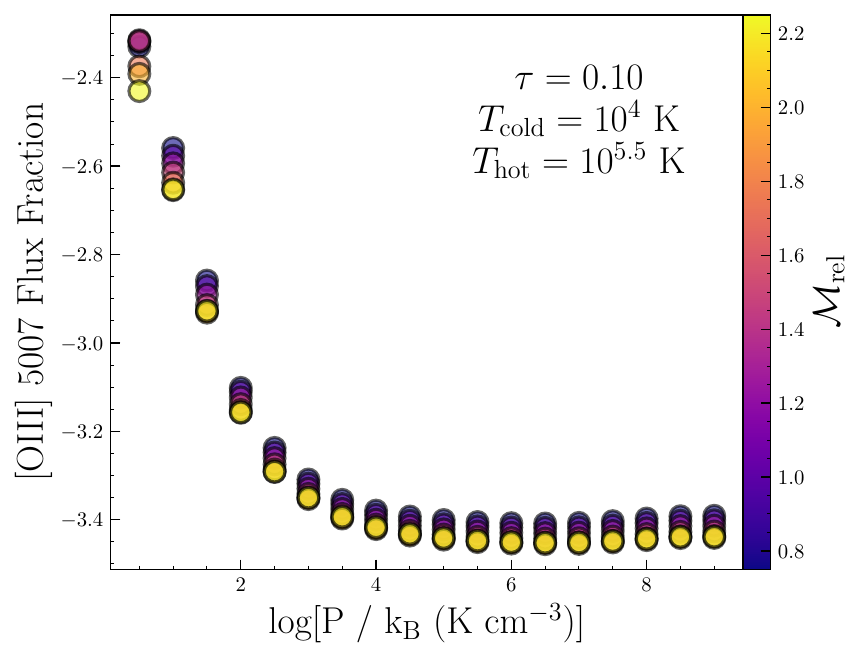}
\caption{Flux fraction grid of [O\,{\footnotesize III}] $\lambda 5007$ as a function of $P$ and $\mathcal{M}_{\rm{rel}}$ within the 1.5-dimensional TRML model of \cite{Chen_2023}. Each model point is color-coded by its $\mathcal{M}_{\rm{rel}}$. We adopt a fiducial value for the dimensionless cooling time factor, $\tau = 10^{-1}$, as defined in Equation (22) of \cite{Chen_2023}. Should $\tau$ be reduced to $10^{-2}$, the flux fraction would decrease approximately by an order of magnitude. Hence, such a reduction would not qualitatively impact the arguments presented in Section \ref{sec:broad_origins}. The hot and cold phases are set at $T_{\rm{hot}} = 10^{5.5} \ \rm{K}$ and $T_{\rm{cold}} = 10^{4} \ \rm{K}$, respectively. The best-fitting dimensionless prefactor, $f_{\nu}$, of the effective turbulent conductivity and the effective turbulent Prandtl number, $Pr$, of each parameter set is determined from Equations (44) and (45) in \cite{Chen_2023}. 
\label{fig:flux_fraction_5007_trml_grid}}
\end{figure}

It is crucial to recognize that the emissivities of various optical emission lines and UV absorption lines, which trace the cold clouds, peak at different temperatures. 
For [O\,{\footnotesize III}] $\lambda 5007$, we set $T_{\rm{min}} = 10^4 \ \rm{K}$ and $T_{\rm{max}} = 10^{5.5} \ \rm{K}$ in Equation \ref{eq:4.7}, where [O\,{\footnotesize III}] $\lambda 5007$ exhibits significant emissivity (see Figure \ref{fig:o3_5007_emissivities}). 
Our subsequent arguments are not sensitive to the choice of $T_{\rm{max}}$, for instance, if we set $T_{\rm{max}}$ to be higher at $10^{6} \ \rm{K}$.

\begin{figure*}[htb]
    \centering
    \includegraphics[width=1.0\textwidth]{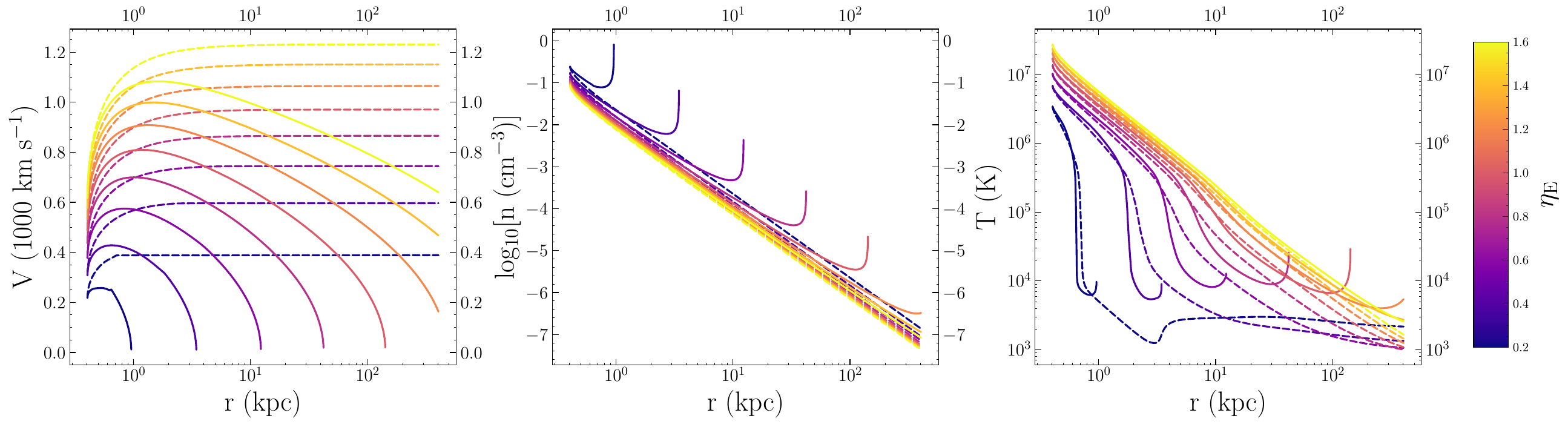}
    \includegraphics[width=1.0\textwidth]{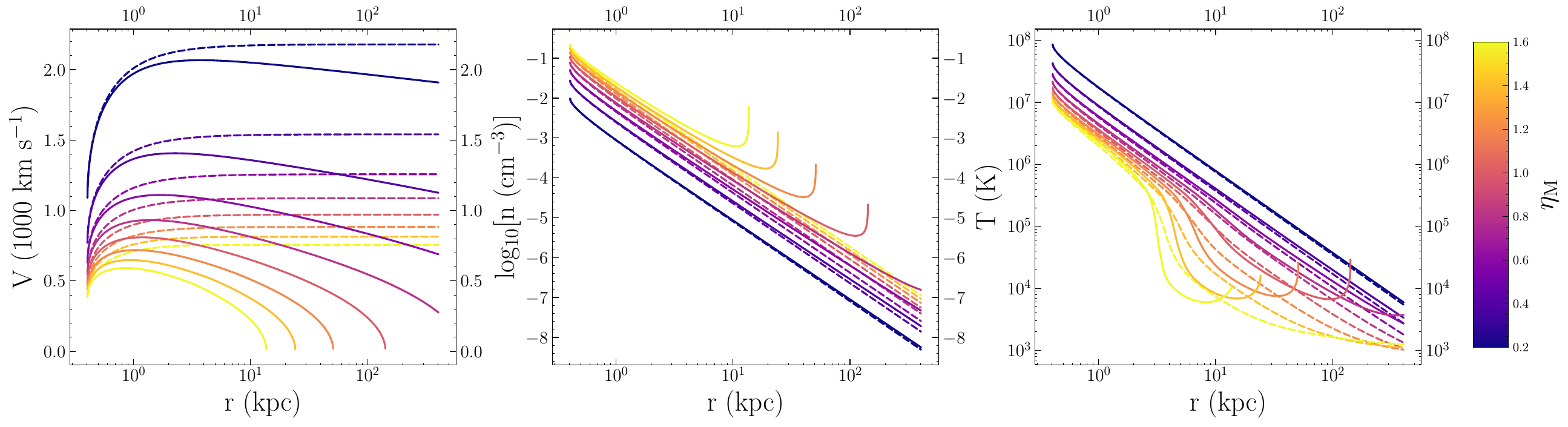}
\caption{Illustrative example of radial profiles of physical properties: \(v(r)\), \(n(r)\), and \(T(r)\), derived from the  \citetalias{Thompson_2016} model. 
The top ($\eta_{\rm{M}} = 1$) and the bottom ($\eta_{\rm{E}} = 1$) panels display solutions with \(\eta_{{\rm{E}}}\) ranging from 0.2 to 1.6 and \(\eta_{{\rm{M}}}\) ranging from 0.2 to 1.6.
Calculations are represented both without gravity (dash lines) and incorporating an assumed isothermal gravitational potential where \(v_e = 400 \ \mathrm{km \ s^{-1}}\) (solid lines). 
Other key parameter values are summarized in Table \ref{table:model_parameters_t16}.
\label{fig:thom16_radial_profiles}}
\end{figure*}

\begin{figure*}[htb]
    \centering
    \includegraphics[width=1.0\textwidth]{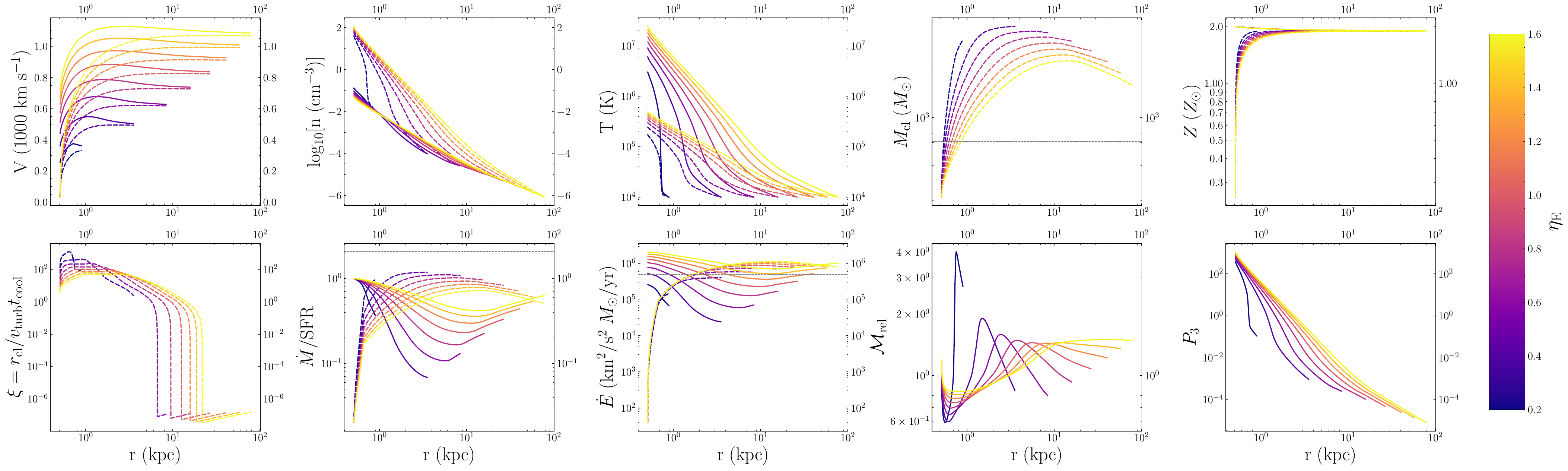}
    \includegraphics[width=1.0\textwidth]{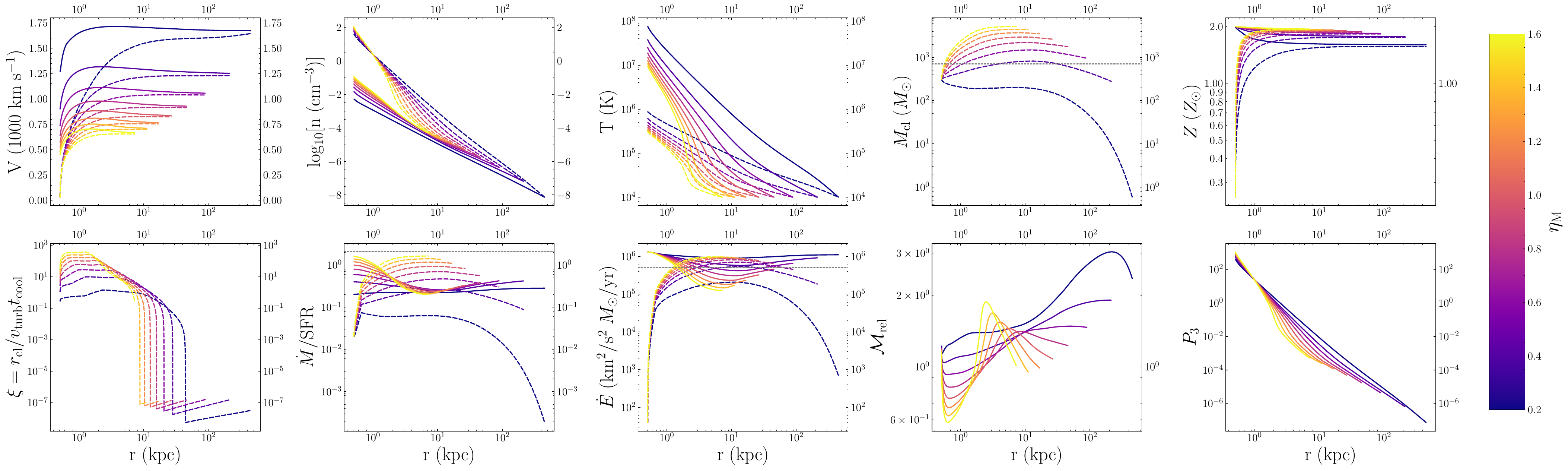}
\caption{Illustrative example of radial profiles of physical properties: \(v(r)\), \(n(r)\), \(T(r)\), \(M_{\rm{cl}}(r)\), \(Z(r)\), $\xi (r)$, \(\dot{M} (r)\), \(\dot{E} (r)\), \(\mathcal{M}_{\rm{rel}} (r)\), and \(P (r)\) derived from the  \citetalias{Fielding_2022} model. 
The top ($\eta_{\rm{M}} = 1$) and the bottom ($\eta_{\rm{E}} = 1$) panels display solutions with \(\eta_{{\rm{E}}}\) ranging from 0.2 to 1.6 and \(\eta_{{\rm{M}}}\) ranging from 0.2 to 1.6.
Calculations are incorporating an assumed isothermal gravitational potential where \(v_{\rm{cir}} = 50 \ \mathrm{km \ s^{-1}}\). 
Other key parameter values are summarized in Table \ref{table:model_parameters_fb22}.
In each panel, the solid lines represent the hot wind, while the dashed lines denote the cold clouds. 
For the \(T(r)\) panel, dashed lines indicate the mixing temperature \(T_{\rm{mix}} = \sqrt{T T_{\rm{cl}}}\) with \(T_{\rm{cl}}\) fixed at \(10^4 \ \rm{K}\) instead. 
The pressure of the hot wind and the cold clouds (i.e., in pressure equilibrium) is expressed as $P_3 = nT / (10^3 \ \rm{cm^{-3} \ K})$.
For the $M_{\rm{cl}}$, $\dot{M}$, and $\dot{E}$ panels, the median measurements of these properties for our sample from UV absorption lines based on \citetalias{Xu_2022} and \citet{Xu_2023} are shown as dotted grey lines.
\label{fig:fb22_radial_profiles}}
\end{figure*}

\subsection{TRMLs} \label{subsec:flux_frac_5007_trml}
Unlike the radiative cooling within the hot wind itself, where the line emissivity and total cooling curves can be integrated over the temperature space (as shown in Figure \ref{fig:o3_5007_emissivities}) to determine the flux fraction of a specific line ($\sim 1 \%$ for [O\,{\footnotesize III}] $\lambda 5007$), the flux fraction of a line within TRMLs is influenced by factors such as pressure $P$ and the relative Mach number, $\mathcal{M}_{\rm{rel}}$, between the hot wind and the cold clouds. To determine the flux fraction of [O\,{\footnotesize III}] $\lambda 5007$ at each radius in \citetalias{Fielding_2022} model, we utilize the 1.5-dimensional model proposed by \cite{Chen_2023}. This approach enables the generation of a flux fraction grid that varies as a function of $P$, and $\mathcal{M}_{\rm{rel}}$.

Based on Equation 25 and Figure 21 from \cite{Chen_2023}, the total cooling flux, $F_{\rm{cool}}$, within a temperature range from $T_{\rm{cold}}$ to $T_{\rm{hot}}$ in the TRMLs is 
\begin{align}
    &\int_{T_{\rm{cold}}}^{T_{\rm{hot}}} \frac{\dd F_{\rm{cool}}}{\dd \log T} \dd T = \int_{T_{\rm{cold}}}^{T_{\rm{hot}}} \frac{\dot{E}_{\rm{cool}} / L_0^2}{\dd z} \left( \frac{\dd T}{\dd z} \right)^{-1} T \dd T \nonumber \\
    &= \int_{T_{\rm{cold}}}^{T_{\rm{hot}}} \frac{\left(\dot{\epsilon} L_0^2 \dd z \right) / L_0^2}{\dd z} \left( \frac{\dd T}{\dd z} \right)^{-1} T \dd T = \int_{T_{\rm{cold}}}^{T_{\rm{hot}}} \frac{\dot{\epsilon}}{\dd T / \dd z} T \dd T. \label{eq:9.1}
\end{align}
In this context, the two phases are presumed to be in pressure equilibrium. Hence, $\dot{\epsilon}$ represents the \citetalias{PS_2020} cooling rate interpolated at constant pressure, ensuring that $P / k_B = n T = \text{constant}$. $\dd T / \dd z$ denotes the temperature gradient in the direction of the mass flux ($\hat{z}$-direction), which is perpendicular to the shear motions occurring between the phases. $L_0$ is the characteristic length of the system defined in Equation (24) from \cite{Chen_2023}.

Similarly, the integrated line flux, $F_{\rm{line}}$, within the TRMLs can be calculated as
\begin{align}
    &\int_{T_{\rm{cold}}}^{T_{\rm{hot}}} \frac{\dd F_{\rm{line}}}{\dd \log T} \dd T = \int_{T_{\rm{cold}}}^{T_{\rm{hot}}} \frac{\dot{e}}{\dd T / \dd z} T \dd T. \label{eq:9.2}
\end{align}
Hence, the flux fraction of a specific line is the ratio between Equations \ref{eq:9.1} and \ref{eq:9.2}
\begin{align}
    \text{Flux Fraction} = \frac{\int_{T_{\rm{cold}}}^{T_{\rm{hot}}} \frac{\dot{e}}{\dd T / \dd z} T \dd T}{\int_{T_{\rm{cold}}}^{T_{\rm{hot}}} \frac{\dot{\epsilon}}{\dd T / \dd z} T \dd T}.
\end{align}
The flux fraction grid of [O\,{\footnotesize III}] $\lambda 5007$ as a function of $P$ and $\mathcal{M}_{\rm{rel}}$ is depicted in Figure \ref{fig:flux_fraction_5007_trml_grid}. Here, we set the temperatures of the hot and cold phases at $T_{\rm{hot}} = 10^{5.5} \ \rm{K}$ and $T_{\rm{cold}} = 10^{4} \ \rm{K}$, respectively, to closely align with the mixing temperature $T_{\rm{mix}}$ in the models of \citetalias{Fielding_2022}. The parameter ranges for $P$ and $\mathcal{M}_{\rm{rel}}$ are roughly consistent with the values derived from the \citetalias{Fielding_2022} models within the ESI effective circular aperture, $R_{\rm{aper}}$ (see Figure \ref{fig:fb22_radial_profiles}).

Compared to $\mathcal{M}_{\rm{rel}}$, the flux fraction of [O\,{\footnotesize III}] $\lambda 5007$ is more sensitive to $P$, displaying a higher flux fraction at the low-$P$ end. Intuitively, one might expect that an increase in $P$ correlates with higher $n$ or augmented radiative cooling across all temperature ranges. 
However, due to the complex interplay between radiative cooling, conduction, and viscous heating, $F_{\rm{cool}}$ tends to show a more uneven distribution in the temperature space at higher $P$, with a strong peak at $\sim 10^4 \ \rm{K}$, where the emissivity of [O\,{\footnotesize III}] $\lambda 5007$ is negligible. 
This feature suggests that although the total cooling flux increases with $P$, the fraction of cooling flux at temperatures relevant for [O\,{\footnotesize III}] $\lambda 5007$ ($\sim 10^5$ K) actually decreases.

\section{Simulated Galactic Wind Emission Line Profile} \label{sec:model}

The emissivities of recombination lines or collisionally-excited lines are proportional to \(n(r)^2\). Thus, to simulate emission line profiles based on a galactic wind model, the key steps are to determine the radial density profile \(n(r)\) and to account for projection effects along the line of sight. Additionally, since the emissivities are given in units of luminosity per unit volume (e.g., \(\mathrm{ergs \ s^{-1} \ cm^{-3}}\) in cgs units), a volume integral is required to evaluate the luminosity of both recombination and collisionally-excited lines emitted from each infinitesimal volume of specific geometries, such as a sphere (considered in this study) or a bicone (addressed in future work). 
We do not account for any photoabsorption or radiative line transfer in the simulated emission line profiles for simplification.
 
\begin{figure*}[htbp]
\centering
\includegraphics[width=0.3\textwidth]{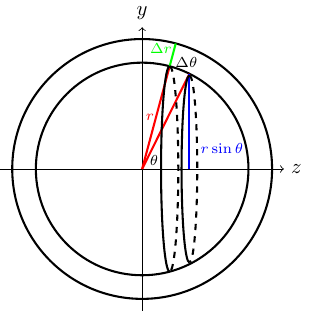}
\includegraphics[width=0.35\textwidth]{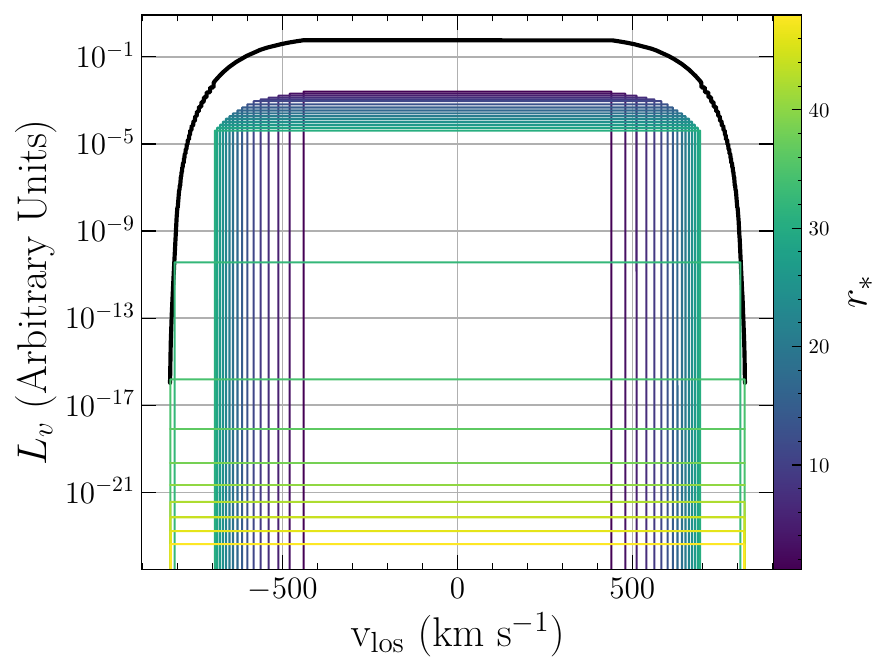}
\includegraphics[width=0.33\textwidth]{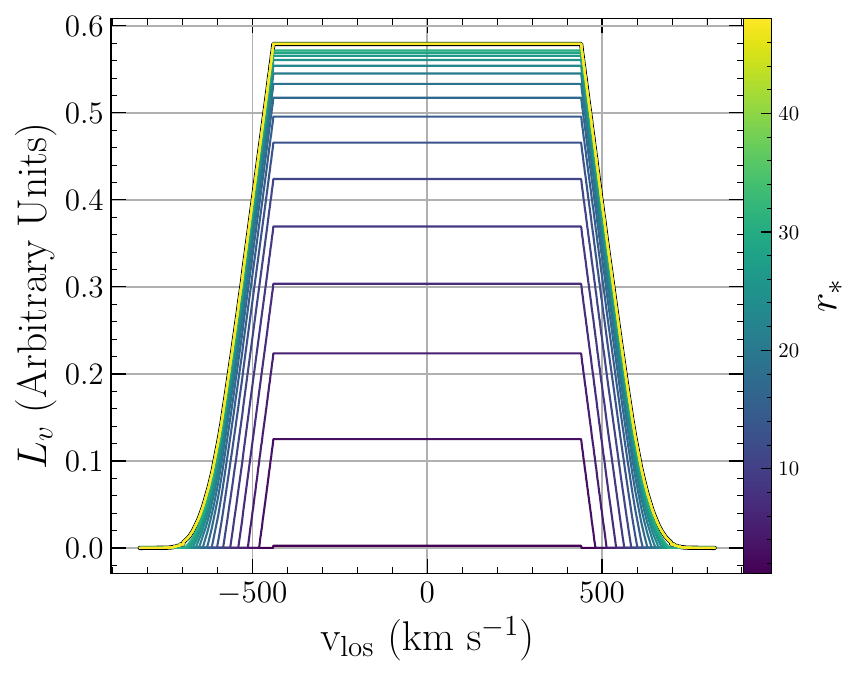}
\caption{{\em Left:} Volume calculation for a thin spherical ring under spherical symmetry. Sightlines are oriented along the $z$-axis, and the cross-section of the spherical shell is depicted in the $y-z$ plane. The ring's radius is \(r \sin{\theta}\), and thus its circumference is \(2 \pi r \sin{\theta}\). The arc length in the \(\theta\) direction is \(r \Delta \theta\), and the shell's thickness in the radial direction is \(\Delta r\). Therefore, the volume, \(\Delta V\), is given by \(2 \pi r^2 \sin{\theta} \Delta r \Delta \theta\). {\em Middle:} Contribution of the luminosity from each spherical shell to a simulated emission line profile based on the \citetalias{Fielding_2022} model that has the same parameter setup as Figure \ref{fig:fb22_radial_profiles} but with $\eta_{{\rm{E}}} = 1.0$ and $\eta_{{\rm{M}}} = 1.0$. The contribution from each shell is color-coded according to its dimensionless radius $r_{\ast} = r / R_{\mathrm{SF}}$ ($1 \leq r_{\ast} \leq 50$). The black solid line represents the luminosity of all spherical shells. The luminosity is depicted on a logarithmic scale for a clearer representation of the contribution from each spherical shell. The luminosity from each spherical shell follows a top-hat function in the $v_{\mathrm{los}}$ space (refer to Equation \ref{eq:2.8} for proof). 
{\em Right:} Cumulative luminosity contributions from spherical shells in the simulated emission line profile. Each cumulative luminosity contribution is color-coded by its corresponding dimensionless radius $r_{\ast} = r / R_{\mathrm{SF}}$. 
\label{fig:spherical_shell_wind_model}}
\end{figure*}


Assuming spherical symmetry, the luminosity of a thin spherical ring (a portion of the spherical shell; see the left panel of Figure \ref{fig:spherical_shell_wind_model}), with radial width ranging from $r$ to $r + \dd r$, polar angle $\theta$ ranging from $\theta$ to $\theta + \dd \theta$, and azimuthal angle $\phi$ ranging from 0 to $2 \pi$ is
\begin{align}
    \label{eq:2.4} \Delta L_{\mathrm{ring}}(r, \theta) &=  \dot{e}(T(r), Z(r), n(r)) \ \Delta V \nonumber \\
    &=  \dot{e}(T(r), Z(r), n(r)) \  (2 \pi \ r^2 \sin{\theta} \ \Delta r \  \Delta \theta  ),
\end{align}
where $\dot{e} (T(r), Z(r), n(r))$ is the interpolated emissivity grid, in units of $\rm{erg \ s^{-1} \ cm^{-3}}$, as a function of radial temperature, gas-phase metallicity, and number density, generated from \citetalias{PS_2020}.

If we choose a coordinate system with the sightlines along the \(z\)-axis, we can relate the line-of-sight velocity to the polar angle \(\theta\) as \(v_{\mathrm{los}} = v(r) \cos{\left(\theta\right)}\). This transformation between \(v_{\mathrm{los}}\) and \(\theta\) can help us simulate the line profile in the \(v_{\mathrm{los}}\) space. Before integration, one important point to mention is that the luminosity of this spherical ring is independent of \(v_{\mathrm{los}}\), which can be proved as follows:
\begin{align}
    \label{eq:2.8} dL_{\mathrm{ring}} (r, \theta) &\propto n(r)^2 \ r^2 \, \sin{\theta} \, d\theta \, dr \nonumber \\ 
    \frac{dL_{\mathrm{ring}} (r, \theta)}{dv_{\mathrm{los}}} &\propto \frac{n(r)^2 \ r^2}{v(r)} \, dr,
\end{align}
indicating the luminosity distribution in the \(v_{\mathrm{los}}\) space is a top-hat function at a specific radius. 
 
To account for the variation of the density profile along the radial direction, \(n(r)\), and the velocity projection effect along the sightline, we need to integrate Equation \ref{eq:2.4} in the radial (from \(r_1\) to \(r_2\)) and the polar (from \(\theta_1\) to \(\theta_2\)) directions as:
\begin{align}
\label{eq:2.6} &L_{\mathrm{ring}} = \int dL_{\mathrm{ring}} (r, \theta) \ dr \ d\theta =  2\pi R_{\rm{SF}}^3  \int_{r_{*1}}^{r_{*2}} \int_{\theta_1 = \cos^{-1}{\left(\frac{v_{\mathrm{los1}}}{v}\right)}}^{\theta_2 = \cos^{-1}{\left(\frac{v_{\mathrm{los2}}}{v}\right)}} \nonumber \\ 
&\dot{e}(T(r_{*}), Z(r_{*}), n(r_{*}))  \,
\rho_{\ast}^2 \ r_{\ast}^2 \, \sin{\left(\theta\right)} \, d\theta \, d\phi \, dr_{*}.
\end{align}
Here we express the polar angle \( \theta \) in terms of \( v_{\mathrm{los}} \) as \( \theta = \cos^{-1}\left(\frac{v_{\mathrm{los}}}{v}\right) \) in the integration bounds of \( \theta \). Thus, \( v_{\mathrm{los1}} \) and \( v_{\mathrm{los2}} \) denote two grid points in the line-of-sight velocity space. 

The middle and right panels of Figure \ref{fig:spherical_shell_wind_model} depict the luminosities of individual and cumulative spherical shells, respectively. Besides the observation that each shell's luminosity remains constant in the \(v_{\mathrm{los}}\) space, it is evident that the inner shells exhibit narrower velocity spreads but contribute more significantly to the overall luminosity compared to the outer shells. This behavior stems from the radial solution of the \citetalias{Fielding_2022} fluid equations, which describes an increasing radial velocity field and a decreasing radial density field of cold clouds.

The assumption of a spherical outflow geometry breaks down for galaxies with a biconical outflow morphology, such as the post-starburst region of J1044+0353 \citep{Peng_2023}. Nevertheless, Equation~\ref{eq:2.6} can still be applied in simulating the line profile of a biconical galactic wind, provided that the half-opening angle and orientation of the bicone can be constrained using integral field spectrographs like KCWI/KCRM. We leave the exploration of the biconical outflow model to future work.

\section{Enthalpy of Galactic Wind Models} \label{sec:enthalpy_model}

\subsection{Enthalpy of \texorpdfstring{\citetalias{Thompson_2016}}{}} \label{subsec:enthalpy_T16}

The advection enthalpy can be written as 
\begin{align}
\frac{\gamma}{\gamma-1} \frac{d(D \dot{M})}{d r}= 
\frac{\gamma}{\gamma-1} \dot{M} \frac{d D}{d r},
\label{eq:8.1}
\end{align}
where $\dot{M}=4 \pi r^{2} \rho v=\text{constant}$ and $D=k_{B} T/ \mu m_{P}$. Based on Equation (29) from \citetalias{Thompson_2016}, we can rewrite $\dd T / \dd r$
\begin{equation}
    \frac{\dd T}{\dd r}=\frac{2 D}{r C_{V}}\left(\frac{v^{2}-v_{e}^{2} / 4}{c_{s}^{2}-v^{2}}\right)+\frac{\dot{q}}{v C_{V}}\left(\frac{c_{T}^{2}-v^{2}}{c_{s}^{2}-v^{2}}\right) \label{eq:8.2}
\end{equation}
as 
\begin{align}
\frac{\dd T}{\dd r}&=
\frac{\gamma-1}{1-\mathcal{M}^{-2}}\left[-2+\frac{v_{e}^{2}}{2 v^{2}}
+\frac{\dot{\varepsilon}}{\rho v^{3} / r}
-\frac{\gamma \dot{\varepsilon}}{\rho c_{s}^{2} v / r}\right] \frac{T}{r}. \label{eq:8.3}
\end{align}
Here $C_{V}=\frac{3 k_{B}}{2 \mu m_{P}}$,  $c_{s}^{2}=\frac{\partial P}{\partial \rho}|_{s}=\gamma \frac{k_{B} T}{\mu m_{p}}=\gamma D$, $c_{T}^{2}=\frac{\partial P}{\partial \rho}|_{T}=\frac{k_{B} T}{\mu m_{p}}= D$, and $ \dot{q}=-\dot{\varepsilon} / \rho$ ($- \dot{\epsilon}$ is the net heating rate).
Therefore, the advection enthalpy is the following 
\begin{align}
\frac{\gamma}{\gamma-1} \dot{M} \frac{\dd D}{\dd r} &= \frac{\gamma-1}{1-\mathcal{M}^{-2}} \left[-2 + \frac{v_{e}^{2}}{2 v^{2}} + \frac{\dot{\varepsilon}}{\rho v / r}\left(\frac{-\gamma}{c_{s}^{2}} + \frac{1}{v^{2}}\right) \right] \nonumber \\
&\quad \frac{D \dot{M}}{r} \frac{\gamma}{\gamma - 1}. \label{eq:8.4}
\end{align}
The advection enthalpy can be classified into three terms with different physical interpretations: 
\begin{enumerate}[leftmargin=*]
\item \underline{\textbf{Radiative Cooling}}: 
\begin{align}
    &\frac{\gamma-1}{1-\mathcal{M}^{-2}} \frac{\dot{\varepsilon}}{\rho v / r}\left(\frac{-\gamma}{c_{s}^{2}}+\frac{1}{v^{2}}\right) \frac{D \dot{M}}{r} \frac{\gamma}{\gamma-1} \nonumber \\
    &= \dot{\varepsilon} 4 \pi r^{2} \frac{1}{1-\mathcal{M}^{-2}}\left(\frac{-\gamma}{c_{s}^{2}}+\frac{1}{v^{2}}\right) c_{s}^{2}=\color{lightblue}{-\frac{\gamma-\mathcal{M}^{-2}}{1-\mathcal{M}^{-2}} \dot{\varepsilon} 4 \pi r^{2}} \label{eq:8.5}
\end{align}
\item \underline{\textbf{Adiabatic Loss}}: 
\begin{align}
    -2\left[\frac{\gamma-1}{1-\mathcal{M}^{-2}}\right] \frac{D \dot{M}}{r} \frac{\gamma}{\gamma-1}= \color{crimson}{\frac{c_s^2}{1-\mathcal{M}^{-2}} \frac{-2 \dot{M}}{r}} \label{eq:8.6}
\end{align}
\item \underline{\textbf{Gravitational Heating}}: 
\begin{align}
\frac{\gamma-1}{1-\mathcal{M}^{-2}}\left[\frac{v_{e}^{2}}{2 v^{2}}\right] \frac{D \dot{M}}{r} \frac{\gamma}{\gamma-1}= \frac{c_{s}^{2}}{1-\mathcal{M}^{-2}} \left(\frac{v_{e}^{2}}{2 v^{2}}\right) \frac{\dot{M}}{r} \label{eq:8.7}
\end{align}
\end{enumerate}
The radiative cooling and adiabatic terms are highlighted in their corresponding colors in Figure \ref{fig:entropy_plots}. We can integrate a specific term in the advection enthalpy over the radial distance, $\dd r$, to obtain the total luminosity of that term within a certain radius.

\subsection{Enthalpy of \texorpdfstring{\citetalias{Fielding_2022}}{}} \label{subsec:enthalpy_fb22}
Following a similar procedure as shown in Section \ref{subsec:enthalpy_T16}, we would obtain similar expressions for the radiative cooling, adiabatic loss, and gravitational heating terms for \citetalias{Fielding_2022}. 
The kinetic energy thermalization (pink) and thermal energy mixing (orange) terms are highlighted in their corresponding colors in Figure \ref{fig:entropy_plots}, and their expressions are shown in Equation (26) of \citetalias{Fielding_2022}.
There are several additional terms due to the mass, momentum, and energy transfer between the hot wind and the cold clouds:
\begin{enumerate}[leftmargin=*]
\item \underline{\textbf{Drag Work}}: 
\begin{align}
\frac{\gamma - \mathcal{M}^{-2}}{1 - \mathcal{M}^{-2}} \dot{p}_{\text{drag}}\left(v-v_{\rm{cl}}\right) 4 \pi r^2
\label{eq:8.9}
\end{align}
\item \underline{\textbf{Mass Transfer}}: 
\begin{align}
\frac{c_s^2}{1-\mathcal{M}^{-2}} \frac{\dot{M}}{r}\left[\frac{\dot{\rho}_{+}-\dot{\rho}_{-}}{\rho v / r}\right] \label{eq:8.10}
\end{align}
\item \underline{\textbf{Momentum Transfer}}: 
\begin{align}
\frac{c_{s}^{2}}{1-\mathcal{M}^{-2}} \frac{\dot{M}}{r} \frac{\dot{\rho}_{+}}{\rho v / r} \frac{v-v_{\rm{cl}}}{v}
\label{eq:8.11}
\end{align}
\item \underline{\textbf{Drag force}}: 
\begin{align}
\frac{c_{s}^{2}}{1-\mathcal{M}^{-2}} \frac{\dot{M}}{r} \frac{\dot{p}_{\rm{drag}}}{\rho v^{2} / r}
\label{eq:8.12}
\end{align}
\end{enumerate}
We refer readers to \citetalias{Fielding_2022} for the detailed definition of each parameter shown from Equations \ref{eq:8.9} to \ref{eq:8.12}.

\section{Entropy of Galactic Wind Models} \label{sec:entropy_model}

\subsection{Entropy of \texorpdfstring{\citetalias{Thompson_2016}}{}}\label{subsec:entropy_model_t16}
Following the entropy $K=P / \rho^{\gamma}$ defined in \citetalias{Fielding_2022}, we have the following logarithmic entropy derivative with respect to radial distance
\begin{equation}
\frac{\dd \log K}{\dd \log r}=\frac{\dd \log P}{\dd \log r}-\gamma \frac{\dd \log \rho}{\dd \log r}. \label{eq:7.1}
\end{equation}
Given that $\frac{\dd \log T}{\dd \log r}=\frac{\dd \log P}{\dd \log r}-\frac{\dd \log \rho}{\dd \log r}$, Equation \ref{eq:7.1} can be written as
\begin{equation}
\frac{\dd \log K}{\dd \log r}=\frac{\dd \log T}{\dd \log r}-(\gamma-1) \frac{\dd \log \rho}{\dd \log r}, \label{eq:7.2}
\end{equation}
where $\frac{\dd \log T}{\dd \log r}=\frac{\dd T}{\dd r} \frac{r}{T}$ and $\frac{\dd \log \rho}{\dd \log r}=\frac{\dd \rho}{\dd r} \frac{r}{\rho}$. Substituting $\dd \rho/ \dd r$ and $\dd T/ \dd r$ with Equations 28 and 29 in \cite{Thompson_2016}, we would obtain 
\begin{align}
&\frac{\dd \log K}{\dd \log r}=\frac{\gamma-1}{1-\mathcal{M}^{-2}}\left[-2+\frac{v_{e}^{2}}{2 v^{2}}
+\frac{\dot{\varepsilon}}{\rho v^{3} / r}
-\frac{\gamma \dot{\varepsilon}}{\rho c_{s}^{2} v / r}\right] \nonumber \\
&-(\gamma-1)\frac{r}{\rho}\left[\frac{2 \rho}{r}(\frac{v^{2}-v_{e}^{2} / 4}{c_{s}^{2}-v^{2}})+\frac{\rho}{v} \frac{D}{C_{V} T} \left(\frac{\dot{\epsilon} / \rho}{c_{s}^{2}-v^{2}}\right)\right]. \label{eq:7.3}
\end{align}
Since we know that $\frac{\gamma D}{v^{2}}-1=\mathcal{M}^{-2}-1$, Equation \ref{eq:7.3} can be simplified as 
\begin{align}
&\frac{\dd \log K}{\dd \log r}=\frac{\gamma-1}{1-\mathcal{M}^{-2}}\left[-2+\frac{v_{e}^{2}}{2 v^{2}}
-\frac{\dot{\varepsilon}}{\rho v / r} \frac{1}{c_{s}^{2}}\left(\gamma-\mathcal{M}^{-2}\right)\right] \nonumber \\
&+\frac{(\gamma-1)}{1-\mathcal{M}^{-2}}\left[2\left(1-\frac{v_{e}^{2}}{4 v^{2}}\right)+\frac{r}{v} \frac{D}{C_{V} T} \frac{\dot{\varepsilon}}{\rho v^{2}}\right]. \label{eq:7.4}
\end{align}
Combining terms in Equation \ref{eq:7.4}, we would get
\begin{align}
&\frac{\dd \log K}{\dd \log r}= \frac{\gamma-1}{1-\mathcal{M}^{-2}}\left[\frac{-\dot{\varepsilon}}{\rho c_{s}^{2} v / r}\left(\gamma-\mathcal{M}^{-2}\right)
+(\gamma-1) \frac{\dot{\varepsilon}}{\rho v^{3} / r}\right] \nonumber \\
&=-(\gamma-1) \gamma \frac{\dot{\varepsilon}}{\rho c_{s}^{2} v / r}=\frac{-(\gamma-1) \dot{\varepsilon}}{P v / r}. \label{eq:7.6}
\end{align}
Inspired by Equation 16 in \citetalias{Fielding_2022}, we could rewrite Equation \ref{eq:7.6} as 
\begin{align}
\frac{\dd \log K}{\dd \log r}=\frac{-(\gamma-1) \dot{\varepsilon}}{P} \frac{r}{v}={\color{lightblue}{-\frac{t_{\rm{adv}}}{t_{\rm{cool}}}}}, \label{eq:7.7}
\end{align}
where $t_{\rm{adv}} = r / v$ and $t_{\rm{cool}}=P/ (\gamma-1) \dot{\varepsilon}$. Therefore, $\dd \log K / \dd \log r$ is the ratio between the wind advection time $t_{\rm{adv}}$ and the wind cooling time $t_{\rm{cool}}$. Here $\dot{\varepsilon}$ is defined the same as the parameter $\mathcal{L}$ in \citetalias{Fielding_2022}.

\subsection{Entropy of \texorpdfstring{\citetalias{Fielding_2022}}{}}\label{subsec:entropy_model_fb22}

Given the equations given in Figure 3 from \citetalias{Fielding_2022},
\begin{align}
&\frac{\partial \log K}{\partial \log r} = \frac{\gamma}{1-\mathcal{M}^{-2}}\left[\frac{\dot{\rho}_{+}}{\rho v / r} \left( \frac{\gamma - 1}{2} \left( \frac{(v - v_{\rm{cl}})^2}{c_s^2} 
\right. \right. \right. \nonumber \\
&- \left. \left.  \frac{(v - v_{\rm{cl}})^2}{v^2} \right)
+ \left( c_s^2 - c_{s,\rm{cl}}^2 \right) \left( \frac{1}{v^2} - \frac{1}{c_s^2} \right) \right) \nonumber \\
&+ (\gamma - 1) \frac{\dot{\epsilon}}{\rho v / r} \left(\frac{1}{v^2} - \frac{1}{c_s^2} \right) \nonumber \\
&+ \left. \frac{(\gamma - 1) \dot{p}_{\text{drag}} (v - v_{\rm{cl}})}{\rho v / r} \left( \frac{1}{c_s^2} - \frac{1}{v^2} \right) \right]. \label{eq:7.8}
\end{align}
Equation \ref{eq:7.8} can be simplified as follows:
\begin{align}
&\frac{\partial \log K}{\partial \log r} = \gamma \frac{\dot{\rho}_{+}}{\rho v / r}\left[\frac{\gamma-1}{2}\left(\frac{\left(v-v_{\rm{cl}}\right)^{2}}{c_{s}{ }^{2}}\right) \right. - \left. \frac{\left(c_{s}{ }^{2}-c_{s, \rm{cl}}{ }^{2}\right)}{c_{s}{ }^{2}}\right] \nonumber \\
&- (\gamma-1) \frac{\dot{\varepsilon}}{\rho v / r} \frac{\rho}{P} + \frac{(\gamma - 1) \dot{p}_{\text{drag}} (v - v_{\rm{cl}})}{P v / r} \nonumber \\
& = {\color{lightblue}{-\frac{t_{\rm{adv}}}{t_{\rm{cool}}}}} + \gamma \frac{\dot{\rho}_{+}}{\rho v / r}\left[{\color{lightpink}{\frac{\gamma-1}{2}\left(\frac{\left(v-v_{\rm{cl}}\right)^{2}}{c_{s}{ }^{2}}\right)}} \right. \nonumber \\
&- \left. {\color{orange}{\frac{\left(c_{s}{ }^{2}-c_{s, \rm{cl}}{ }^{2}\right)}{c_{s}{ }^{2}}}}\right] + \frac{(\gamma - 1) \dot{p}_{\text{drag}} (v - v_{\rm{cl}})}{P v / r}.
\label{eq:7.9}
\end{align}
Compared to Equation \ref{eq:7.7}, the ratio between $t_{\rm{adv}}$ and $t_{\rm{cool}}$ appears in both entropy equations; however, the \citetalias{Fielding_2022} model includes three additional terms, kinetic energy thermalization (pink), thermal energy mixing (orange), and drag work, due to the turbulent radiative mixing between hot winds and cold clouds (see Figure \ref{fig:entropy_plots}).

\begin{figure*}[ht]
\centering
\includegraphics[width=1.0\textwidth]{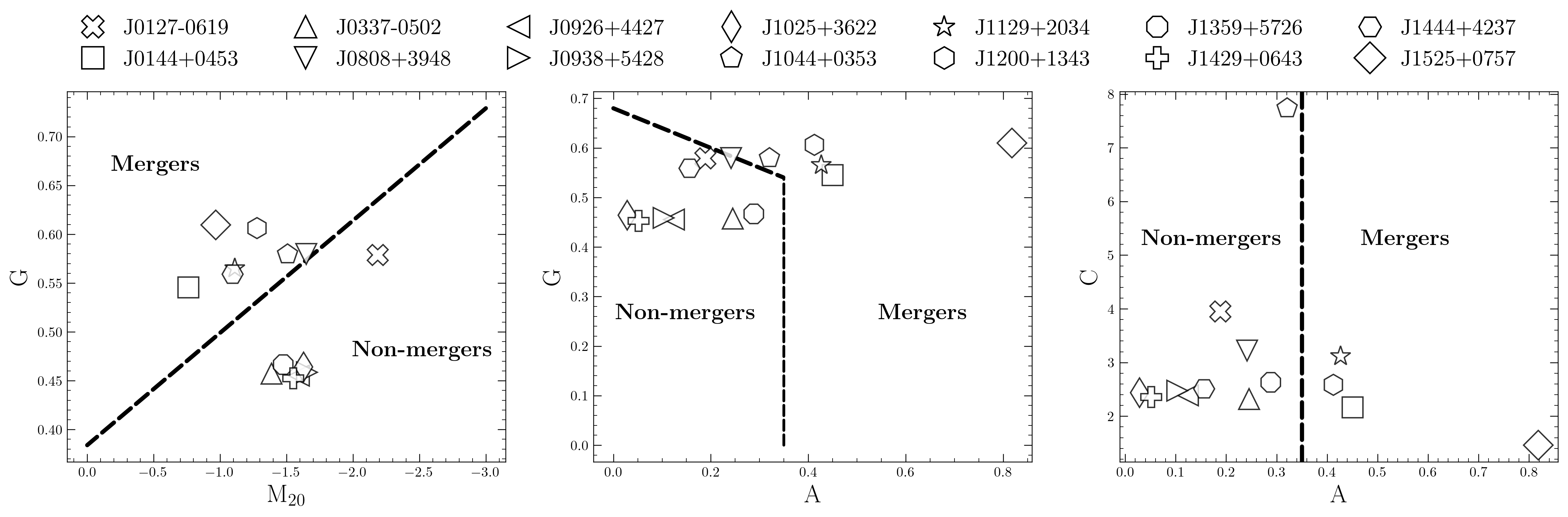}
\caption{Merger classification using the $G - M_{20}$, $G - A$, and $C - A$ relations for local galaxies \citep{Lotz_2004, Lotz_2008}. Galaxies are considered mergers if they lie above the black dashed lines in the $G - M_{20}$ panel and to the right of the black dashed lines in the $G - A$ and $C - A$ panels, respectively. 
\label{fig:merger_classify}}
\end{figure*} 

\section{Merger-Morphology Indicators} \label{sec:merger_indicators}
We leverage the \textsc{Python} package \texttt{statmorph} \citep{statmorph} to calculate the merger-sensitive statistics, including the Concentration, Asymmetry, and Smoothness (CAS) \citep{Conselice_2003} and the Gini-$M_{20}$ \citep{Lotz_2004} statistics, on the DECalS g band image of each galaxy to help us classify our sample into the (1) merger-like and (2) non-merger categories. 

We summarize the essential steps involved in morphological analysis using \texttt{statmorph} and refer readers to \citet{statmorph} for detailed information. We initially define the region of interest around the center of each galaxy (see Figure \ref{fig:esi_aperture}) with a size of approximately \(24\farcs0 \times 24\farcs0\) in the DECaLS g-band image. Subsequently, we generate a segmentation map by establishing the detection threshold for identifying galaxy signals at 3$\sigma$ using \texttt{Photutils} \citep{larry_bradley_2022_6825092}. Upon obtaining the Point Spread Function (PSF) and the corresponding error image for each galaxy from the DESI Sky Viewer, \texttt{statmorph} is utilized to quantify the morphology of the sources from the segmentation map.

We consider galaxies as potential mergers if they satisfy at least one of the demarcation relations ($G - M_{20}$, $G - A$, and $C - A$) for local galaxies as outlined in \citet{Lotz_2004} and \citet{Lotz_2008}. As shown in Figure \ref{fig:merger_classify}, this criterion results in exactly half of our sample (7 out of 14) being categorized as potential mergers. We confirm that this classification remains consistent when using the DECaLS r-band images instead. 

To be cautious, J0337+0502 (alternatively known as SBS 0335–052E in the literature), which is identified as forming a galaxy pair with SBS 0335–052W with a projected separation of $\sim$ 22 kpc \citep{Ekta_2009, Herenz_2023}, is classified as a ``non-merger'' galaxy in our study. This inconsistency could be due to (1) our morphological analysis, detailed in Appendix \ref{sec:merger_indicators}, focuses solely on a zoom-in 24\farcs0 $\times$ 24\farcs0 region around J0337+0502 of the strongly-interacting system and (2) morphology disturbances will not be clearly visible if the galaxy pair is not in the first-pericenter pass or the final coalescence stage of interaction \citep{Lotz_2008}. Since \cite{Ekta_2009} suggest this interacting system has surpassed its initial close encounter—evidenced by merger features such as short ``tails'' and a faint ``bridge'', and supported by comparisons with collisionless N-body merger simulations—we argue that J0337+0502 may currently be transitioning from the first pericenter to the maximal separation \cite[the positions of its morphology indicators also support this point; see Figure \ref{fig:merger_classify} and Figure 5 of][]{Lotz_2008}. 

\bibliography{sample631}{}
\bibliographystyle{aasjournal}



\end{document}